\RequirePackage{lineno} 
\documentclass[aps,prc,superscriptaddress,nofootinbib,showpacs,floatfix,singlecolumn]{revtex4}
\usepackage{graphicx} 
\usepackage{float} 
\usepackage{amssymb}
\usepackage{url}
\usepackage[colorlinks,linkcolor=blue,citecolor=blue,filecolor=black,urlcolor=blue]{hyperref}
\usepackage{cancel}
\usepackage{MnSymbol}
\usepackage{multirow}
\usepackage{bm}

\newcommand{\lr}[1]{\left\langle #1\right\rangle}
\newcommand{\lrave}[1]{\left[ #1\right]}

\newcommand{\pT} {\ensuremath{p_{\mathrm{T}}}}

\newcommand{\pp}{\mbox{$pp$}}

\newcommand{\sca}{\mathrm{SC}(2,3)}
\newcommand{\scb}{\mathrm{SC}(2,4)}
\newcommand{\sumet}{\Sigma E_{\textrm T}}
\newcommand{\nch}{\mbox{$N_{\mathrm{ch}}$}}

\newcommand{\nchrec}{\mbox{$N_{\mathrm{ch}}^{\mathrm{rec}}$}}
\newcommand{\ns}{N_{\mathrm{s}}}
\newcommand{\npart}{N_{\mathrm{w}}}
\newcommand{\npartf}{N_{\mathrm{w}}^{\mathrm{F}}}
\newcommand{\npartb}{N_{\mathrm{w}}^{\mathrm{B}}}
\newcommand{\ntwoc}{N_{\mathrm{an}}}
\newcommand{\cobs}{\mathrm{cent}_{\mathrm{obs}}}

\newcommand{\nqp}{N_{\mathrm{qp}}}
\newcommand{\pnbd}{p_{\mathrm{nbd}}}
\newcommand{\ncoll}{N_{\mathrm{bin}}}
\newcommand{\sqrtnn}{\mbox{$\sqrt{s_{\mathrm{NN}}}$}}

\usepackage{color}
\usepackage{harpoon}
\definecolor{my}{rgb}{1, 0, 0}
\begin{document} 
\title{Centrality fluctuations in heavy-ion collisions}
 \newcommand{\sunysb}{Department of Chemistry, Stony Brook University, Stony Brook, NY 11794, USA}
 \newcommand{\bnl}{Physics Department, Brookhaven National Laboratory, Upton, NY 11796, USA}
\author{Mingliang Zhou}\affiliation{\sunysb}
\author{Jiangyong Jia}\email[Correspond to\ ]{jjia@bnl.gov} \affiliation{\sunysb}\affiliation{\bnl}
\date{\today}
\begin{abstract}
Volume or centrality fluctuations (CF) is one of the main uncertainties for interpreting the centrality dependence of many experimental observables. The CF is constrained by centrality selection based on particle multiplicity in a reference subevent, and contributes to observables measured in another subevent. Using a Glauber-based independent source model, we study the influence of CF on several distributions of multiplicity $N$ and eccentricities $\epsilon_n$: $p(N)$, $p(\epsilon_n)$, $p(\epsilon_n,\epsilon_m)$ and $p(N,\epsilon_n)$, where the effects of CF is quantified using multi-particle cumulants of these distributions. In mid-central collisions, a general relation is established between the multiplicity fluctuation and resulting CF in the reference subevent. In ultra-central collisions, where distribution of particle production sources is strongly distorted, we find these cumulants exhibit rich sign-change patterns, due to observable-dependent non-Gaussianity in the underlying distributions. The details of sign-change pattern change with the size of the collision systems. Simultaneous comparison of these different types cumulants between model prediction and experimental data can be used to constrain the CF and particle production mechanism in heavy-ion collisions. Since the concept of centrality and CF are expected to fluctuate in the longitudinal direction within a single event, we propose to use pseudorapidity-separated subevent cumulant method to explore the nature of intra-event fluctuations of centrality and collective dynamics. The subevent method can be applied for any bulk observable that is sensitive to centrality, and has the potential to separate different mechanisms for multiplicity and flow fluctuations happening at different time scales. The forward detector upgrades at RHIC and LHC will greatly enhance such studies in the future.
\end{abstract}
\maketitle
\section{Introduction}\label{sec:1}
Centrality is an important concept for heavy-ion collisions, which characterizes the amount of overlap or size of the fireball in the collision region. Conceptually, the definition of centrality is not unique, people often use~\cite{Miller:2007ri,Loizides:2014vua,Adler:2013aqf} 1) the number of nucleons $\npart$ in the overlap region, also known as participants or wounded nucleons, 2) the two-component model where the event activity is theorized to be proportional to a linear combination of $\npart$ and number of binary nucleon-nucleon collisions $\ncoll$: $\ntwoc\equiv (1-x)\npart/2+x\ncoll$, with $x$ being a tunable constant,  3) the number of constituent-quark participants $\nqp$ in the overlap region. Since these quantities, generally referred to as the number of sources $\ns$, are not directly measurable, a Glauber model that include nuclear geometry and particle production is often used to connect the $\ns$ with the experimentally measured event activity $\cobs$, such as the number of charge particles $\nch$ or the total transverse energy $\sumet$ in a given rapidity range. Glauber model also provides estimates for many other parameters that describe the initial collision geometry, such as eccentricities $\epsilon_n$, which describe the azimuthal asymmetry in the distribution of the sources in the transverse plane. 

In data analysis, the centrality estimator is usually defined as the reference particle multiplicity $N_{\mathrm{A}}$ in a forward pseudorapidity window A, $\cobs\equiv N_{\mathrm{A}}$, and the observables of interest are measured using particles in a different pseudorapidity window B, usually around mid-rapidity. Due to fluctuations in particle production, events with the same $\ns$ may have different values of $\cobs$. Conversely, events selected with the same $\cobs$ can have different values of $\ns$. If a physics observable measured in the subevent B changes with $\ns$, its fluctuation would be affected by $\ns$ fluctuation associated with centrality selection defined on $N_{\mathrm{A}}$. The fluctuation of $\ns$ for fixed $\cobs$ value is commonly referred to as ``volume fluctuations''~\cite{Jeon:2003gk,Skokov:2012ds,Luo:2013bmi}, which is an irreducible ``centrality fluctuations'' (CF). The CF is large in peripheral collisions or small collision systems where it often dominates the uncertainties in $\ns$ estimation. The CF is expected to be strongly distorted in ultra-central collisions (UCC) due to the steeply falling distribution of $p(\ns)$~\cite{Skokov:2012ds,Xu:2016qzd,Bzdak:2016jxo}. The CF also contributes to the measurement of event-by-event fluctuations of conserved quantities, and is one of the main source of model uncertainty for extraction of the final state dynamical fluctuations~\cite{Aggarwal:2010wy,Adamczyk:2013dal,Adamczyk:2014fia} associated with the critical end point in the QCD phase diagram~\cite{Luo:2017faz,Li:2017via}. Experimental measurement of the CF helps to clarify the meaning of centrality and provide insights on the sources for particle production in heavy-ion collisions.

Any observable that is sensitive to $p(\ns)$ fluctuation can be used to study the CF through the multi-particle cumulants of this observable. Besides the multiplicity fluctuation $p(N)$, we also use the fluctuation of harmonic flow $v_n$ to probe the CF effects. The basic idea is the following: hydrodynamical simulations show that $v_n$ are driven by the eccentricity $\epsilon_n$ of the initial collision geometry, $v_n\propto \epsilon_n$ for $n=2$ and 3~\cite{Qiu:2011iv,Gardim:2011xv,Niemi:2012aj}. Since the sources determining the event centrality also control $\epsilon_n$ of the event, the fluctuation of $\ns$ from CF gives rise to additional fluctuations in $\epsilon_n$, which in turn generates additional fluctuations in $v_n$. 

This paper presents a detailed study of the centrality fluctuations and their influences on multiplicity and eccentricity fluctuations. Several types of cumulants are constructed for various fluctuation observables, including $p(N)$, $p(\epsilon_n)$, correlation between eccentricities of different order $p(\epsilon_n,\epsilon_m)$, as well as correlation between multiplicity and eccentricity $p(N,\epsilon_n)$. An independent source model based on standard Glauber model is used to generate reference multiplicity used for centrality selection, as well as particles used to calculate physics observables. The CF is induced by placing a narrow selection on reference multiplicity, and its influences on multi-particle cumulants in another subevent are calculated. Special focus is given to UCC collisions, where the CF and subsequently $p(\epsilon_n)$ are expected to be non-Gaussian due to the boundary effects imposed by steeply falling $p(\ns)$ distribution.

There has been extensive studies of CF using multiplicity cumulants (see Ref.\cite{Asakawa:2015ybt,Luo:2017faz} for a review). Skokov et.al.\cite{Skokov:2012ds} first pointed out the importance of CF for multiplicity fluctuation measurement, and derived a general formula relating multiplicity cumulants to the CF within an independent source model framework. Most studies focused on the impact of CF on cumulants for conserved charge, e.g. net-proton, for the search of CEP in the RHIC beam energy scan program\cite{Luo:2013bmi,Braun-Munzinger:2016yjz,Kitazawa:2017ljq}. Extending earlier work of Refs.~\cite{Xu:2016qzd,Xu:2016jaz}, we derive a general relation between multiplicity cumulants and CF in subevent A, and discuss how this CF contributes to the cumulants in subevent B. We find that the second-order cumulant or scaled-variance of the CF is not sensitive to the particle distribution for each source $p(n)$ and is mainly determined by $p(\ns)$. In contrast, the higher-order cumulants of CF are also affected by $p(n)$, except in UCC collisions where they are mostly controlled by $p(\ns)$.

Recently, ATLAS Collaboration observed a characteristic sign-change of four-particle cumulant for $v_2$, $c_2\{4\}$, in ultra-central Pb+Pb collisions~\cite{ATLAS:2017zcm}. The location and magnitude of the positive $c_2\{4\}$ depend on the rapidity range used to define centrality. We show that this sign-change behavior is related to the centrality smearing effects from CF. We predict similar sign-change behavior for $c_3\{4\}$ and symmetric cumulants $\sca$ and $\scb$, as well as more complex sign-change pattens for higher-order cumulants $c_2\{6\}$ and $c_2\{8\}$. We find that the magnitude and detailed pattern of sign-change are sensitive to the fluctuation of $\ns$. We also carry out a study of the mixed correlation between multiplicity and eccentricity, and predict significant positive correlation between the two. We argue that by exploring the UCC in different collision systems and as a function of $\eta$, one can use flow and multiplicity cumulants to constrain the longitudinal dynamics of particle production.

The structure of the paper is as follows. Section~\ref{sec:2} introduces the independent source model. Section~\ref{sec:3} presents the results on the multiplicity cumulants, focusing on the CF arising from the centrality definition and its limiting behaviors in UCC. Section~\ref{sec:4} presents results on the eccentricity cumulants, which probes the probability distribution $p(\epsilon_n)$ and $p(\epsilon_n,\epsilon_m)$, and are found to be very sensitive to the CF. Results on the multiplicity-eccentricity mixed cumulants for $p(N,\epsilon_n)$ are discussed in Section~\ref{sec:5}. In section~\ref{sec:6}, we extended above studies to other smaller collision systems. In Section~\ref{sec:7}, we introduce the subevent cumulant method, recently successfully applied for flow correlations~\cite{Jia:2017hbm,Aaboud:2017blb,Huo:2017nms,Nie:2018xog}, to multiplicity cumulants and mixed cumulants. We argue that the subevent methods are very useful in probing the longitudinal structure and particle production mechanism in heavy-ion collisions. We then summarize and discuss the main results in Section~\ref{sec:8}. Details on derivation of some of the formulas are given in Appendix.
\section{Independent source model and centrality fluctuations}\label{sec:2}

Particle production is simulated with a simple independent source model, where the total particle multiplicity in each A+A collision is calculated as a sum of particles from each source via a common probability distribution. The sources could be participating nucleons, those given by two-component model, or participating constituent quarks and they are generated using a Glauber model framework~\cite{Miller:2007ri}. Quantities describing the collision geometry, such as the transverse area or the eccentricities $\epsilon_n$, can be obtained from transverse positions $(x,y)$ of the sources. The present study does not model explicitly the sub-nucleonic degree-of-freedom, and the number of sources $\ns$ therefore represents either the $\npart$ (the wounded nucleon or WN model) or $\ntwoc=(1-x)\npart/2+x\ncoll$ (the two-component model). However, as pointed out in Ref.~\cite{Adler:2013aqf}, the $\ntwoc$ and associated nuclear geometry is a good proxy for the number of participating constituent quarks and their associated nuclear geometry. We leave the explicit study of sub-nucleonic degree-of-freedom to future work.

The particle production from each source is assumed to follow a negative binomial distribution (NBD):
\begin{eqnarray}
\label{eq:1}
\pnbd(n;m,p) = \frac{(n+m-1)!}{(m-1)!n!} p^n(1-p)^m,\; p = \frac{\bar{n}}{\bar{n}+m}
\end{eqnarray}
where $\bar{n}$ is the average number of particle in the acceptance, and $p$ is the probability of a particle falling into the acceptance. This form has been widely used to describe the multiplicity distributions in $pp$ collisions~\cite{Giovannini:1985mz,Ghosh:2012xh}. One important property of the NBD for our study is its relative width $\hat{\sigma}$:
\begin{eqnarray}
\label{eq:5}
\hat{\sigma}^2 \equiv \frac{\lr{(n-\bar{n})^2}}{\bar{n}^2} = \frac{1}{\bar{n}}+\frac{1}{m}\;.
\end{eqnarray}
which controls the strength of the fluctuation for each source. 

The distribution of total multiplicity is obtained as a superposition of the NBD distributions from all sources:
\begin{eqnarray}
\label{eq:2}
p(N;\ns) =\pnbd(n_1;m,p)\otimes \pnbd(n_2;m,p)\otimes \ldots \otimes \pnbd(n_{_{\ns}};m,p)  = \pnbd(N; m\ns,p)\;, N \equiv \sum_{i=1}^{\ns} n_i.
\end{eqnarray}
where we used the additive nature of the NBD distributions for convolution. One interesting consequence of this feature is that we can subdivide the multiplicity distribution into sources with smaller number of particles:
\begin{eqnarray}
\label{eq:3}
\pnbd(N; m\ns,p)  = \pnbd(n_1;m/k,p)\otimes \pnbd(n_2;m/k,p)\otimes \ldots \otimes \pnbd(n_{k_{\ns}};m/k,p)\;,
\end{eqnarray}
where each source is subdivided into $k$ identical sources with smaller average $\bar{n}/k$ but the same $p$, without changing the total multiplicity distribution. In this case, a Glauber model with and without explicit treatment of sub-nucleonic degree-of-freedom would be identical to each other, unless $k$ is allowed to fluctuate for each wounded nucleon.

The distribution of sources in the collision zone is described by a standard Glauber model for various collision systems. The nucleons are assumed to have a hard-core of 0.3 fm in radii, their transverse positions are generated according to the Woods-Saxon distribution as provided by Ref.~\cite{Loizides:2014vua}.  A nucleon-nucleon cross-section of $\sigma=68$~mb is used to simulate the collisions at $\sqrtnn=5.02$ TeV. The usual geometric quantities, including $\npart$, $\ncoll$, $\epsilon_n$ are calculated for each event. The $\ntwoc$ in the two-component model is given by choosing $x=0.09$, very close to those used at the top RHIC energy~\cite{Adler:2013aqf}, which was shown to approximately describe the multiplicity distribution in A+A collisions~\cite{Adler:2013aqf,Abelev:2013qoq}. Combined with particle production from Eq.~\ref{eq:1}, we obtain the distributions for multiplicity and eccentricities: $p(N)$, $p(\epsilon_n)$, $p(\epsilon_n,\epsilon_m)$, and $p(N,\epsilon_n)$. The shape of these distributions are characterized by multi-particle cumulant observables, which are described in later sections.

\begin{table}[!t]
\centering
\caption{The various parameters sets for the NBD (Eq.~\ref{eq:1}) used for modeling the particle production in the wounded nucleon model (left) and two-component model (right). The parameters with apostrophe have the same $\hat{\sigma}$ but smaller $\bar{n}$ than the corresponding ones without apostrophe (e.g. Par0 vs Par0$'$).}
\label{tab:1}
\begin{tabular}{|c|c|c|c|c|}\hline 
\multicolumn{5}{|c|}{Wounded nucleon model}\\\hline
         &   $p$ &   $m$ & mean $\bar{n}$& RMS/mean $\hat{\sigma}$ \\\hline
Par0     & 0.688 & 3.45  &      7.6     & 0.65\\\hline
Par1     & 0.831 & 1.55  &      7.6     & 0.88\\\hline
Par2     & 0.928 & 0.593 &      7.6     & 1.35\\\hline
Par0$'$ & 0.351 & 6.77  &      3.6     & 0.65\\\hline
Par1$'$ & 0.644 & 2.00  &      3.6     & 0.88\\\hline
Par2$'$ & 0.849 & 0.647 &      3.6     & 1.35\\\hline
\end{tabular}
\begin{tabular}{|c|c|c|c|c|}\hline 
\multicolumn{5}{|c|}{Two-component model}\\\hline
         &   $p$ &   $m$ & mean $\bar{n}$& RMS/mean $\hat{\sigma}$ \\\hline
Par0     & 0.391 & 13.7  &      8.7     & 0.43\\\hline
Par1     & 0.738 & 3.10  &      8.7     & 0.66\\\hline
Par2     & 0.909 & 0.878 &      8.7     & 1.12\\\hline
Par0$'$ & 0.063 & 85.7  &      5.7     & 0.43\\\hline
Par1$'$ & 0.596 & 3.84  &      5.7     & 0.66\\\hline
Par2$'$ & 0.860 & 0.927 &      5.7     & 1.12\\\hline
\end{tabular}
\end{table}
\begin{figure}[h!]
\begin{center}
\includegraphics[width=0.48\linewidth]{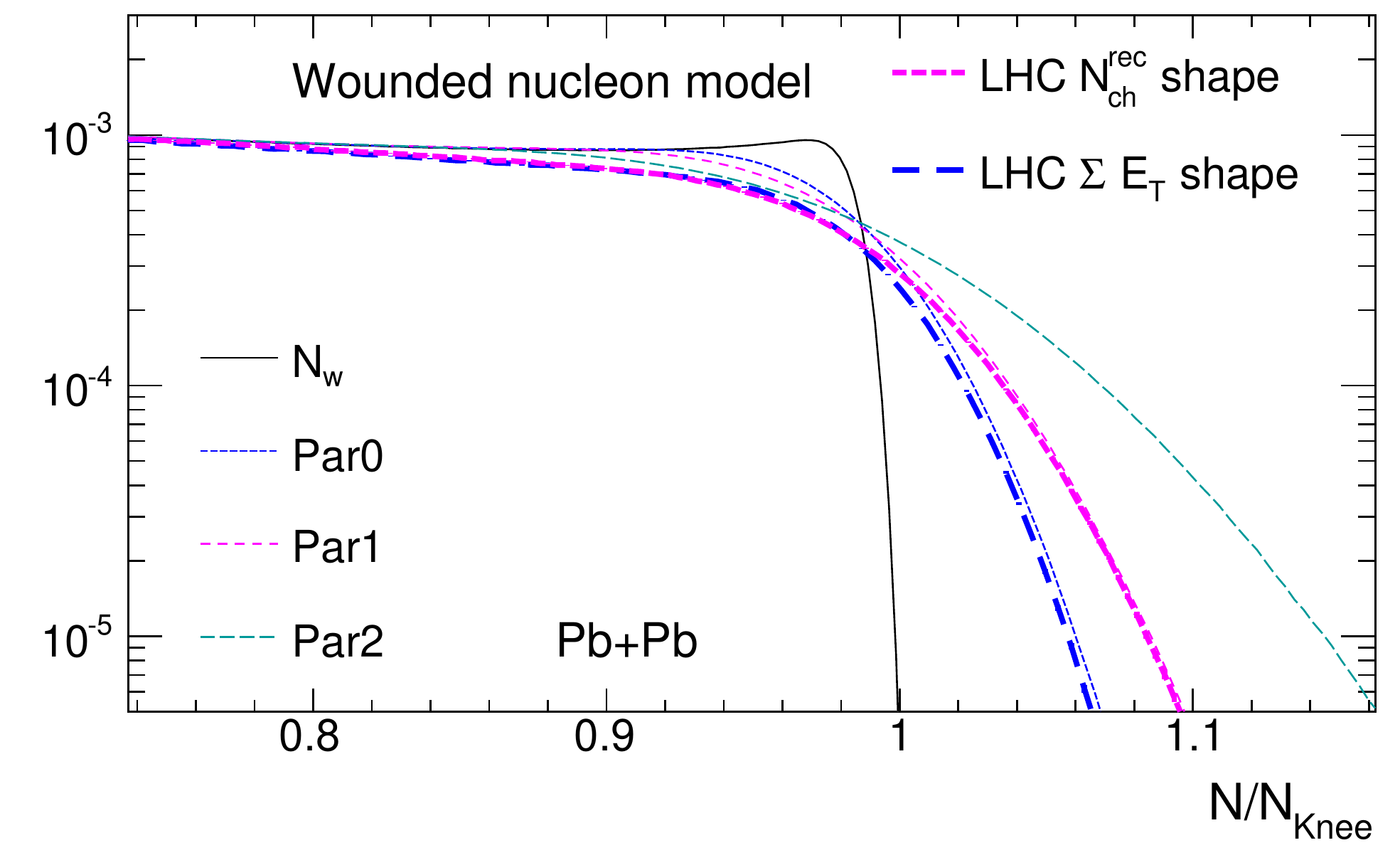}\includegraphics[width=0.48\linewidth]{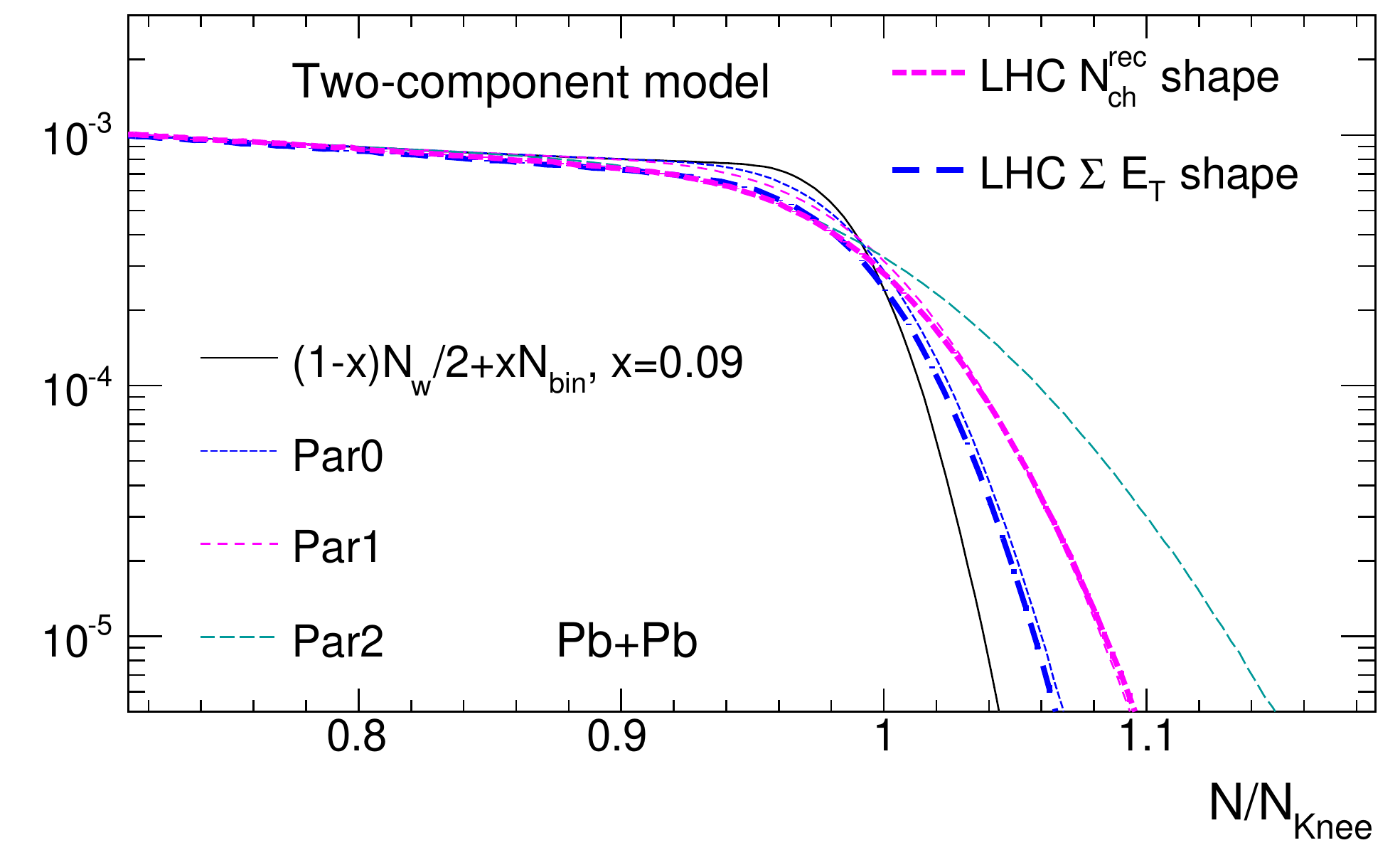}
\end{center}
\caption{\label{fig:1} The distributions of sources and produced particles based on Par0-Par1 in Table~\ref{tab:1} rescaled by their knee values as described in the text for the wounded nucleon model (left panel) and two-component model (right panel). They are compared with the shapes of the experimental $\nchrec$ ($|\eta|<2.5$) and $\sumet$ ($3.2<|\eta|<4.9$) distributions from Ref.~\cite{ATLAS:2017zcm}.}
\end{figure}

Table~\ref{tab:1} lists the NBD parameters for the wounded nucleon model and two-component model. The three parameter sets ``Par0'', ``Par1'', ``Par2'' have  the same $\bar{n}$ but different $\hat{\sigma}$. The Par0 and Par1 are adjusted to approximately describe the shapes of the experimental $\nchrec$ ($|\eta|<2.5$) and $\sumet$ ($3.2<|\eta|<4.9$) distributions from the ATLAS Collaboration~\cite{ATLAS:2017zcm}, while the Par2 corresponds to a case with much larger fluctuation. The distributions generated from the three parameter sets are rescaled horizontally by the knee, defined as the average multiplicity for $2A=416$ nucleons for Pb+Pb collisions, $N_{\textrm{knee}}=2A\bar{n}$. Figure~\ref{fig:1} shows the three rescaled distributions for the wounded nucleon model (left panel) and two-component model (right panel), respectively. They are compared to the $\nchrec$ or $\sumet$ distributions, which are also rescaled by the knee values obtained for Par0 and Par1, respectively. The two-component model slightly better describes the shape of the data, as shown by previous studies~\cite{Adler:2013aqf,Abelev:2013qoq}. This is because the source distribution from the two-component model $p(\ntwoc)$ has a smoother and broader knee than that given by the wounded nucleon model $p(\npart)$. The relative widths $\hat{\sigma}$ for each source, therefore, are also much smaller for the two-component model than the wounded nucleon model. One important consequence is that once the generated $p(N)$ distribution is tuned to have similar shape (i.e. by matching to the same experimental measured $p(\nchrec)$ distribution), the centrality fluctuations are typically larger in the wounded nucleon model than the two-component model at the same total multiplicity.

In the independent source picture, particle production for each source in wounded nucleon model can be loosely related to one half of the particles from one $pp$ collision. The charged particle multiplicity distributions in $pp$ collisions at LHC are known to be approximately described by NBD fits. The extracted $\hat{\sigma}$ depends on the $\eta$ window, $\sqrt{s}$ and $\pT$. For $\pT$ integrated charged particles, it is within the range of $0.5\lesssim\hat{\sigma}\lesssim1$~\cite{Ghosh:2012xh}, but increases by 10--20\% from $\pT>0.1$ GeV to $\pT> 0.5$ GeV~\cite{Aad:2016xww}. Therefore the choice of the parameters in Table~\ref{tab:1} covers a reasonable range for $\hat{\sigma}$.

Table~\ref{tab:1} also shows three parameter sets with apostrophe, Par0$'$, Par1$'$, and Par2$'$, which have the same $\hat{\sigma}$ but smaller $\bar{n}$, and therefore larger $m$ than the corresponding ones without apostrophe. These parameter sets are used to study whether the multiplicity cumulants are controlled by $\hat{\sigma}$ or whether they depend also on $m$. 

As was discussed in the introduction, to avoid auto-correlations effects, the particles used to define event centrality should not be included in the measurement of physics observables. Therefore, measurements performed in subevent B are affected by the centrality fluctuations in subevent A associated with the centrality selection. The second-order cumulant of the multiplicity fluctuation, for example, can be written as the sum of the fluctuations in each source in subevent B and fluctuation in the number of sources $\ns$ induced by centrality selection in subevent A~\cite{Heiselberg:2000fk}:
\begin{eqnarray}
\label{eq:4}
\lr{(N-\bar{N})^2} = \bar{N}_{\mathrm {s,A}}\lr{(n_B-\bar{n}_B)^2} + \bar{n}_B^2\lr{(N_{\mathrm {s,A}}-\bar{N}_{\mathrm {s,A}})^2},
\end{eqnarray}
where we have associated each quantity with its subevent via subscripts. Subevents A and B are generated using parameter sets Par0--Par2 listed in the Table~\ref{tab:1}. One main focus of this paper is to understand how the multiplicity fluctuation in subevent A induces the CF, which then influence the fluctuation of physics observables in subevent B.

The relative width $\hat{\sigma}$ of the particle production for each source plays an important role. It largely determines how the $p(N)$ is smeared relative to $p(\ns)$. For large enough $\ns$, $p(N;\ns)$ is expected to approach a Gaussian with relative width of $\hat{\sigma}/\sqrt{\ns}=1/\sqrt{\ns (\frac{1}{\bar{n}}+\frac{1}{m})}$. Therefore, in ultra-central collisions, where the $\ns$ is sufficiently large, the shape of $p(N)$ is expected to approach the shape of $p(\ns)$. Due to the fast dropping $p(\ns)$, the $\ns$ distributions in UCC have significant non-Gaussian shape. This non-Gaussian fluctuation is expected to lead to non-zero values of higher-oder multiplicity and flow cumulants.  Therefore the study of the centrality fluctuations in UCC events provides an unique opportunity to understand the nature of the sources in the early stage of heavy-ion collisions and how they drive the fluctuations of collective flow. Note that the shape of $p(\ns)$ was also parameterized as impact parameter distribution $p(b)$ convoluted with fluctuation of sources at fixed $b$~\cite{Das:2017ned}, in this approach, however, inferring the impact parameter of the event does not necessarily constrain the centrality/volume fluctuation. 

Obviously, the independent source model based on Glauber and NBD has certain limitations in its predictive power. It does not model the interaction between different sources, which clearly is important in the final state. These interactions may modify the particle correlations in each source or create new sources of fluctuations. Our model also assumes explicitly that $\ns$ is the same independent of rapidity. In reality, the $\ns$ and the length of the source in rapidity are expected to have strong fluctuations~\cite{Bozek:2015bna,Pang:2015zrq,Schenke:2016ksl,Ke:2016jrd,Shen:2017bsr}. For example, the sub-nucleonic degree-of-freedom may evolve with rapidity, such that the number of sources for each nucleon is not the same between mid-rapidity and forward-rapidity~\cite{Schenke:2016ksl,Jia:2015jga}. These longitudinal fluctuations tend to weaken the centrality correlation between the forward and mid-rapidity, such that the CF in forward rapidity may not be the same as that at mid-rapidity. Nevertheless, our model serves as a useful baseline. It can be considered as a first step towards a more realistic simulation that include the full space-time dynamics of the heavy-ion collisions. 

\section{centrality fluctuations and multiplicity cumulants}\label{sec:3}
We use the following definition of multiplicity cumulants for distributions of total multiplicity $p(N)$, multiplicity distribution for each source $p(n)$, and distribution of total number of sources $p(\ns)$:
\begin{eqnarray}
\label{eq:6}
&&K_2 = \frac{\lr{(\delta N)^2}}{\bar{N}}\;, K_3 = \frac{\lr{(\delta N)^3}}{\bar{N}}\;, K_4 = \frac{\lr{(\delta N)^4}-3\lr{(\delta N)^2}^2}{\bar{N}}\;, \delta N = N-\bar{N}\\
&&k_2 = \frac{\lr{(\delta n)^2}}{\bar{n}}\;,\;\; k_3 = \frac{\lr{(\delta n)^3}}{\bar{n}}\;,\;\;\; k_4 = \frac{\lr{(\delta n)^4}-3\lr{(\delta n)^2}^2}{\bar{n}}\;, \;\;\;\delta n = n-\bar{n}\\
&&k_2^{\mathrm {v}} = \frac{\lr{(\delta \ns)^2}}{\bar{\ns}}\;, k_3^{\mathrm {v}} = \frac{\lr{(\delta \ns)^3}}{\bar{\ns}}\;, k_4^{\mathrm {v}} = \frac{\lr{(\delta \ns)^4}-3\lr{(\delta \ns)^2}^2}{\bar{\ns}}\;, \delta \ns = \ns-\bar{\ns}\;.
\end{eqnarray}
These quantities are related to each other via following well-known formula~\cite{Skokov:2012ds}:
\begin{eqnarray}
\label{eq:7}
K_2 = k_2+\bar{n}k_2^{\mathrm {v}}\;, K_3 = k_3+3k_2\bar{n}k_2^{\mathrm {v}}+\bar{n}^2k_3^{\mathrm {v}},\; K_4 = k_3+(4k_3+3k_2^2)\bar{n}k_2^{\mathrm {v}}+6k_2\bar{n}^2k_3^{\mathrm {v}}+\bar{n}^3k_4^{\mathrm {v}}\;.
\end{eqnarray}
The second-order cumulant defined this way is the same as the scaled-variance $\omega$ used in many previous studies~\cite{Heiselberg:2000fk,Aggarwal:2001aa,Adare:2008ns,Konchakovski:2005hq}. The second-order cumulant for each source is also related to the relative width: $\hat{\sigma}^2=k_2/\bar{n}$. For the NBD distribution used to describe the particle production for each source, the cumulants depend only on the $p$ parameter:
\begin{eqnarray}
\label{eq:8}
k_2(p) = \frac{1}{1-p}=1+\frac{\bar{n}}{m},\; k_3(p) = \frac{1+p}{(1-p)^2},\; k_4(p) = \frac{1}{1-p}+\frac{6p}{(1-p)^3}\;.
\end{eqnarray}

We now discuss the relation between the fluctuation of $N$ for events with a fixed $\ns$ and the fluctuation of $\ns$ for events with the same $N$. The moments of these two fluctuations can be expressed as:
\begin{eqnarray}
\label{eq:9}
\lr{(\delta N)^k}_{\ns} = \int (\delta N)^k p(N;\ns) dN,\; \lr{(\delta\ns)^k}_{N} = \int (\delta\ns)^k p(N;\ns) p(\ns) d\ns\;.
\end{eqnarray}
In mid-central collisions where the $p(\ns)$ can be treated as a constant, a precise relation can be derived for NBD distribution (details in Appendix~\ref{sec:a1}). The $\lr{(\delta N)^k}_{\ns}$ and $k_m$ calculated for NBD parameter $p$ is directly related to the $\lr{(\delta\ns)^k}_{N}$ and $k_m^{\mathrm{v}}$ calculated for $1-p$:
\begin{eqnarray}
\label{eq:10}
\bar{n}^{m-1} k_m^{\mathrm{v}} (1-p) = \frac{p^{m-1}}{(1-p)^{m-1}}k_m (p)
\end{eqnarray}
From this, we obtain several useful relations between the multiplicity fluctuation and centrality fluctuation:
\begin{eqnarray}
\label{eq:11}
&&r_2 = \frac{\bar{n}k_2^{\mathrm{v}}}{k_2} =1\\\label{eq:11b}
&&r_3 = \frac{\bar{n}^2k_3^{\mathrm{v}}}{k_3} = \frac{2-p}{1+p}\;, \frac{1}{2} \leq r_3\leq2\\\label{eq:11c}
&&r_4 = \frac{\bar{n}^3k_4^{\mathrm{v}}}{k_4} =\frac{p^2+6(1-p)}{(1-p)^2+6p}\;, \frac{1}{6} \leq r_4\leq6
\end{eqnarray}

In ultra-central collisions where the boundary effect on $\ns$ is important, one could approximate multiplicity for large $\ns$ with a narrow Gaussian $p(N;\ns) \approx \frac{1}{\sqrt{2\pi \hat{\sigma}^2 \ns}} e^{-\frac{(N-\bar{n}\ns)^2}{2\hat{\sigma}^2\ns}}$ via the central-limit theorem~\cite{Begun:2006uu}. In this case, the moments of centrality fluctuations can be estimated as:
\begin{eqnarray}
\label{eq:12}
\lr{(\delta\ns)^k}\approx \int (\delta\ns)^k\frac{1}{\sqrt{2\pi\hat{\sigma}^2\ns}}  e^{-\frac{(\ns-\bar{\ns})^2}{2\hat{\sigma}^2\ns}} p(\ns) d\ns\;.
\end{eqnarray}
This shows that the centrality fluctuations are only sensitive to the relative width $\hat{\sigma}$ of the $p(n)$, not its functional form. One important consequence of Eq.~\ref{eq:12} is that ${k_2}\approx\bar{n}k_2^{\mathrm{v}}$ in Eq.~\ref{eq:11} is generally valid for independent source model in central collisions, even if $p(n)$ is not NBD.

The calculation of cumulants follows the standard procedure. Each Pb+Pb event is divided into two subevents: subevent A for centrality selection and subevent B for the calculation of multiplicity cumulants. The particle multiplicities in these two subevents, $N_{\mathrm{A}}$ and $N_{\mathrm{B}}$, are generated independently from the same $\ns$ in each event. The events are divided into narrow centrality classes according to $N_{\mathrm{A}}$.  The cumulants are first calculated from $p(N_{\mathrm{B}})$ for events with the same $N_{\mathrm{A}}$, which are then combined into broader $N_{\mathrm{A}}$ ranges to reduce the statistical uncertainty. For each $N_{\mathrm{A}}$ range, the average number of sources, $\lr{\ns}$, is calculated based on the 2D correlation between $\ns$ and $N_{\mathrm{A}}$. The multiplicity cumulants, denoted as $K_{m,B|A}$, are then presented as a function of $\lr{\ns}$. In this setup,  the centrality fluctuations arise from the subevent A, and we rewrite Eq.~\ref{eq:8} as:
\begin{eqnarray}
\nonumber
K_{2,B|A} &=& k_{2,B}+\bar{n}_{B}k_{2,A}^{\mathrm {v}}\;, \\\nonumber
K_{3,B|A} &=& k_{3,B}+3k_{2,B}\bar{n}_Bk_{2,A}^{\mathrm {v}}+\bar{n}_B^2k_{3,A}^{\mathrm {v}},\;\\\label{eq:13}
K_{4,B|A} &=& k_{4,B}+(4k_{3,B}+3k_{2,B}^2)\bar{n}_Bk_{2,A}^{\mathrm {v}}+6k_{2,B}\bar{n}_B^2k_{3,A}^{\mathrm {v}}+\bar{n}_B^3k_{4,A}^{\mathrm {v}},
\end{eqnarray}
The $K_{m,B|A}$ is the observed multiplicity fluctuation in subevent B, which has contributions from $k_{m,B}$, multiplicity fluctuation within each source in subevent B, and $k_{m,A}^{\mathrm {v}}$, the centrality fluctuation from subevent A. The $k_{m,A}^{\mathrm {v}}$ is calculated from $p(\ns)$ for events selected with fixed $N_{\mathrm{A}}$, then averaged over a finite $N_{\mathrm{A}}$ range. In the following, we first discuss the behavior of $k_{m,A}^{\mathrm {v}}$ and its relation to $k_{m,A}$, then discuss results for the total multiplicity fluctuation $K_{m,B|A}$.

\begin{figure}[h!]
\begin{center}
\includegraphics[width=0.9\linewidth]{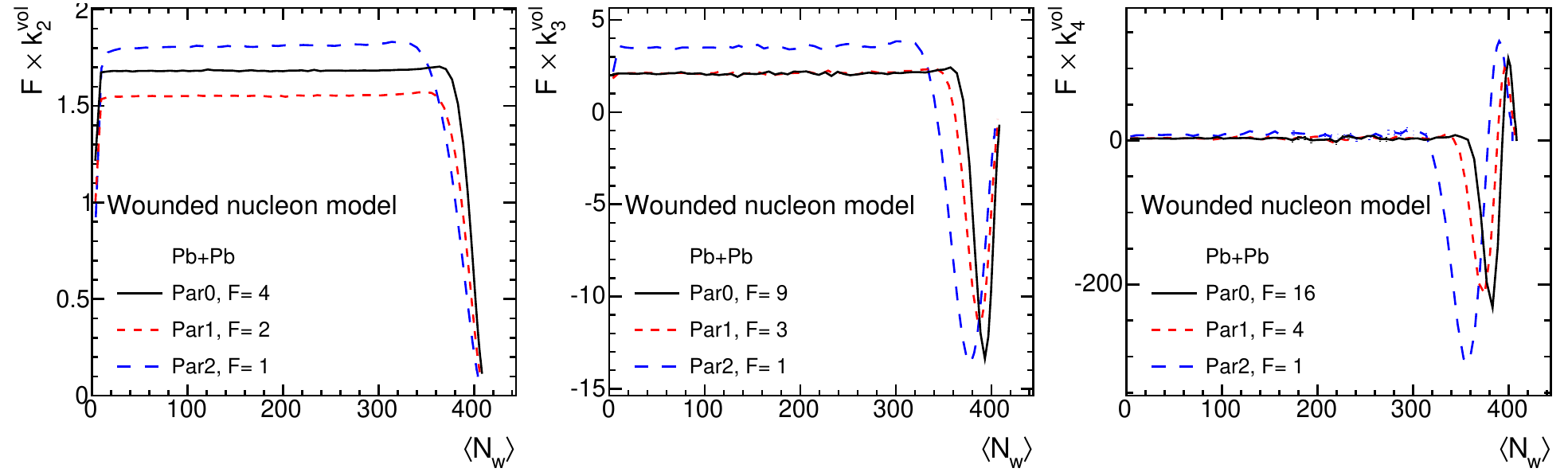}
\includegraphics[width=0.9\linewidth]{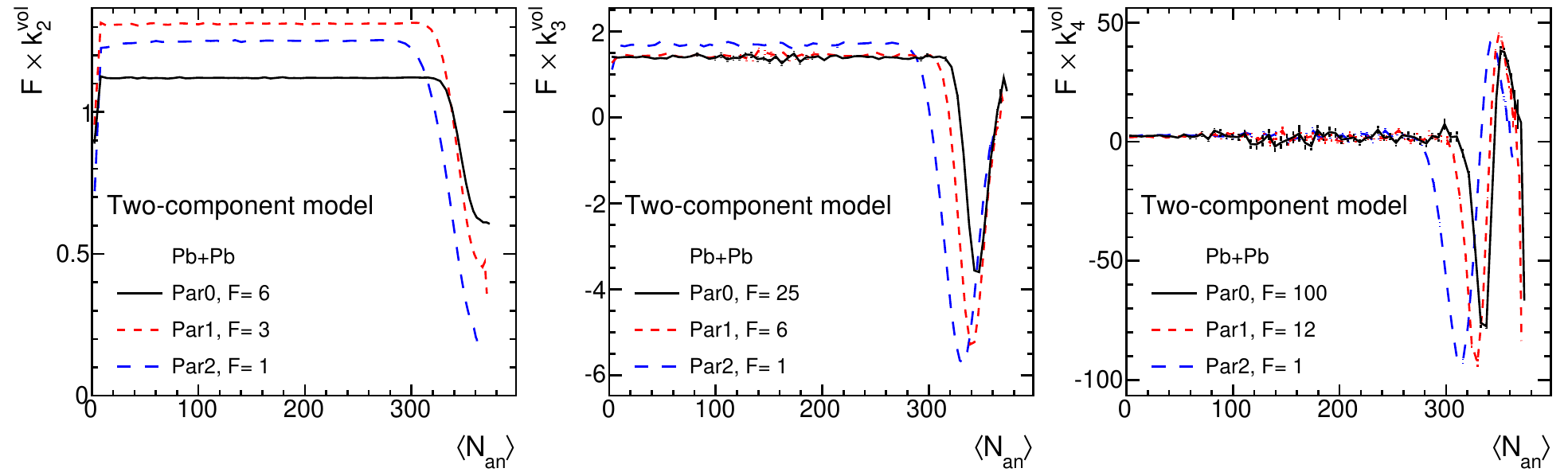}
\end{center}
\caption{\label{fig:2} Cumulants for centrality fluctuations of different order,  $k_{2}^{\mathrm {v}}$ (left column),  $k_{3}^{\mathrm {v}}$ (middle column) and $k_{4}^{\mathrm {v}}$ (right column) from wounded nucleon model (top row) and two-component model (bottom row) as a function of $\lr{\ns}$. They are calculated using parameter set Par0, Par1 and Par2 for NBD distributions from Table~\ref{tab:1}. The $k_{m}^{\mathrm {v}}$ values are scaled by a factor as indicated in the legend, so they can be shown in the same panel.}
\end{figure}

The top-row of Figure~\ref{fig:2} shows the second, third and fourth order cumulants for centrality fluctuations in the wounded nucleon model. Each cumulant has been calculated for the three NBD parameter sets, Par0, Par1 and Par2, from Table~\ref{tab:1}. The values of the second-order cumulant (or scaled variance) $k_{2}^{\mathrm {v}}$ are constant in mid-central collisions, but decreases toward very peripheral and very central collisions. These decreases are due to the boundary effects on $\ns$ that reduces the width of the centrality fluctuations. The higher-order cumulants $k_{3}^{\mathrm {v}}$  and $k_{4}^{\mathrm {v}}$ show strong oscillating behavior toward central collisions, which reflects the strong non-Gaussianity of $p(\ns)$ for events required to have similar $N_{\mathrm{A}}$ (see Eq.~\ref{eq:12}). Over the full centrality range, the value of $k_{m}^{\mathrm {v}}$ depends strongly on the relative width $\hat{\sigma}$ of the NBD parameters. Larger $\hat{\sigma}$ leads to stronger smearing and larger multiplicity fluctuation, and the impact is stronger for higher-order cumulants. These results are qualitatively similar to those obtained in earlier studies~\cite{Skokov:2012ds,Xu:2016qzd,Bzdak:2016jxo}.

The calculations are repeated for the two-component model, and the results are shown in the bottom-row of Figure~\ref{fig:2}. In general, the cumulants for a given parameter set is smaller than its counterpart in the wounded nucleon model, as indicated by the larger multiplicative factor in the legends, although they are tuned to have similar $p(N)$ distribution (comparing the left and right panels of Figure~\ref{fig:1}). We also see that the $k_{2}^{\mathrm {v}}$ values in ultra-central collisions do not decrease to zero as seen for the wounded nucleon model case. These differences are due to a weaker constraint from the boundary effects due to a broader $p(\ns)$ distribution than that from the wounded nucleon model in central collisions.

\begin{figure}[h!]
\begin{center}
\includegraphics[width=1\linewidth]{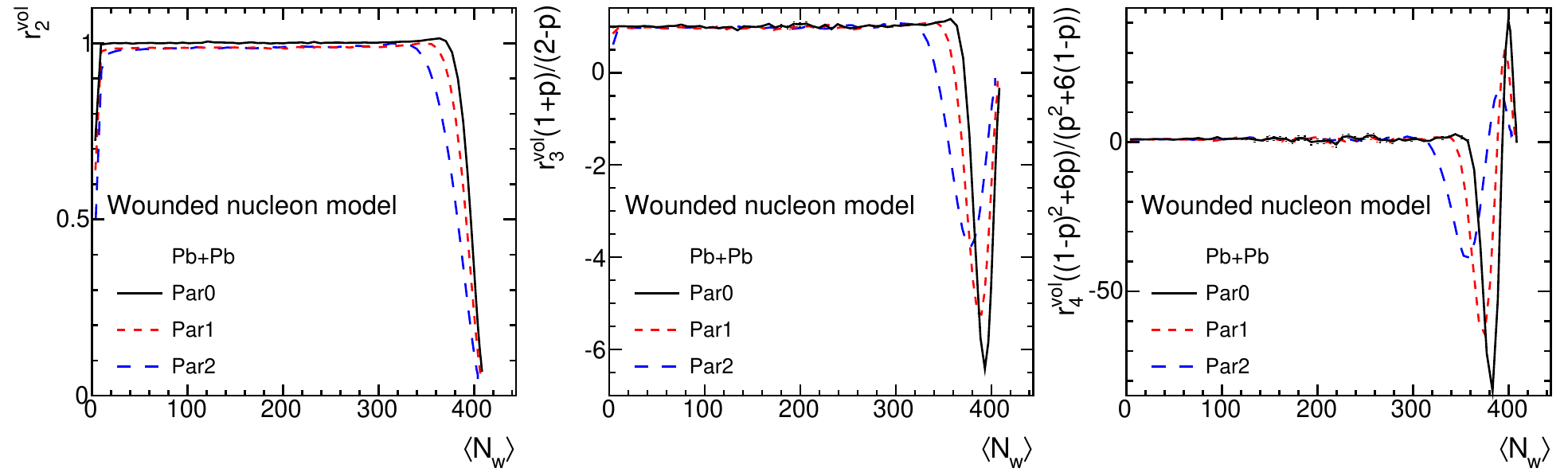}
\end{center}
\caption{\label{fig:3} The $r_2$ (left), $r_3\frac{1+p}{2-p}$ (middle), and $r_4\frac{(1-p)^2+6p}{p^2+6(1-p)}$ (right) defined in Eqs.~\ref{eq:11}--\ref{eq:11c} as a function of $\lr{\npart}$ in the wounded nucleon model.}
\end{figure}

In mid-central region where the $p(\ns)$ is a slowly varying function, we derived Eqs.~\ref{eq:10}--\ref{eq:11c} that relate the multiplicity cumulants for each source to centrality fluctuation cumulants. The validity of these relations are verified in Figure~\ref{fig:3} for the wounded nucleon model. The scaled ratios $r_2$, $r_3\frac{1+p}{2-p}$ and $r_4\frac{(1-p)^2+6p}{p^2+6(1-p)}$ are plotted for Par0, Par1 and Par2, which are expected to be one if these relations are valid. We found this indeed is the case for mid-central collisions. These relations breaks down in most peripheral and central collisions.

\begin{figure}[h!]
\begin{center}
\includegraphics[width=1\linewidth]{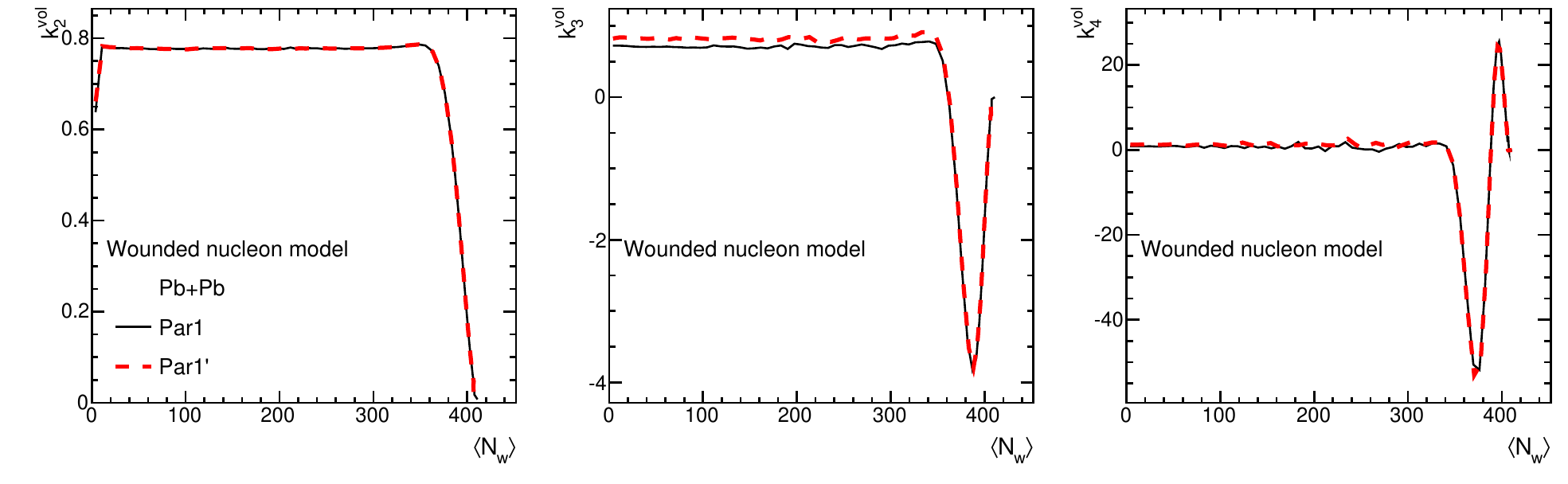}
\end{center}
\caption{\label{fig:4} The cumulants for centrality fluctuation, $k_2^{\mathrm{v}}$ (left), $k_3^{\mathrm{v}}$ (middle) and $k_4^{\mathrm{v}}$ (right) as a function of $\lr{\npart}$. They are calculated in the wounded nucleon model for Par1 and Par1$'$ which has different $\bar{n}$ but the same $\hat{\sigma}$.}
\end{figure}

Using the Gaussian approximation Eq.~\ref{eq:12}, we have argued that the properties of $k_m^{\mathrm{v}}$ is controlled by the relative width $\hat{\sigma}$ in the full centrality for $m=2$ and in central collisions for $m>2$. This is verified by comparing two NBD parameterizations with the same $\hat{\sigma}$. Figure~\ref{fig:4} shows the volume cumulants calculated for Par1 and Par1$'$, whose $\bar{n}$ values differ by more than factor of two. We found that the $k_2^{\mathrm{v}}$  are nearly identical between the two in the full centrality range. For higher-order cumulants, they are very close to each other in the ultra-central region where $p(\ns)$ dominates the properties of cumulants, but differ significantly otherwise. 

\begin{figure}[h!]
\begin{center}
\includegraphics[width=1\linewidth]{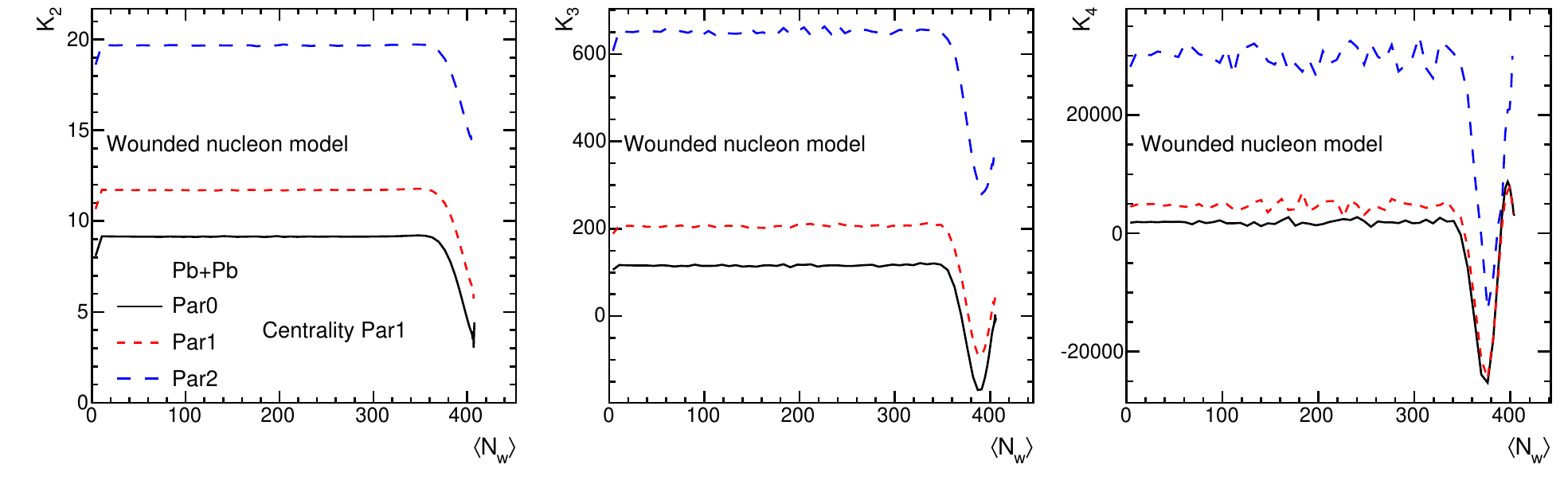}
\end{center}
\caption{\label{fig:5} The total multiplicity cumulant $K_{2,B|A}$ (left panel), $K_{3,B|A}$ (middle panel) and $K_{4,B|A}$ (right panel) as a function of $\lr{\npart}$ obtained in the wounded nucleon model. Centrality is defined in subevent A with parameter set Par1, and three curves in each panels are cumulants calculated with parameter set Par0, Par1 and Par2. }
\end{figure}

Now we focus our attention on the behavior of total multiplicity fluctuation in subevent B when centrality is defined in subevent A, as described by Eq.~\ref{eq:13}. In our model, the property of each source is assumed to be independent of centrality, i.e. $k_{m,B}$ is constant. Therefore the shape of the multiplicity cumulant $K_{m,B|A}$ as a function of $\lr{\ns}$ must reflect the centrality dependence of $k_{m,A}^{\mathrm{v}}$. Figure~\ref{fig:5} shows the $K_{m,B|A}$ for $m=2,3,4$ calculated with the Par1 for subevent A but all three different parameter sets for subevent B. Indeed the three curves are different from each other by a constant offset, which reflects their different $k_{m,B}$ values, while the non-flatness is due to a common $k_{m,A}^{\mathrm{v}}$~\footnote{These results implies the cumulant ratios, for instance $K_4/K_2$, are much larger than those involved in the net-proton fluctuations~\cite{Luo:2017faz} in the flat region where the CF is not important. In this region,  $K_4/K_2\approx k_4/k_2=k\sigma^2$, where $k$ is the excess kurtosis and $\sigma^2$ is the variance of the fluctuation for each source. The large $K_4/K_2$ is mainly because the $\sigma^2$ for total multiplicity is much larger than that for the net-proton.}.

In ultra-central collisions where the cumulants are dominated by the shape of $p(\ns)$, if the cumulants for each source is not very large, i.e. $p$ is not very close to one, we found that the contribution of the cross-terms in Eq.~\ref{eq:13} are subdominant and the higher-order cumulants can be approximated as $K_{3,B|A} \approx k_{3,B}+\bar{n}_B^2k_{3,A}^{\mathrm {v}},\;K_{4,B|A} \approx k_{4,B}+\bar{n}_B^3k_{4,A}^{\mathrm {v}}$, which imply that the shape of $K_{m,B|A}$ should be similar to $k_{m,A}^{\mathrm {v}}$ but is rescaled by a constant $\bar{n}_B^{m-1}$. 

One could also select centrality in subevent B and study multiplicity cumulant in subevent A, and ask what is the relation between $K_{m,B|A}$ and $K_{m,A|B}$. Since Eq.~\ref{eq:11} is valid in mid-central collisions independent of the underlying $p(n)$, the relation for scaled variance is particular useful. Using Eq.\ref{eq:13}, we have $K_{2,B|A} \approx k_{2,B}+\frac{\bar{n}_{B}}{\bar{n}_{A}}k_{2,A}$ and similarly for $K_{2,A|B}$, from which we obtain:
\begin{eqnarray}
\label{eq:14}
K_{2,B|A}-K_{2,A|B} = (\bar{n}_B-\bar{n}_A)\left(\frac{k_{2,B}}{\bar{n}_{B}}+\frac{k_{2,A}}{\bar{n}_{A}}\right)=(\bar{n}_B-\bar{n}_A)(\hat{\sigma}_B+\hat{\sigma}_A)\;,
\end{eqnarray}
This implies that if subevent A and B are used for centrality selection for each other, the scale variance for subevent B is larger than subevent A if it also has large $\bar{n}$. Furthermore, if the two subevents have the same relative width for each source, i.e. $\hat{\sigma}_{A}=\hat{\sigma}_{B}$, then $K_{2,B|A}-K_{2,A|B}= (\bar{n}_B-\bar{n}_A)\left(\hat{\sigma}+k_{2}^{\mathrm{v}}\right)$, and it is valid over the entire centrality range.

Let's consider another situation where the subevent A$'$ is used to provide centrality selection for subevent B. If $\hat{\sigma}_{A}=\hat{\sigma}_{A'}$, cumulants in subevent B is independent of whether A or A$'$ is used for centrality selection: $K_{2,B|A}=K_{2,B|A'}$. For higher-order cumulant, this statement is only true in ultra-central collisions. On the other hand, if subevent B is used for centrality selection, relation $K_{2,A|B}/K_{2,A'|B} = \bar{n}_A/\bar{n}_{A'}$ is valid over the full centrality range. For the third-order cumulants, we find that the terms involving centrality fluctuations are related to each other by a constant:
\begin{eqnarray}
\label{eq:15}
\frac{3k_{2,A}\bar{n}_Ak_{2,B}^{\mathrm {v}}+\bar{n}_{A}^2k_{3,B}^{\mathrm {v}}}{3k_{2,A'}\bar{n}_{A'}k_{2,B}^{\mathrm {v}}+\bar{n}_{A'}^2k_{3,B}^{\mathrm {v}}} = \frac{\bar{n}_{A}^2}{\bar{n}_{A'}^2}
\end{eqnarray}

\section{centrality fluctuations and eccentricity cumulants}\label{sec:4}
In heavy-ion collisions, it is commonly believed that the flow vector ${\bm v}_n=v_ne^{in\Psi_n}$ is driven by hydrodynamic response to the eccentricity vector ${\bm \epsilon}_n$, calculated from the transverse position $(r,\phi)$ of the particle production sources 
\begin{eqnarray}
\label{eq:c1}
{\bm \epsilon}_n = {\epsilon}_n e^{in\Phi_n} \equiv  -\frac{\langle r^n e^{in\phi}\rangle}{\langle r^n\rangle},
\end{eqnarray}
where we have used complex number to encode the magnitude and phase. Hydrodynamic model studies have show ${\bm v}_2$ and ${\bm v}_3$ correlate almost linearly with corresponding eccentricity vectors, ${\bm v}_n = b_n {\bm \epsilon}_n$~\cite{Qiu:2011iv,Gardim:2011xv,Niemi:2012aj}, where $b_n$ is a constant within a given centrality class. The quadrangular flow ${\bm v}_4$, on the other hand, has a large non-linear contribution proportional to ${\bm v}_2^2$, on top of a linear contribution associated with the fourth-order eccentricity ${\bm \epsilon}_4$. Measurements~\cite{Aad:2014fla,Aad:2015lwa,ATLAS:2017zcm} show that the linear contribution dominates the 0-10\% centrality range, where ${\bm \epsilon}_4\propto {\bm v}_4$ is still a good approximation. The higher-order flow harmonics (${\bm v}_5$ ..) are always dominated by non-linear terms, so we won't discuss them here.

The flow probability distribution $p(v_n)$ is characterized by the multi-particle cumulants defined as:
\begin{eqnarray}\nonumber
c_n\{2\}&=&\left\langle v_n^2\right\rangle\\\nonumber
c_n\{4\}&=&\left\langle v_n^4\right\rangle-2\left\langle v_n^2\right\rangle^2\\\nonumber
4c_n\{6\}&=&\left\langle v_n^6\right\rangle-9\left\langle v_n^4\right\rangle\left\langle v_n^2\right\rangle+12\left\langle v_n^2\right\rangle^3\\\label{eq:c3}
33c_n\{8\}&=&\left\langle v_n^8\right\rangle-16\left\langle v_n^6\right\rangle\left\langle v_n^2\right\rangle-18\left\langle v_n^4\right\rangle^2+144\left\langle v_n^4\right\rangle\left\langle v_n^2\right\rangle^2-144\left\langle v_n^2\right\rangle^4.
\end{eqnarray}
The pre-factors 4 and 33 are introduced, such that $c_n\{2k\}=\left(-1\right)^{k+1}v_n^{2k}$ if $v_n$ is a constant. If the scaling between $v_n$ and $\epsilon_n$ were exactly linear, $p(v_n)$ has the same shape as $p(\epsilon_n)$ up to a constant scale factor. Therefore the nature of flow fluctuations can be studied via the corresponding eccentricity cumulants, $c_{n,\epsilon}\{2k\}$, which are obtained by replacing the $v_n$ with $\epsilon_n$ in Eq.~\eqref{eq:c3}, for example $c_{n,\epsilon}\{4\}=\left\langle \epsilon_n^4\right\rangle-2\left\langle \epsilon_n^2\right\rangle^2$. The following relation should be valid for any integers $k$:
\begin{eqnarray}
\label{eq:c4}
\frac{c_n\{2k\}}{c_n\{2\}^k}=\frac{c_{n,\epsilon}\{2k\}}{c_{n,\epsilon}\{2\}^k},k=2,3...\;
\end{eqnarray}
Based on this, we shall define normalized cumulants similar to Ref.~\cite{Giacalone:2017uqx}:
\begin{eqnarray}
\label{eq:c5}
\hat{c}_n\{2k\} \equiv \frac{c_n\{2k\}}{c_n\{2\}^k}, \; \hat{c}_{n,\epsilon}\{2k\} \equiv \frac{c_{n,\epsilon}\{2k\}}{c_{n,\epsilon}\{2\}^k}.
\end{eqnarray}
The normalized flow cumulant $\hat{c}_n\{2k\}$ for a event class can be approximated by corresponding normalized eccentricity cumulants $\hat{c}_{n,\epsilon}\{2k\}$, which can be easily calculated from the Glauber model.

Cumulant observables can also be defined to study correlation between different flow harmonics $p({\bm v}_n,{\bm v}_m)$. Three interesting examples are:
\begin{eqnarray}
\label{eq:c6}
\mathrm{nsc}(2,3)=\frac{\lr{v_2^2v_3^2}}{\lr{v_2^2}\lr{v_3^2}}-1\;,\mathrm{nsc}(2,4)=\frac{\lr{v_2^2v_4^2}}{\lr{v_2^2}\lr{v_4^2}}-1\;, \mathrm{nac}(2,4)=\frac{\lr{{\bm v}_2^2{\bm v}_4^*}}{\sqrt{\lr{v_2^4}\lr{v_4^2}}}\;.
\end{eqnarray}
The first two, $\mathrm{nsc}(2,3)$ and $\mathrm{nsc}(2,4)$ are known as normalized ``symmetric cumulant''~\cite{Bilandzic:2013kga}, which measure the correlation of the magnitudes of two flow harmonics. The third one $\mathrm{nac}(2,4)$ is the analogous normalized ``asymmetric cumulant'' or event-plane correlator~\cite{Jia:2012ma,Aad:2014fla}. Similarly, these observables can also be defined based on the eccentricities:
\begin{eqnarray}
\label{eq:c7}
\mathrm{nsc}_\epsilon(2,3)=\frac{\lr{\epsilon_2^2\epsilon_3^2}}{\lr{\epsilon_2^2}\lr{\epsilon_3^2}}-1\;,\mathrm{nsc}_\epsilon(2,4)=\frac{\lr{\epsilon_2^2\epsilon_4^2}}{\lr{\epsilon_2^2}\lr{\epsilon_4^2}}-1\;, \mathrm{nac}_\epsilon(2,4)=\frac{\lr{{\bm \epsilon}_2^2{\bm \epsilon}_4^*}}{\sqrt{\lr{\epsilon_2^4}\lr{\epsilon_4^2}}}\;.
\end{eqnarray}
Since $\epsilon_2$ and $\epsilon_3$ are linearly proportional to the corresponding harmonic flow, it is expected that $\mathrm{nsc}_\epsilon(2,3)=\mathrm{nsc}(2,3)$. However $\mathrm{nsc}_\epsilon(2,4)$ and $\mathrm{nac}_\epsilon(2,4)$ can be used to estimate the corresponding flow correlations only in very central collisions.  In the following we shall study the influence of centrality fluctuations on eccentricity fluctuations. 

The same events generated for multiplicity analysis are used for calculating the eccentricity cumulants. The events are divided into narrow bins according to the generated multiplicity distribution $p(N)$. Since eccentricity is a global event property, there is no need to distinguish the subevent used for centrality and subevent used for eccentricity cumulants. The cumulants are first calculated in each bin, which are combined to give the values for broader multiplicity bins. These cumulants are then used to obtained the normalized cumulants according to Eqs.~\ref{eq:c4}--\ref{eq:c7}. The final results are presented as a function of $\lr{\ns}$, which is calculated based on the 2D correlation between $N$ and $\ns$, but can be mapped to any other $x$-axis such as $N/N_{\mathrm{knee}}$. This procedure is repeated for each parameter set listed in Table~\ref{tab:1}, and separately for the wounded nucleon model and the two-component model. For comparison, we also calculate the eccentricity cumulants for events binned directly on $\ns$, for which the effects of centrality fluctuations are minimized. This case is equivalent to modeling $p(n)$ as a delta function.

\begin{figure}[h!]
\begin{center}
\includegraphics[width=0.9\linewidth]{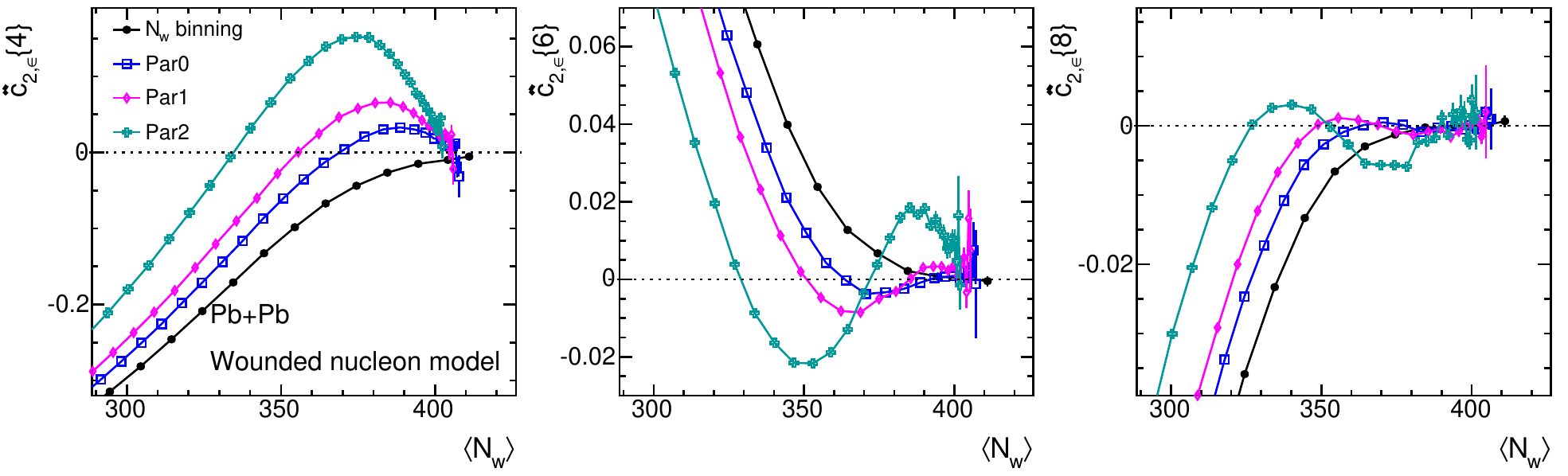}
\includegraphics[width=0.9\linewidth]{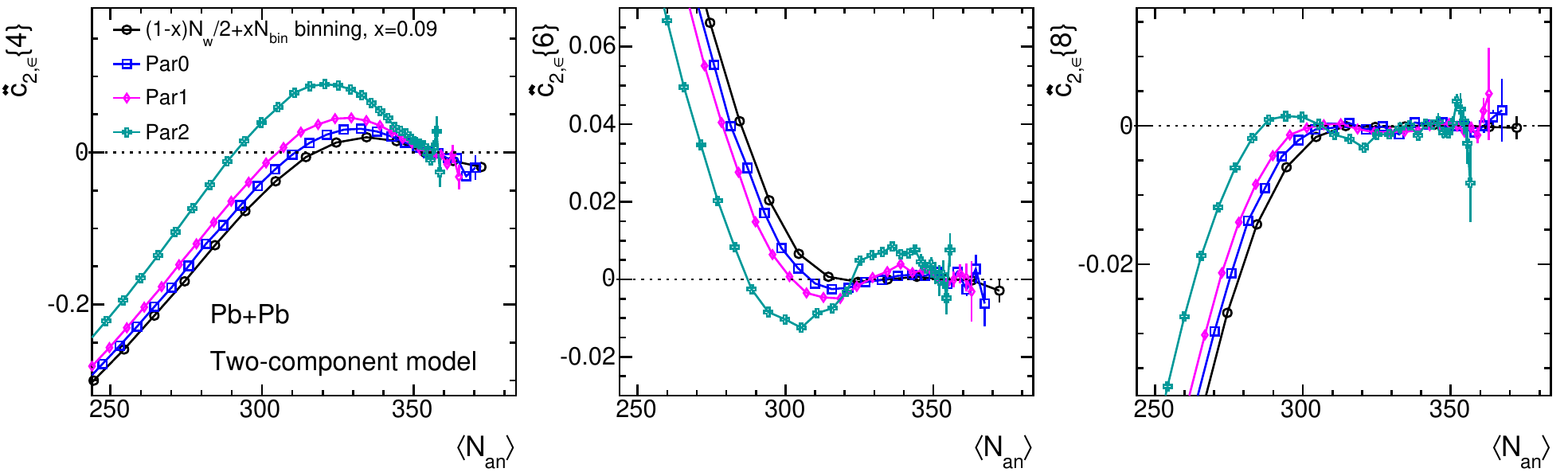}
\end{center}
\caption{\label{fig:c1} The normalized cumulants $\hat{c}_{2,\epsilon}\{4\}$ (left) and $\hat{c}_{2,\epsilon}\{6\}$ (middle) and $\hat{c}_{2,\epsilon}\{8\}$ for the three parameter sets in Table.~\ref{tab:1} for the wounded nucleon model (top row) and two-component model (bottom row). They are calculated in narrow particle multiplicity bins then combined and mapped to average number of sources.}
\end{figure}
The top row of Figure~\ref{fig:c1} show the normalized cumulants $\hat{c}_{2,\epsilon}\{4\}$, $\hat{c}_{2,\epsilon}\{6\}$ and $\hat{c}_{2,\epsilon}\{8\}$ as a function of $\lr{\npart}$ in the wounded nucleon model. When eccentricity cumulants are calculated for events binned directly on $\npart$, the magnitudes of normalized cumulants decrease toward more central collisions, but they never change sign, i.e. $\hat{c}_{2,\epsilon}\{4\}<0$, $\hat{c}_{2,\epsilon}\{6\}>0$ and $\hat{c}_{2,\epsilon}\{8\}<0$ over the entire centrality range.  However when events are binned on generated particle multiplicity distribution $p(N)$, which is broader than $p(\npart)$ due to particle production, a characteristic sign-change is observed in central collisions. In particular, $\hat{c}_{2,\epsilon}\{4\}$ becomes positive, reaches a maximum and then decreases to zero toward more central collisions. This finding is qualitatively similar to the sign-change behavior of $c_2\{4\}$ observed in the ATLAS data~\cite{ATLAS:2017zcm}. Furthermore, as the relative width $\hat{\sigma}$ increases from that for Par0 to that for Par2, the location where $\hat{c}_{2,\epsilon}\{4\}$ crosses zero, is shifted toward less central collisions and the maximum $\hat{c}_{2,\epsilon}\{4\}$ value increases. Our study also predicts more complex pattens of sign-change for higher-order cumulants when events are binned on generated particle multiplicity $p(N)$. For large enough $\hat{\sigma}$, the $\hat{c}_{2,\epsilon}\{6\}$ shows double sign-change, and $\hat{c}_{2,\epsilon}\{8\}$ shows triple sign-change in the central collision region.

The bottom row of Figure~\ref{fig:c1} shows similar results calculated for the two-component model. It is interesting to see that the cumulants calculated for events binned directly on $\ntwoc$ already exhibit sign changes in central collisions. This implies that the sign-change is also sensitive to the nature of the fluctuations of the sources that drive the collective flow, not only on the smearing of sources by $p(n)$. After including particle production, the behavior of $\hat{c}_{2,\epsilon}\{2k\}$ show qualitatively similar trends as those observed for the wounded nucleon model, but the magnitudes of $\hat{c}_{2,\epsilon}\{2k\}$ in the sign-change region are smaller. 

\begin{figure}[h!]
\begin{center}
\includegraphics[width=0.9\linewidth]{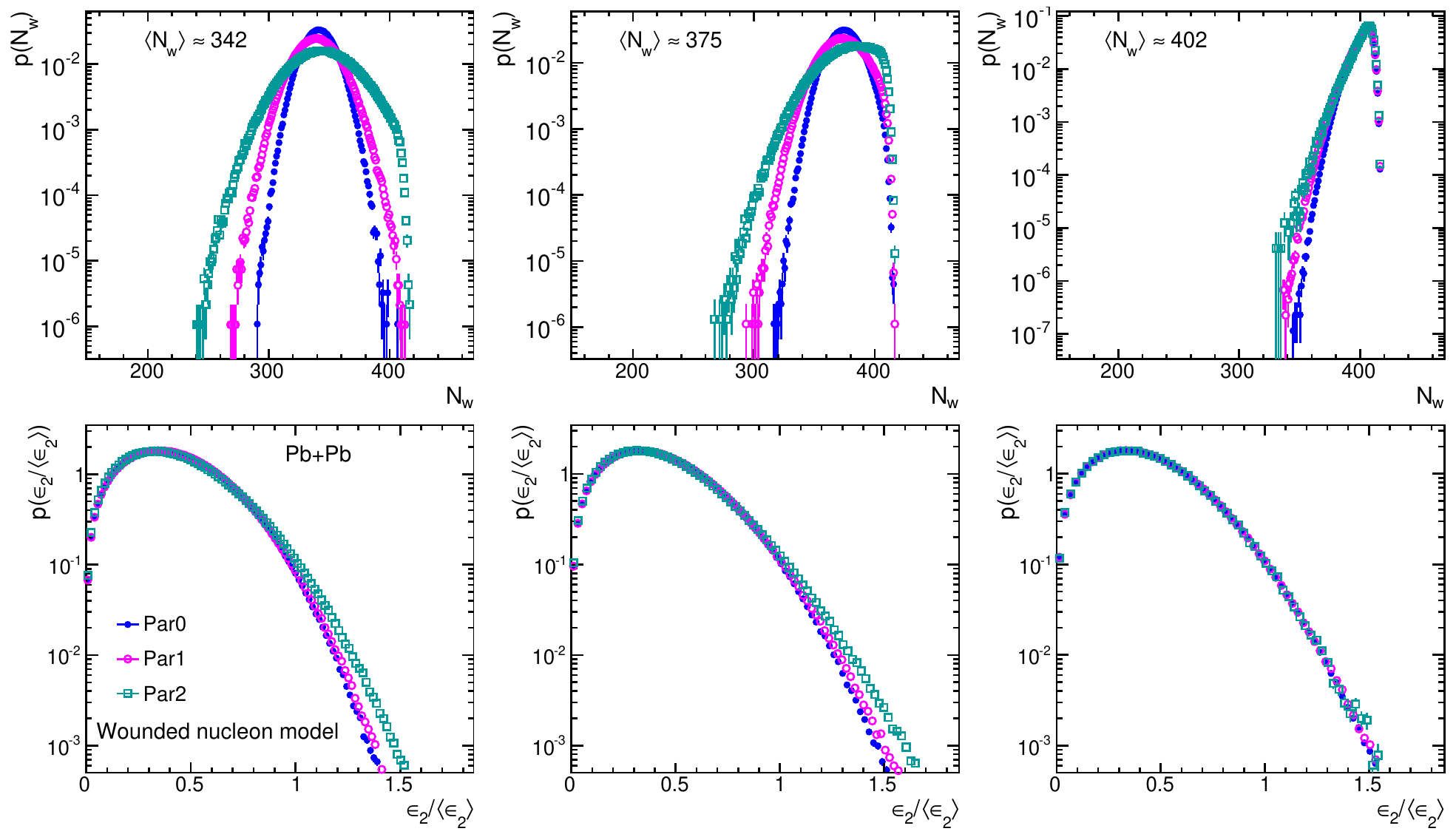}
\end{center}
\caption{\label{fig:c2} Distribution of $\npart$ (top row) and corresponding scaled eccentricity $\epsilon_2/\lr{\epsilon_2}$ (bottom row) for events selected in three range of particle multiplicity $N$ in the central collision region. The three curves in each panel corresponds to the three parameter sets for the wounded nucleon model.}
\end{figure}
To further understand the origin of the sign-change behavior, we focus on $\hat{c}_{2,\epsilon}\{4\}$ shown in the top-left panel of Figure~\ref{fig:c1}. We choose a particular range of multiplicity distribution $p(N)$ for the three parameter sets, corresponding to roughly the same $\lr{\npart}$. We then calculate the corresponding distributions of $\npart$ and scaled eccentricity $\epsilon_2/\lr{\epsilon_2}$ for the selected events. We repeat this procedure in three different ranges, and plot the corresponding distributions $p(\npart)$ and $p(\epsilon_2/\lr{\epsilon_2})$ in the three columns of Figure~\ref{fig:c2}. The distributions $p(\npart)$ and $p(\epsilon_2/\lr{\epsilon_2})$ are different between the three parameter sets, even though they correspond to similar $\lr{\npart}$. This observation suggests that the sign-change of $\hat{c}_{2,\epsilon}\{2k\}$ reflects the non-Gaussianity of $p(\epsilon_2/\lr{\epsilon_2})$, which arises due to combining events with different $\npart$ and therefore different $p(\epsilon_2)$ shape.

\begin{figure}[h!]
\begin{center}
\includegraphics[width=1\linewidth]{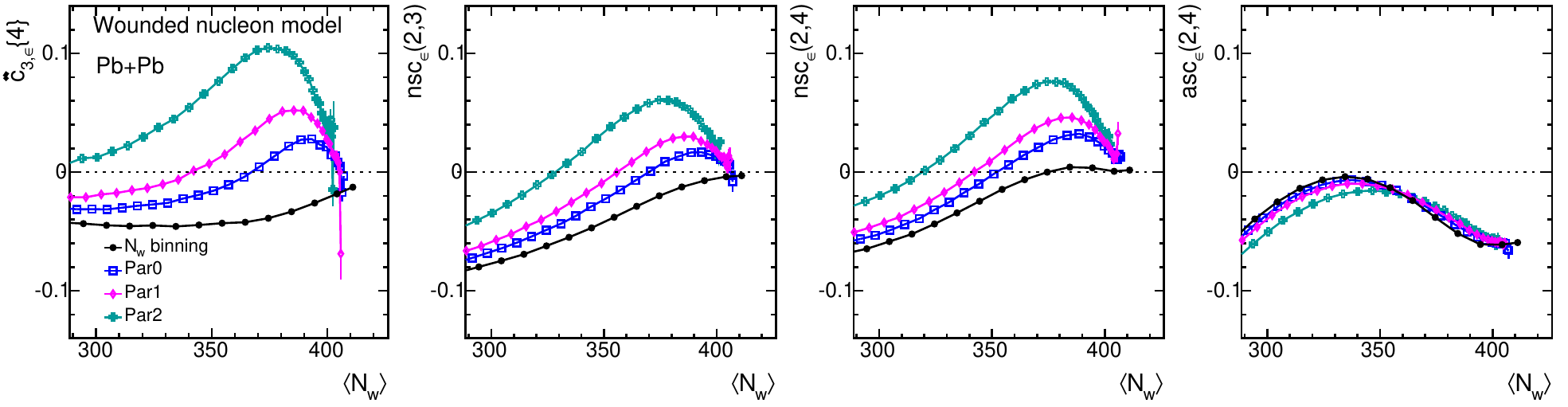}
\end{center}
\caption{\label{fig:c3} The normalized cumulants $\hat{c}_{3,\epsilon}\{4\}$ (left), $\mathrm{nsc}_\epsilon(2,3)$ (second to the left), $\mathrm{nsc}_\epsilon(2,4)$ (second to the right) and  $\mathrm{asc}_\epsilon(2,4)$ (right) for the three parameter sets in Table.~\ref{tab:1} for the wounded nucleon model. They are calculated in narrow particle multiplicity bins then combined and mapped to average number of $\npart$.}
\end{figure}
Figure~\ref{fig:c3} shows the centrality dependence of other normalized cumulant observables, $\hat{c}_{3,\epsilon}\{4\}$, $\mathrm{nsc}(2,3)$, $\mathrm{nsc}_\epsilon(2,4)$ and $\mathrm{asc}_\epsilon(2,4)$ calculated in the wounded nucleon model. The characteristic sign-charge patterns are observed in central collisions except for $\mathrm{asc}_\epsilon(2,4)$. ATLAS measurement seems to suggest a sign-change for $\hat{c}_{3,\epsilon}\{4\}$ when event class is defined using the charge particle multiplicity at mid-rapidity~\cite{ATLAS:2017zcm}. However, the uncertainties of the present measurement are too large for a definite conclusion. We emphasize that it would be very interesting to measure $\mathrm{nsc}_\epsilon(2,3)$, which should have better statistical precision than $\hat{c}_{3,\epsilon}\{4\}$. This observable is relatively insensitive to the final state effects~\cite{Huo:2013qma,Bilandzic:2013kga,Aaboud:2017tql}, and therefore can be used to probe the initial centrality fluctuations. We have also repeated such studies for two-component model, the sign-change pattens are similar but with smaller magnitude (see Appendix~\ref{sec:a3}).

Figure~\ref{fig:c4} compares $\hat{c}_{2,\epsilon}\{4\}$, $\hat{c}_{3,\epsilon}\{4\}$  and $\mathrm{nsc}_\epsilon(2,3)$ calculated for Par0, Par1 and Par2, with those calculated for Par0$'$, Par1$'$ and Par2$'$. A good consistency is observed, suggesting the eccentricity cumulants depend only on $\hat{\sigma}$, similar to the multiplicity cumulants discussed in Section~\ref{sec:3}. Particle distributions $p(n)$ other than NBD are also studied, and the results are found to be insensitive to the functional form of the $p(n)$ (see Appendix~\ref{sec:a3}). 
\begin{figure}[h!]
\begin{center}
\includegraphics[width=1\linewidth]{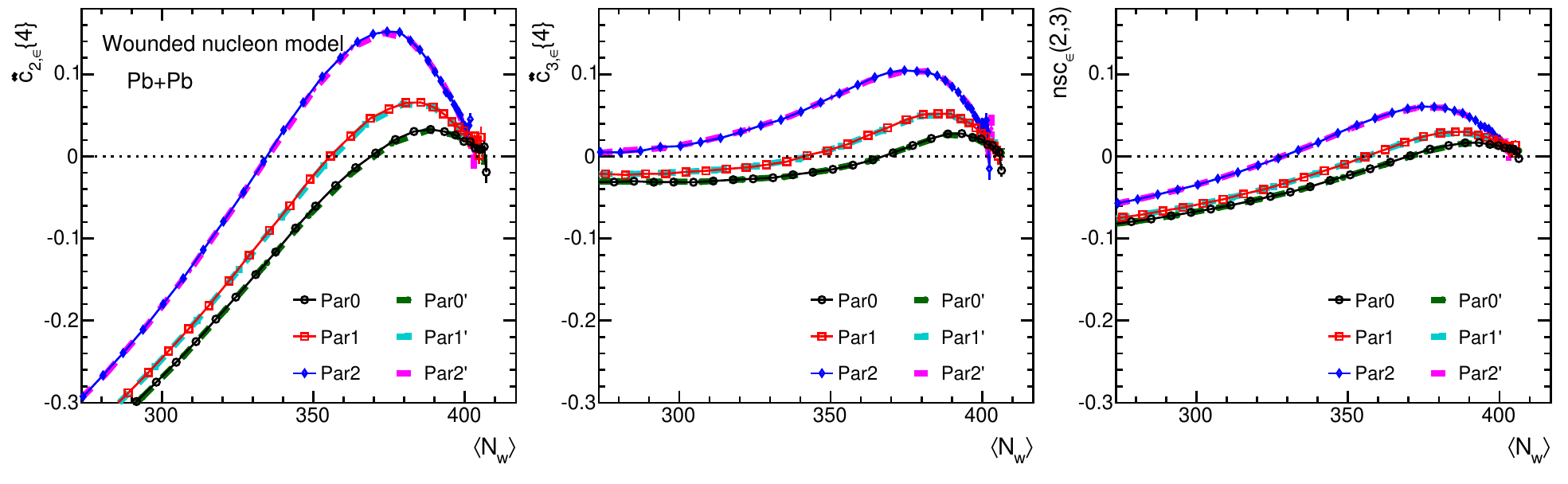}
\end{center}
\caption{\label{fig:c4} The normalized eccentricity cumulant $\hat{c}_{2,\epsilon}\{4\}$ (left panel), $\hat{c}_{3,\epsilon}\{4\}$ (middle panel) and $\mathrm{nsc}_\epsilon(2,3)$ (right panel) compared between ParX and ParX$'$ from Table~\ref{tab:1} for the wounded nucleon model.}
\end{figure}

Our study suggests that normalized cumulants, just like multiplicity cumulants discussed in previous section, are sensitive to the underlying $p(\ns)$ and $\hat{\sigma}$. Therefore, by fitting the measured $p(N)$ and flow cumulants such as $\hat{c}_{2}\{4\}$, one could simultaneously constrain the $p(\ns)$ and $\hat{\sigma}$ in a given model.

\section{centrality fluctuations and Multiplicity-eccentricity mixed cumulants}\label{sec:5}
\begin{figure}[h!]
\begin{center}
\includegraphics[width=1\linewidth]{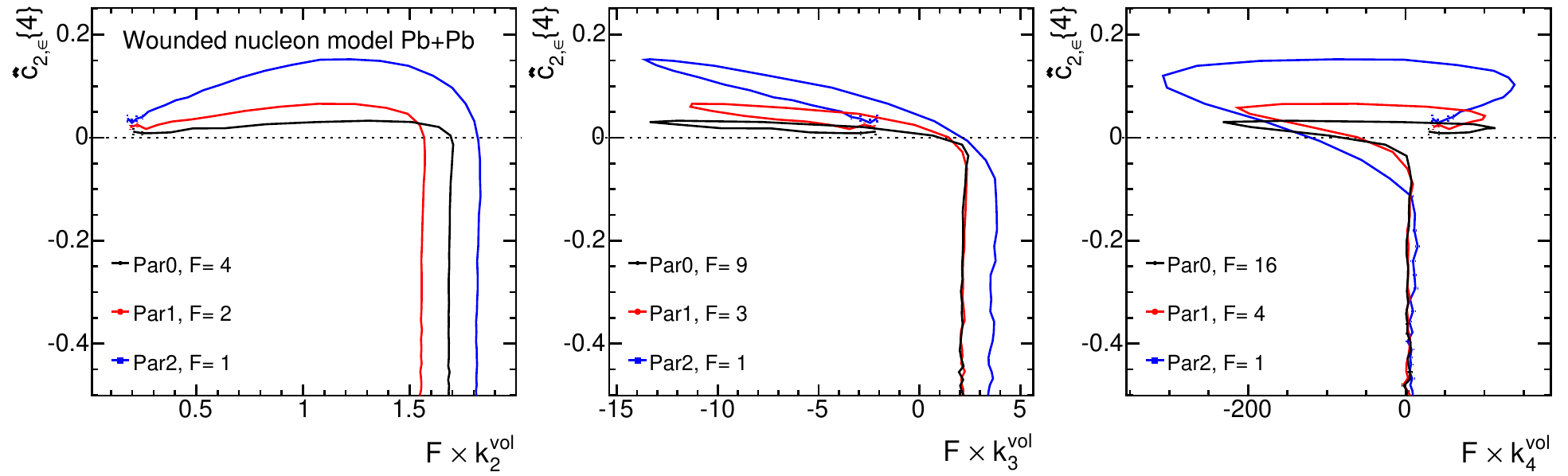}
\end{center}
\caption{\label{fig:d1} The correlation between $\hat{c}_{2,\epsilon}\{4\}$,  and cumulants for centrality fluctuations $k_m^\mathrm{v}$ obtained from the wounded nucleon model for the three parameter sets from the Table~\ref{tab:1}.}
\end{figure}

Since both multiplicity and eccentricity fluctuations are sensitive to the CF effects, it would be interesting to correlate them together. One way to do this is to measure the multiplicity cumulant and eccentricity cumulant using the same event class definition and plot one against the other. One such example is shown in Figure~\ref{fig:d1}, where the $\hat{c}_{2,\epsilon}\{4\}$ from Figure~\ref{fig:c1} and CF cumulant $k_m^\mathrm{v}$ from Figure~\ref{fig:2} are plotted against each other over the full range of $\npart$. The advantage of this correlation is that both axes represent physical quantities instead of using $\lr{\ns}$ for one axis, therefore such correlation can be directly compared to experimental measurements. The $k_m^\mathrm{v}$ is nearly independent of $\hat{c}_{2,\epsilon}\{4\}$ until reaching central collisions, where both observables are affected by CF. Such pattern could be searched for in the experimental data analysis by correlating the multiplicity cumulants with the flow cumulants.

Another, more direct way to study the multiplicity-eccentricity correlation is to construct mixed correlators between sources and eccentricity such as:
\begin{eqnarray}
\label{eq:d1}
F(k_1^{\mathrm{v}},\epsilon_n)=\frac{\lr{\ns\epsilon_n^2}}{\lr{ \ns}{\lr{\epsilon_n^2}}}-1,\;F(k_2^{\mathrm{v}},\epsilon_n)=\frac{\lr{(\delta \ns)^2\epsilon_n^2}}{\lr{(\delta \ns)^2}{\lr{\epsilon_n^2}}}-1\;,
\end{eqnarray}
which are related to the multiplicity eccentricity correlation that is easier to access experimentally:
\begin{eqnarray}
\label{eq:d2}
F(K_1,\epsilon_n)=\frac{\lr{N\epsilon_n^2}}{\lr{ N}{\lr{\epsilon_n^2}}}-1,\;F(K_2,\epsilon_n)=\frac{\lr{(\delta N)^2\epsilon_n^2}}{\lr{(\delta N)^2}{\lr{\epsilon_n^2}}}-1,
\end{eqnarray}
Expression for higher-order mixed cumulants and associated discussions are given in Appendix~\ref{sec:a2}. Assuming linear response between $\epsilon_n$ and $v_n$, these quantities should be equal to  experimentally measurable quantities $F(K_m,v_n)$, which are obtained by simply replacing $\epsilon_n$ by $v_n$ in Eq.~\ref{eq:d2}. In this section, we discuss the behaviors of $F(K_m,\epsilon_n)$ and their relations to the centrality fluctuations $F(k_m^{\mathrm{v}},\epsilon_n)$, which can be regarded as predictions for $F(K_m,v_n)$.

In the independent source model, eccentricities depend only on the positions of sources and are not correlated with multiplicity fluctuation within each source. Therefore, if centrality is selected on subevent A and particle multiplicity is calculated at subevent B, it is easy to show that:
\begin{eqnarray}\label{eq:d3}
F(K_{1,B|A},\epsilon_n)=F(k_{1,A}^{\mathrm{v}},\epsilon_n)\;,\;F(K_{2,B|A},\epsilon_n)=  \frac{k_{2,B}F(k_{1,A}^{\mathrm{v}},\epsilon_n)+\bar{n}_{B}k_{2,A}^{\mathrm{v}}F(k_{2,A}^{\mathrm{v}},\epsilon_n)}{k_{2,B}+\bar{n}_{B}k_{2,A}^{\mathrm{v}}}\;,
\end{eqnarray}
where we have added explicit subscripts to indicate that the CF cumulants $k_m^{\mathrm{v}}$ arise from subevent A, and $K_{m,B|A}$ is the multiplicity cumulants in subevent B when centrality is defined in subevent A.  If subevents A and B are replaced with subevents A$'$ and B$'$ with the same relative widths $\hat{\sigma}_A=\hat{\sigma}_{A'}$ and $\hat{\sigma}_B=\hat{\sigma}_{B'}$, one can also show that $F(K_{2,B|A},\epsilon_n)=F(K_{2,B'|A'},\epsilon_n)$.

\begin{figure}[h!]
\begin{center}
\includegraphics[width=1\linewidth]{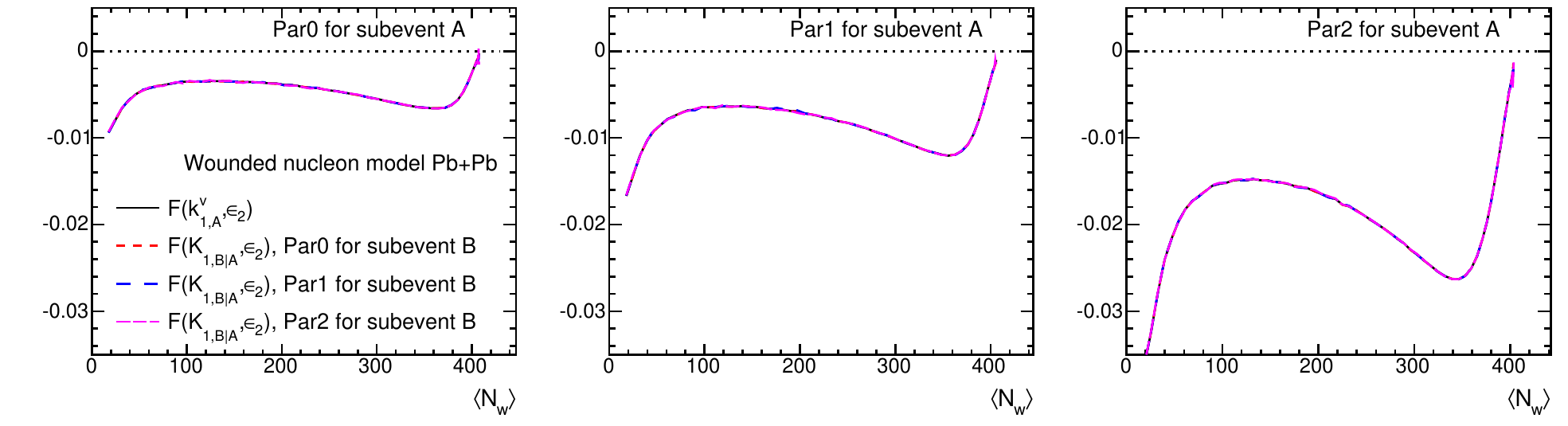}
\end{center}
\caption{\label{fig:d2} The multiplicity-eccentricity mixed cumulants $F(K_{1,B|A},\epsilon_2)$ calculated with three parameter sets for subevent B for centrality defined in subevent A, and compared with $F(k_{1,A}^{\mathrm{v}},\epsilon_2)$.  All four curves are found to be on top of each other. The three panels corresponds three different NBD parameter sets for subevent A. They are all calculated for the wounded nucleon model. }
\end{figure}

Figure~\ref{fig:d2} shows the lowest-order multiplicity-eccentricity correlation from the wounded nucleon model. The values of $F(K_{1,B|A},\epsilon_2)$ always agree with $F(k_{1,A}^{\mathrm{v}},\epsilon_2)$ independent of the parameter sets used in the subevent B. However, the effects of centrality fluctuation from subevent A has a strong influence on $F(k_{1,A}^{\mathrm{v}},\epsilon_n)$. Larger $\hat{\sigma}$ associated with parameter set Par2 gives the largest signal (right panel). The values of these correlators are slightly negative indicating an weak anti-correlation between multiplicity and eccentricity fluctuations. Note that the magnitudes of these correlators are very small, on the order of a few percent or less. Therefore they could be easily affected by finite dynamical effects not included in our simulation. 

\begin{figure}[h!]
\begin{center}
\includegraphics[width=1\linewidth]{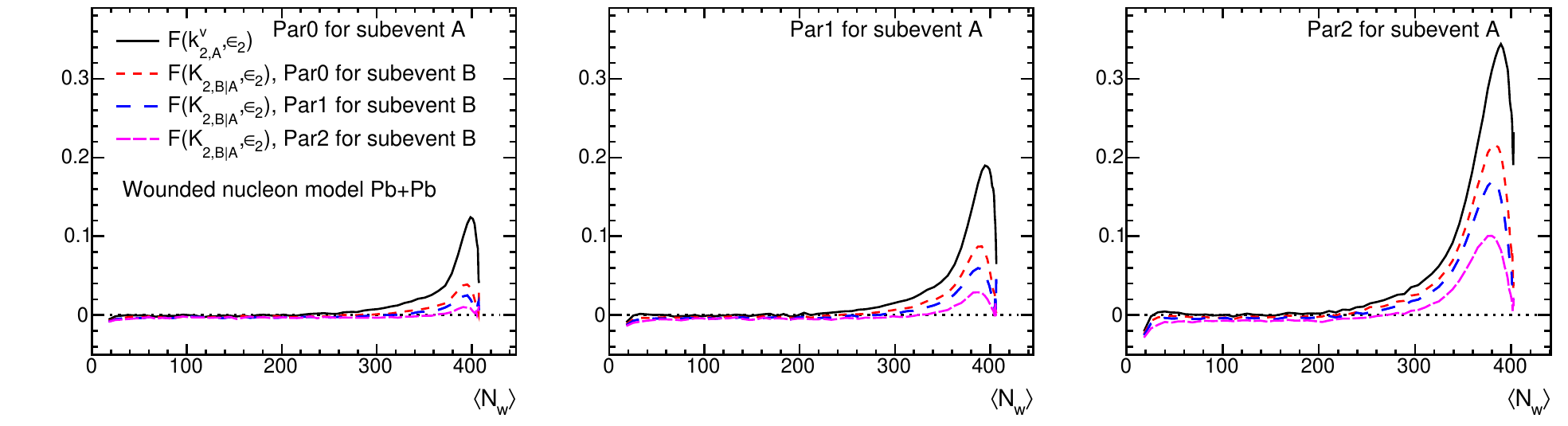}
\end{center}
\caption{\label{fig:d3} The multiplicity-eccentricity mixed cumulants $F(K_{2,B|A},\epsilon_2)$ calculated with three parameter sets for subevent B for centrality defined in subevent A, and compared with $F(k_{2,A}^{\mathrm{v}},\epsilon_2)$. The three panels corresponds three NBD parameter sets for subevent A. They are all calculated for the wounded nucleon model.}
\end{figure}

Figure~\ref{fig:d3} shows the correlation between multiplicity variance and eccentricity from the wounded nucleon model. The values of $F(K_{2,B|A},\epsilon_2)$ and $F(k_{2,A}^{\mathrm{v}},\epsilon_2)$ are small in most of the centrality range, but increase dramatically towards very central collisions, where $F(K_{2,B|A},\epsilon_2)$ show similar shape as $F(k_{2,A}^{\mathrm{v}},\epsilon_2)$ but with smaller magnitudes. Results for Par2 in subevent B have smallest magnitudes, implying that the correlation between $\ns$ and eccentricity tends to be diluted by a broader $p(n)$.

Equation~\ref{eq:d3} shows that the $F(K_{2,B|A},\epsilon_n)$ has contributions from the first- and second-order CF cumulants. Combining with Eq.~\ref{eq:13}, the second-order volume-eccentricity correlation can be expressed as:
\begin{eqnarray}\label{eq:d4}
F(k_{2,A}^{\mathrm{v}},\epsilon_n) &=&\frac{K_{2,B|A}}{K_{2,B|A}-k_{2,B}}F(K_{2,B|A},\epsilon_n)-\frac{k_{2,B}}{K_{2,B|A}-k_{2,B}}F(K_{1,B|A}^{\mathrm{v}},\epsilon_n)
\end{eqnarray}
All quantities on the r.h.s of the equation, except $k_{2,B}$, can be directly measured. The value of $k_{2,B}$ could be estimated from minimum bias $pp$ collisions.

\section{Dependence on the size of the collision system}\label{sec:6}
In the independent source model framework, the centrality fluctuations depend on the distribution of sources $p(\ns)$ and smearing from particle production for each source $p(n)$. The shape of $p(\ns)$ also changes with the size of the collision system, leading to different amount of centrality fluctuations. 
\begin{figure}[h!]
\begin{center}
\includegraphics[width=1\linewidth]{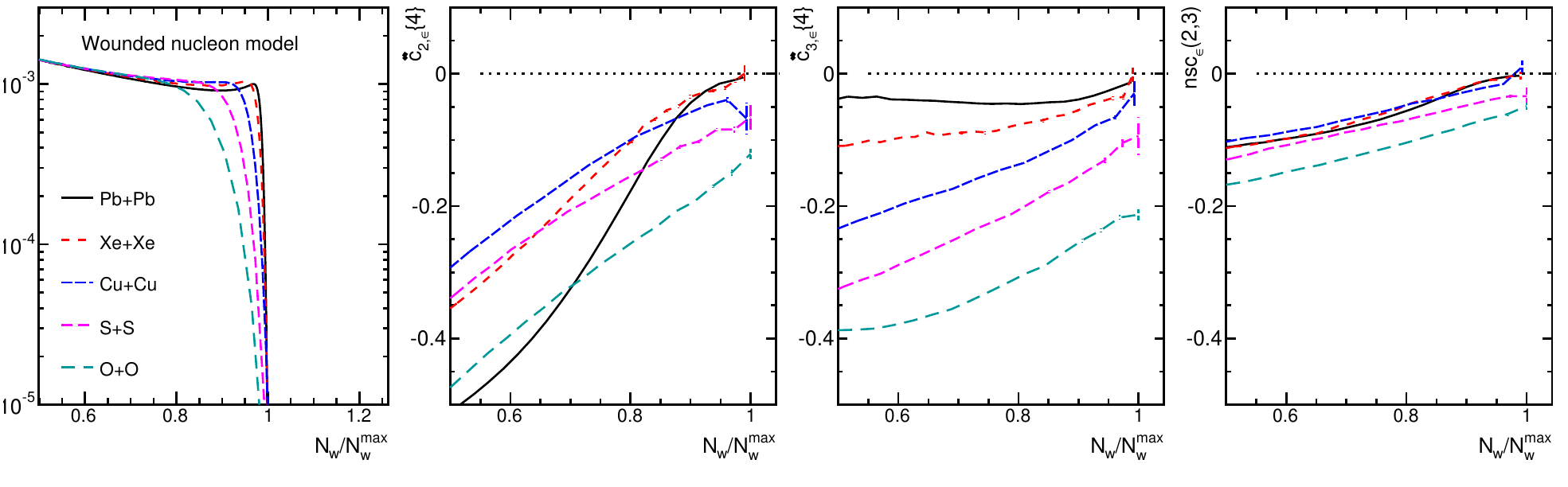}
\end{center}
\caption{\label{fig:f1} The distribution of $\npart$ (left panel), $\hat{c}_{2,\epsilon}\{4\}$ (second to the left) and $\hat{c}_{3,\epsilon}\{4\}$ (second to the right) and $\mathrm{nsc}_\epsilon(2,3)$ (right panel) as a function of $\npart/2A$. They are obtained using the wounded nucleon model from the sources without particle productions.}
\end{figure}

We studied several nuclei covering a broad range of the $\ns$: Xe+Xe~\footnote{We haven't consider the nuclear deformation of Xe, which could be important for $\epsilon_2$ in central collisions. However it is expected to not change the overall system size dependence.}, Cu+Cu, S+S and O+O collisions, with total number of nucleons $2A=$ 258, 126, 64, 32 respectively. The Glauber simulation is carried out using the same NBD parameters from Table~\ref{tab:1}, and experimental observables discussed in previous sections are calculated for each collision system. For simplicity, only results obtained with Par1 are discussed. 

Figure~\ref{fig:f1} shows the distribution of $\ns$ and eccentricity cumulants from these collision systems in central collision region prior to particle production. In order to compare them properly across different collision systems, the $x$-axes for these quantities have been rescaled by $\npart^{\mathrm{max}}=2A$. The eccentricity cumulants $\hat{c}_{n,\epsilon}\{4\}$ and $\mathrm{nsc}_\epsilon(2,3)$ remain negative over the entire $\npart$ range, and is more negative for smaller collision system.

\begin{figure}[t!]
\begin{center}
\includegraphics[width=1\linewidth]{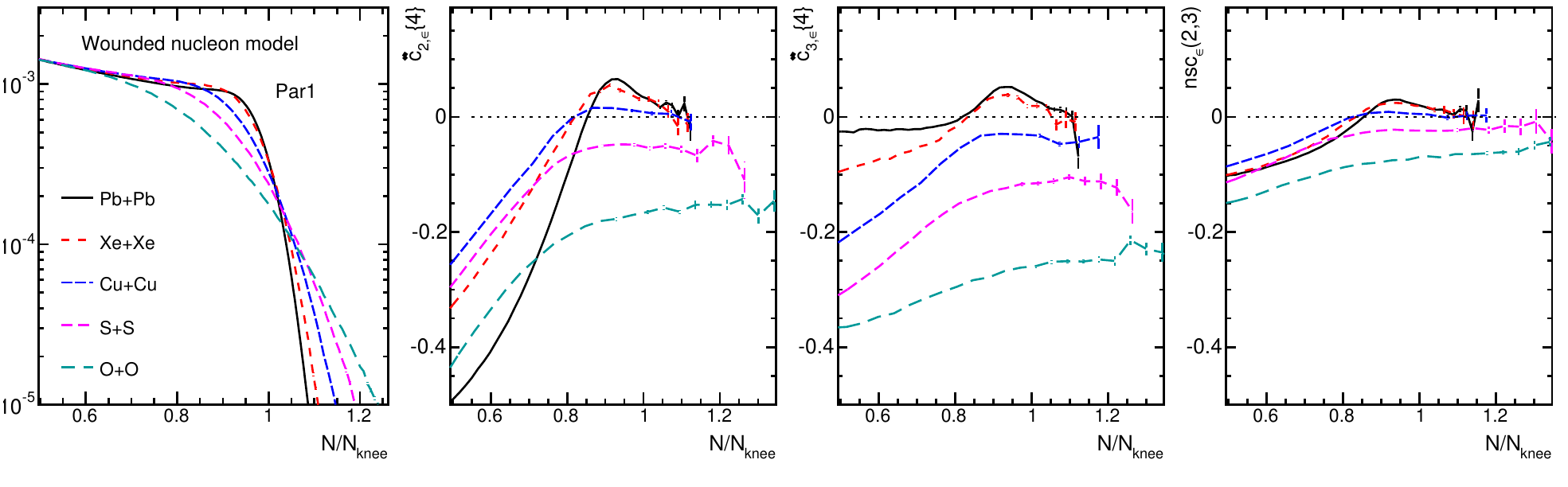}
\end{center}
\caption{\label{fig:f2}  The distributions of $\npart$ (left panel), $\hat{c}_{2,\epsilon}\{4\}$ (second to the left) and $\hat{c}_{3,\epsilon}\{4\}$ (second to the right) and $\mathrm{nsc}_\epsilon(2,3)$ (right panel) as a function of particle multiplicity scaled by the knee. The eccentricity cumulants are obtained from events binned in the final particle multiplicity $p(N)$. The calculation is done using the wounded nucleon model and Par1 from Table~\ref{tab:1}.}
\end{figure}

Figure~\ref{fig:f2} shows the same quantities for event class binned in particle multiplicity produced with Par1 from Table~\ref{tab:1}. The $x$-axes have been rescaled by the average multiplicity for maximum number of sources, $N_{\mathrm{knee}}=2A\bar{n}$. The $p(N)$ distributions show significant tails in smaller systems comparing to large systems, which are expected to affect the centrality fluctuations and eccentricity fluctuations. Indeed, the $\hat{c}_{n,\epsilon}\{4\}$ values show significant convex shape in the ultra-central region, which leads to a sign-change when the collision systems are large enough. It is interesting to note that the sign for O+O system never changes. This is because the $\hat{c}_{n,\epsilon}\{4\}$ values are significantly negative in Figure~\ref{fig:f1}, any additional smearing from $p(n)$ apparently is not sufficient to make it to flip sign.  
\begin{figure}[t!]
\begin{center}
\includegraphics[width=1\linewidth]{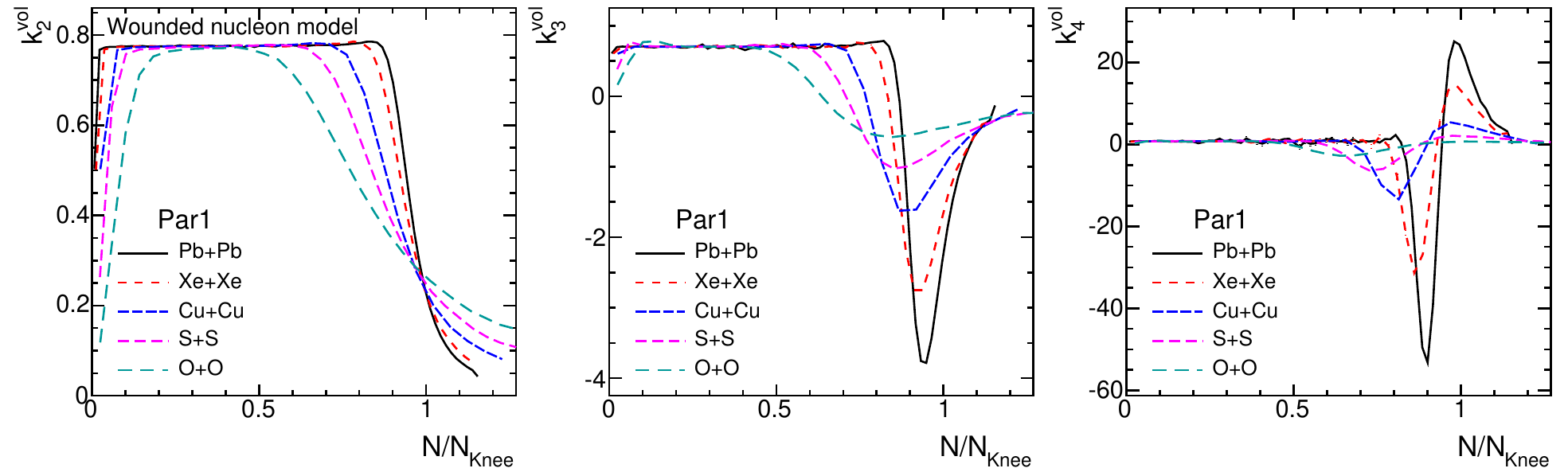}
\end{center}
\caption{\label{fig:f3} Cumulants for centrality fluctuations of different order, $k_1^{\mathrm{v}}$ (left), $k_2^{\mathrm{v}}$ (middle) and $k_3^{\mathrm{v}}$ (right) as a function of scaled multiplicity for different collision systems. They are calculated for parameter set Par1 of the NBD distribution in the wounded nucleon model.}
\end{figure}

\begin{figure}[h!]
\begin{center}
\includegraphics[width=1\linewidth]{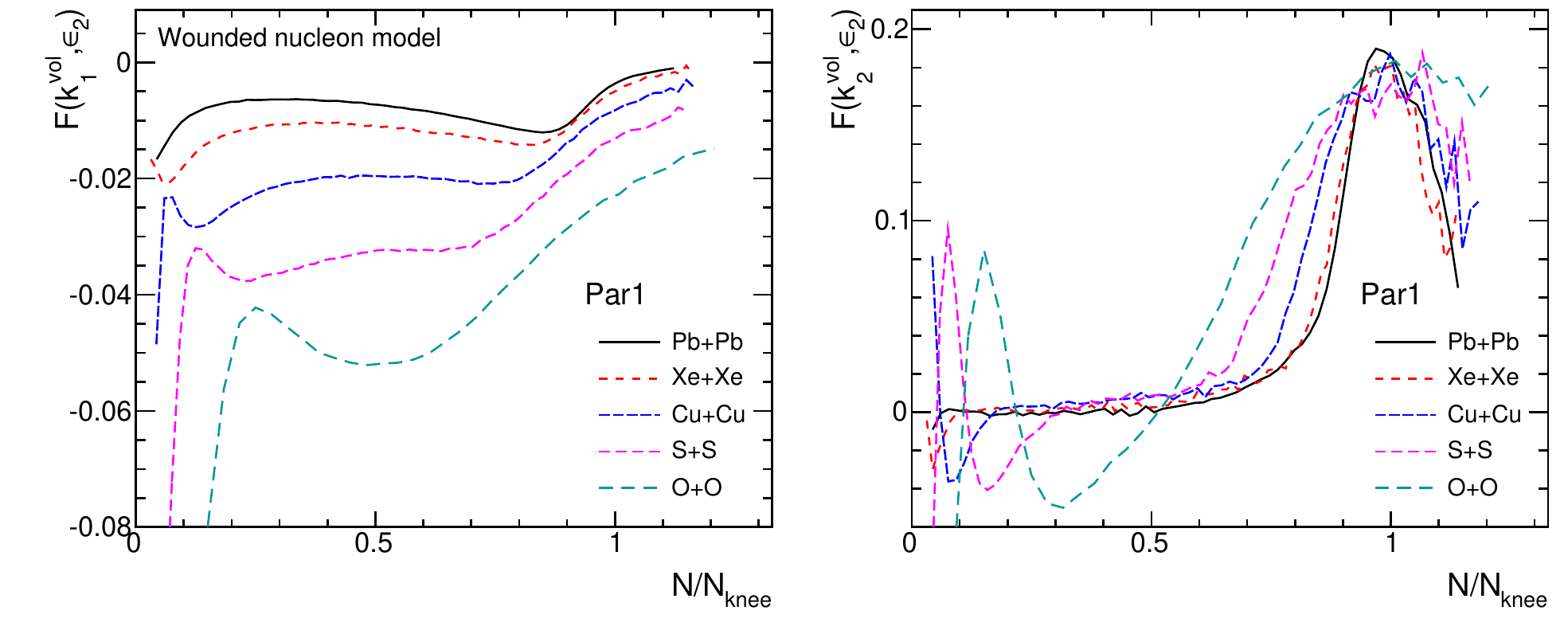}
\end{center}
\caption{\label{fig:f4} The multiplicity-eccentricity mixed cumulants $F(k_1^{\mathrm{v}},\epsilon_2)$ (left) and $F(k_2^{\mathrm{v}},\epsilon_2)$ (right) as a function of scaled multiplicity for different collision systems. They are calculated for parameter set Par1 of the NBD distribution in the wounded nucleon model.}
\end{figure}

Figure~\ref{fig:f3} shows the cumulants for centrality fluctuations. As the collision system becomes smaller, the centrality range where centrality fluctuations play a significant role becomes much wider. However, the maximum magnitudes of these cumulants in central collisions become smaller. Figure~\ref{fig:f4} shows the multiplicity-eccentricity mixed cumulants $F(k_1^{\mathrm{v}},\epsilon_2)$ and $F(k_2^{\mathrm{v}},\epsilon_2)$, the magnitude of $F(k_1^{\mathrm{v}},\epsilon_2)$ increases dramatically for smaller collision system, but more or less similar between different systems for $F(k_2^{\mathrm{v}},\epsilon_2)$. These results suggest that study of multiplicity and flow mixed cumulants for different collision system are sensitive to the nature of the sources and particle production mechanism.  Measurement of system-size dependence of cumulants in ultra-central collisions could provide detailed information on the nature of coupling between centrality fluctuations and flow fluctuations, and help us to understand whether and how the sources driving particle production and sources driving the flow fluctuation are related to each other.

\begin{figure}[h!]
\begin{center}
\includegraphics[width=1\linewidth]{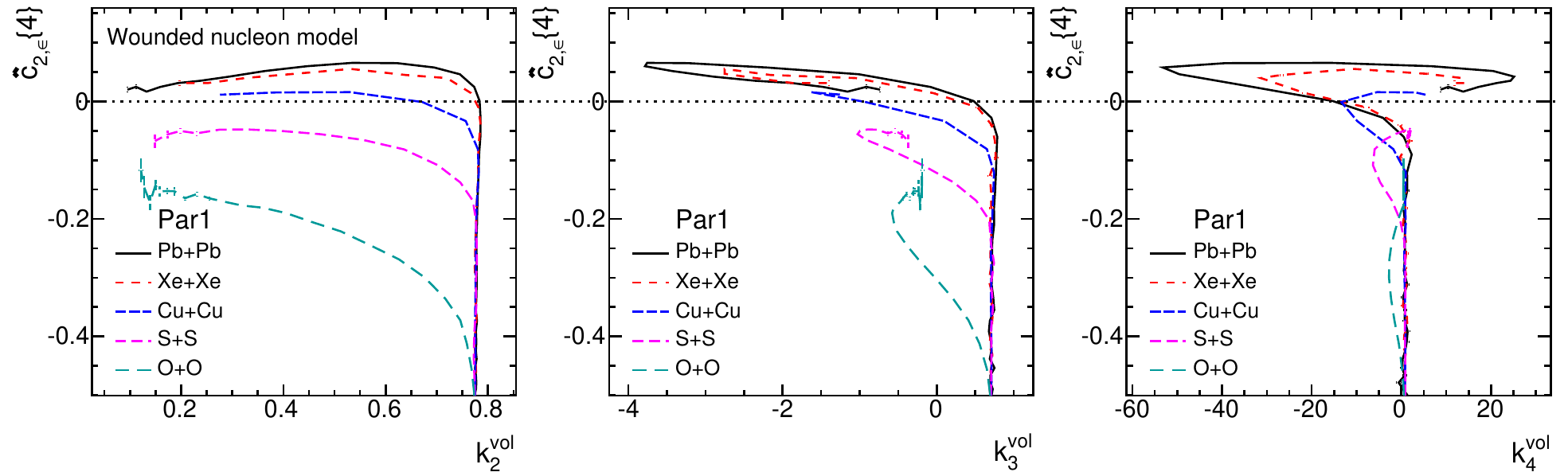}
\end{center}
\caption{\label{fig:f5}  The correlation between $\hat{c}_{2,\epsilon}\{4\}$ and cumulants for CF cumulants $k_m^\mathrm{v}$ obtained from the wounded nucleon model for the parameter set Par1 from Table~\ref{tab:1}.}
\end{figure}

Last but not the least, Figure~\ref{fig:f5} shows the correlation between centrality fluctuation cumulant $k_m^\mathrm{v}$ and eccentricity cumulant $\hat{c}_{2,\epsilon}\{4\}$, both obtained using the same event centrality definition according to Par1. Interesting system-size dependent behavior is observed in central collisions.  Such pattern could be searched for in the experimental data analysis.

\section{Subevent multiplicity cumulants and eccentricity cumulants}\label{sec:7}

In the standard cumulant method discussed before, all particles passing the selection criteria are treated as a single event, and all combinations, pairs, triplets... among these particles are considered. In this section, we consider cumulants calculated from subevents, i.e. particles in the combination are taken from different regions in $\eta$.~\footnote{This should be distinguished from the subevent A for centrality selection and subevent B used for cumulant analysis discussed earlier. Subevent A is always assumed for centrality selection in all cumulant methods. The key of subevent cumulant is to subdivided particles used for analysis (subevent B) into smaller subevents and only take combinations across these different smaller subevents.} This so-called subevent cumulant method has been used in flow correlation studies and has the advantage to suppress short-range correlations, and to expose long-range dynamics of the event~\cite{Jia:2017hbm,Aaboud:2017blb,Huo:2017nms,Nie:2018xog}. Since centrality is a global property of the event, multiplicity cumulants based on subevent method is a useful tool to study the centrality fluctuations.

Lets consider several subevents $a$,$b$, $c$..., each from a unique $\eta$ range, and we construct the following ``normalized'' two- and three-particle cumulants:
\begin{eqnarray}\label{eq:e1}
C_{2,ab}=\frac{\lr{\delta N_a\delta N_a}}{\bar{N}_a\bar{N}_b}\;,C_{3,abc} = \frac{\lr{\delta N_a\delta N_b\delta N_c}}{\bar{N}_a\bar{N}_b\bar{N}_c},\;C_{3,a^2b} = \frac{\lr{(\delta N_a)^2\delta N_b}}{\bar{N}_a^2\bar{N}_b}\;.
\end{eqnarray}
$C_{3,a^2b}$ corresponds to three-particle cumulants where two particles are chosen from subevent $a$ and one particle from subevent $b$. This notation is similar to the reduced correlation functions of Ref.~\cite{Bzdak:2016sxg}. We can identify the following four types of subevent four-particle cumulants,
\begin{eqnarray}\nonumber
C_{4,abcd} &=& \frac{\lr{\delta N_a\delta N_b\delta N_c\delta N_d}-\lr{\delta N_a\delta N_b}\lr{\delta N_c\delta N_d}-\lr{\delta N_a\delta N_c}\lr{\delta N_b\delta N_d}-\lr{\delta N_a\delta N_d}\lr{\delta N_b\delta N_c}}{\bar{N}_a\bar{N}_b\bar{N}_c\bar{N}_d}\\\nonumber
C_{4,a^2b^2} &=& \frac{\lr{(\delta N_a)^2(\delta N_b)^2}-\lr{(\delta N_a)^2}\lr{(\delta N_b)^2}-2\lr{\delta N_a\delta N_b}^2}{\bar{N}_a^2\bar{N}_b^2}\\\nonumber
C_{4,a^2bc} &=& \frac{\lr{(\delta N_a)^2\delta N_b\delta N_c}-\lr{(\delta N_a)^2}\lr{\delta N_b\delta N_c}-2\lr{\delta N_a\delta N_b}\lr{\delta N_a\delta N_c}}{\bar{N}_a^2\bar{N}_b\bar{N}_c}\\\label{eq:e2}
C_{4,a^3b} &=& \frac{\lr{(\delta N_a)^3\delta N_b}-3\lr{(\delta N_a)^2}\lr{\delta N_a\delta N_b}}{\bar{N}_a^3\bar{N}_b}
\end{eqnarray}
Same definitions can be obtained for correlation within each source by replacing upper-case letters to lower-case letters. Some examples are given below:
\begin{eqnarray}\nonumber
&&c_{2,ab} = \frac{\lr{\delta n_a\delta n_b}}{\bar{n}_a\bar{n}_b}\;,\;c_{3,abc} = \frac{\lr{\delta n_a\delta n_b\delta n_c}}{\bar{n}_a\bar{n}_b\bar{n}_c}\;,\\\label{eq:e3}
&&c_{4,abcd} = \frac{\lr{\delta n_a\delta n_b\delta n_c\delta n_d}-\lr{\delta n_a\delta n_b}\lr{\delta n_c\delta n_d}-\lr{\delta n_a\delta n_c}\lr{\delta n_b\delta n_d}-\lr{\delta n_a\delta n_d}\lr{\delta n_b\delta n_c}}{\bar{n}_a\bar{n}_b\bar{n}_c\bar{n}_d}
\end{eqnarray}

For comparison purpose, we also define the normalized cumulants without using subevents:
\begin{eqnarray}\nonumber
&&C_{2}=\frac{\lr{(\delta N)^2}}{\bar{N}^2}\;,C_{3} = \frac{\lr{(\delta N)^3}}{\bar{N}^3}\;,C_{4} = \frac{\lr{(\delta N)^4}-3\lr{(\delta N)^2}^2}{\bar{N}^4}\\\label{eq:e4}
&&c_{2}=\frac{\lr{(\delta n)^2}}{\bar{n}^2}\;,c_{3} = \frac{\lr{(\delta n)^3}}{\bar{n}^3}\;,c_{4} = \frac{\lr{(\delta n)^4}-3\lr{(\delta n)^2}^2}{\bar{n}^4}
\end{eqnarray} 
They are related to the previously defined cumulants by a normalization factor: $C_m\equiv K_m/\bar{N}^{m-1}$ and $c_m\equiv k_m/\bar{n}^{m-1}$.

Just like for the standard cumulants, the subevent multiplicity cumulants contain contributions from correlations within each source and fluctuation of the $\ns$. In the independent source model, one can show that:
\begin{eqnarray}\nonumber
&&C_{2,ab} = \frac{1}{\lr{\ns}}\frac{\lr{\delta n_a\delta n_b}}{\bar{n}_a\bar{n}_b}+\frac{\lr{\delta \ns^2}}{\lr{\delta \ns}^2}= \frac{c_{2,ab}+k_2^{\mathrm{v}}}{\lr{\ns}}\\\nonumber
&&C_{3,abc} = \frac{c_{3,abc}+(c_{2,ab}+c_{2,ac}+c_{2,bc})k_{2}^{\mathrm{v}}+k_{3}^{\mathrm{v}}}{\lr{\ns}^2}\\\label{eq:e5}
&&C_{4,abcd}=\frac{c_{4,abcd}+(c_{2,ab}c_{2,cd}+pe.+c_{3,abc}+pe.)k_2^{\mathrm{v}}+(c_{2,ab}+pe.)k_3^{\mathrm{v}}+k_4^{\mathrm{v}}}{\lr{\ns}^3}
\end{eqnarray}
``pe.'' is a short-hand notation for other terms obtained from permutation of $abcd$, e.g. $c_{2,ab}c_{2,cd}+pe.\equiv c_{2,ab}c_{2,cd}+c_{2,ac}c_{2,bd}+c_{2,ad}c_{2,bc}$, $c_{3,abc}+pe.\equiv c_{3,abc}+ c_{3,abd}+ c_{3,acd}+ c_{3,bcd}$, and $c_{2,ab}+pe.\equiv c_{2,ab}+c_{2,ac}+c_{2,ad}+c_{2,bc}+c_{2,bd}+c_{2,cd}$. From these we can see normalized cumulants $C_m$ scales as $1/\lr{\ns}^{m-1}$.

One advantage of subevent cumulants is that they are much less susceptible to the statistical bias present in the standard cumulants~\cite{Chatterjee:2017mhc}. To see this, let consider events with very small acceptance. In this case, $p(n)$ follows a Poisson distribution, and $c_m=1/\bar{n}^{m-1}$ which diverges when $\bar{n}\rightarrow 0$. Therefore, the non-vanishing values of $c_m$ reflect pure statistical effect and do not carry any dynamical information. In contrast, since the statistical fluctuations are uncorrelated between different subevents, the subevent cumulants $c_{2,ab}$, $c_{3,abc}$ and $c_{4,abcd}$ carry only dynamical correlations, therefore one expects $c_{2,ab}\ll c_2$, $c_{3,abc}\ll c_3$ and $c_{4,abcd}\ll c_4$ for small detector acceptance. This conclusion is also true for $C_m$. 

ATLAS has previously measured $C_{2,ab}(\eta_a,\eta_b)$ in $pp$ collisions~\cite{Aaboud:2016jnr}, where the two subevents are chosen differentially in $\eta$ over $|\eta|<2.4$. The $C_{2,ab}$ signal is decomposed into a short-range and a long-range component. The short-range component has approximately a Gaussian shape in $\Delta\eta=\eta_a-\eta_b$ with a width of about one unit, while the long-range component scales as $C_{2,ab}\approx a\eta_a\eta_b$. The magnitude of long-range component decreases as a function of $\nchrec$, and the value of the coefficient $a$ is about 0.015 in minimum bias $pp$ collisions (see Figure 7 in the auxiliary figure of Refs.\cite{Aaboud:2016jnr}). Therefore with a reasonable rapidity gap requirement, subevent cumulants for each source is expected to have very small values, assuming each source can be approximated by the $pp$ collision.  This is expected as both the short-range correlation and statistical bias has been suppressed. In this case, the subevent cumulants for total multiplicity are expected to be dominated by centrality fluctuations:
\begin{eqnarray}\label{eq:e6}
C_{2,ab} \approx \frac{k_2^{\mathrm{v}}}{\lr{\ns}}, C_{3,abc} \approx \frac{k_{3}^{\mathrm{v}}}{\lr{\ns}^2},\;C_{4,abcd}\approx\frac{k_4^{\mathrm{v}}}{\lr{\ns}^3}\;.
\end{eqnarray}
This is especially true if the subevent used for centrality selection has small acceptance and therefore larger centrality fluctuations.

The subevent method also provide a natural way to use mixed event technique to correct for detector effects, such as non-binomial detector response~\cite{Bzdak:2016qdc,He:2018mri}. Take $C_{3,abc}$ as an example, it can be measured in narrow pseudorapidity bins and then integrated to broader $\eta$ range to recover the standard cumulant result~\cite{Bzdak:2016sxg}. The differential distribution also provide a way to handle correlated detector effects such as track splitting or merging effects, for example by smoothing the non-physical structures in the differential distribution $C_{3,abc}(\eta_a,\eta_b,\eta_c)$ before the integration.

Another property of subevent cumulants is that for $C_{2,ab}$, $C_{3,abc}$ and $C_{4,abcd}$, where one particle is taken from each subevent in constructing each pair, triplet or quadruplet, their factorial cumulant/moment are the same as the cumulant/moment. This feature simplifies some technical difficulties in experimental analysis, such as efficiency correction.

One can also generalize the subevent cumulant to the multiplicity-eccentricity mixed cumulants. For example, mixed cumulant between multiplicity variance and eccentricity in Eq.~\ref{eq:d2} can be written as:
\begin{eqnarray}
\label{eq:e7}
F(K_{2,ab},\epsilon_n)=\frac{\lr{\delta N_a\delta N_b\epsilon_n^2}}{\lr{\delta N_a\delta N_b}{\lr{\epsilon_n^2}}}-1=\frac{c_{2,ab}F(k_{1}^{\mathrm{v}},\epsilon_n)+k_{2}^{\mathrm{v}}F(k_{2}^{\mathrm{v}},\epsilon_n)}{c_{2,ab}+k_{2}^{\mathrm{v}}}\approx \frac{k_{2}^{\mathrm{v}}}{c_{2,ab}+k_{2}^{\mathrm{v}}}F(k_{2}^{\mathrm{v}},\epsilon_n),
\end{eqnarray}
where we have used the approximation that both $c_{2,ab}$ and $F(k_{1}^{\mathrm{v}},\epsilon_n)$ are small numbers. Therefore, the subevent cumulants probe more directly the correlations between centrality fluctuation and eccentricity fluctuation.

Our independent source model does not contain longitudinal dynamics within each source, therefore it can not be used to make precise prediction for the behavior of the subevent cumulants. It would be interesting to extend the current model framework to use $\pp$ collisions to approximate the correlations within each source, as well as to calculate these observables using dynamic model of heavy-ion collisions, such as HIJING~\cite{Gyulassy:1994ew}, AMPT~\cite{Lin:2004en} and event-by-event hydrodynamics models with full 3D dynamical initial state~\cite{Shen:2017bsr}. We leave this to a future work.

The standard cumulant and subevent cumulant method can be generalized to any bulk observables that is sensitive to the global fluctuations of the system. One example that has been studied recently is the transverse momentum fluctuations, which is sensitive the transverse size of the fireball. A event with smaller transverse size accumulates stronger radial flow, and therefore a larger average transverse momentum, $\lrave{\pT}$~\cite{Broniowski:2009fm}. This physics is beyond the scope of our independent source model, but has been studied in event-by-event hydrodynamics models. The authors of Refs.~\cite{Chatterjee:2017mhc,Bozek:2016yoj} have studied the correlation of the $\lrave{\pT}$ between two pseudorapidity bins, as well as mixed-correlation between $\lr{\pT}$ and harmonic flow $v_n$ from several pseudorapidity bins. These correlations have been shown to be less affected by non-flow backgrounds and more sensitive to the transverse size fluctuations of the fireball.

\section{Summary and discussion}\label{sec:8}
In heavy-ion collisions, due to fluctuations in the particle production process, the centrality or the volume of the fireball for events selected to have the same final-state particle multiplicity fluctuates from event to event. This so-called volume or centrality fluctuations lead to significant uncertainties in interpreting centrality dependence of experimental observables. This paper investigates the effects of centrality fluctuations on multiplicity and flow fluctuations in an independent source model framework, which simulates the particle multiplicity as a superposition of particles from $\ns$ uncorrelated sources in each event, $N=\sum_{i=1}^{\ns} n_i$, where the particle multiplicity in each source $n_i$ follows a common negative binomial probability distribution $p(n)$. A Glauber model is used to simulate the transverse distribution of sources in each event, and to calculate the eccentricity $\epsilon_n$. Following the standard experimental centrality selection procedure, the centrality fluctuations are imposed by selecting events with fixed $N$, which contribute to event-by-event fluctuations of source multiplicity and eccentricities: $p(\ns)$, $p(\epsilon_n)$, $p(\epsilon_n,\epsilon_m)$, and $p(\ns,\epsilon_n)$. These distributions are directly related to experimentally measurable distributions: $p(N)$, $p(v_n)$, $p(v_n,v_m)$, and $p(N,v_n)$. The main goal of this paper is to propose a set of cumulant observables related to these distributions and study their sensitivities to centrality fluctuations. 

We first study the multiplicity cumulants $K_m$, which describe multiplicity fluctuation $p(N)$ in Pb+Pb collisions. In experimental data analysis, the centrality is typically defined in one subevent A, and multiplicity cumulants are calculated in a different subevent B for events in a narrow centrality range. The total multiplicity cumulant $K_{m,B|A}$, therefore receives contributions from cumulant for each source in subevent B $k_{m,B}$ and centrality cumulants for the sources $\ns$ in subevent A, $k_{m,A}^{\mathrm{v}}$ (see Eq.~\ref{eq:13}). Since our model assumes $k_m$ to be independent of $\ns$, the centrality dependence of $K_{m,B|A}$ is mainly controlled by $k_{m,A}^{\mathrm{v}}$. We also studied relation between the multiplicity cumulant for each source $k_{m,A}$ and the resulting centrality cumulants $k_{m,A}^{\mathrm{v}}$ in subevent A. In mid-central collisions where $p(\ns)$ is a slow-varying function, we derived a general formula Eq.~\ref{eq:10} relating $k_{m|A}$ and $k_{m,A}^{\mathrm{v}}$. This formula is valid independent of the functional form of $p(n)$ for the second-order cumulant or scaled variance ($m=2$), but is valid for higher-order cumulants when $p(n)$ is negative binomial distribution (NBD). In central collisions where $p(\ns)$ is a rapidly-changing function, the $k_{m,A}^{\mathrm{v}}$ is found to be dependent on the relative width of the $p(n)$, $\hat{\sigma}$.

Next we study the influence of the fluctuation of sources on the eccentricities $\epsilon_n$, which characterize the shape of the collision zone and drive the final state harmonic flow $v_n$. We found that the centrality fluctuations for a given centrality selection criteria influence significantly $p(\epsilon_n)$ and $p(\epsilon_n,\epsilon_m)$. This is especially true in central collisions, where eccentricity fluctuations are very sensitive to any non-Gaussianity introduced by centrality fluctuations. Indeed, we found that the four-, six- and eight-particle cumulants for $\epsilon_2$ and $\epsilon_3$ exhibit rather complex sign-change patterns in central collisions, indicative of significant non-Gaussianity in $p(\epsilon_n)$. Similar sign-change patterns are also observed for four-particle symmetric cumulants between $\epsilon_2$ and $\epsilon_3$, and between $\epsilon_2$ and $\epsilon_4$, consistent with significant non-Gaussianity of $p(\epsilon_2,\epsilon_3)$ and $p(\epsilon_2,\epsilon_4)$ in central collisions. We found these eccentricity cumulants are sensitive to the underlying $p(\ns)$; they are also sensitive to the $\hat{\sigma}$ of $p(n)$ but not its functional form.  We also studied mixed cumulants between multiplicity and eccentricity to probe $p(\ns,\epsilon_n)$. We found a small anti-correlation between $\ns$ and $\epsilon_n^2$, but a rather strong correlation between variance of $\ns$, $(\delta\ns)^2$, and $\epsilon_n^2$ in central collisions. Note that the current studies ignores the fact that impact parameter of nucleon-nucleon maybe correlated with A+A centrality, which may lead to small centrality dependent biases on the calculated eccentricity values.

These studies are also repeated for smaller collisions systems. We found the non-Gaussian behavior of centrality fluctuations due to boundary effect of $p(\ns)$ plays a more important role: it influences $k_{m}^{\mathrm{v}}$ in a wider centrality range but the overall magnitudes of $k_{m}^{\mathrm{v}}$ are smaller. The sign-change of eccentricity cumulants become less dramatic or disappear in small systems. The anti-correlation between $\ns$ and $\epsilon_n^2$ becomes stronger while the correlation between $(\delta\ns)^2$ and $\epsilon_n^2$ remains approximately the same for smaller systems. Repeating these studies in experimental data and comparing with model predictions could improve our understanding of the particle production mechanism, as well as the nature of sources and their fluctuations as function of system size.

We also explore the possibility of using pseudorapidity-separated subevents to study the multiplicity and flow fluctuations. This so-called subevent cumulants avoid the statistical bias arising from auto-correlations which complicate the interpretation of the experimental results; they also allow a systematic separation of the short-range final-state dynamical effects from long-range global event characteristics which often can be associated with centrality fluctuations. The mathematical expression is simplified in the subevent method, and the results are expected to be closer to the expectation from global centrality fluctuations (see Eqs.~\ref{eq:e6} and \ref{eq:e7}). The subevent method also provides a natural way to use mixed-event technique to correct for detector effects, such as non-binomial detector response. They can be measured in narrow $\eta$ bins and then integrated to broader $\eta$ range to recover the standard cumulant result~\cite{Bzdak:2016sxg}. The differential distribution provides a handle for the correlated detector effects such as track splitting or merging effects, for example by smoothing non-physical structures in the differential distribution before integration. Recently, a lot of experimental efforts have been devoted to studies of fluctuation of conserved charges, such as net-charge or net-proton fluctuations~\cite{Luo:2017faz}. Since the centrality is defined in a subevent separated from cumulant measurement, the centrality fluctuations could be very important non-critical flucutation background for these analyses~\cite{Li:2017via}. 

Our present study assumes that the sources are boost-invariant in the longitudinal direction. In reality, the number of sources $\ns$ as well as their distributions in the transverse plane may fluctuate in rapidity: 1) in models based on string picture~\cite{Bozek:2015bna,Pang:2015zrq,Shen:2017bsr}, the number of strings, their lengths and end-points in rapidity fluctuates. 2) The sub-nucleonic degree-of-freedoms are expected to evolve with rapidity~\cite{Schenke:2016ksl}: in the forward rapidity, the projectile nucleons are dominated by a few large-$x$ partons, while the target nucleons are expected to contribute mainly low-$x$ soft gluons. 3) The number of forward-going and backward-going participating nucleons, $\npartf$ and $\npartb$, are not the same in a given event~\cite{Jia:2015jga,Jia:2014ysa}.  Due to these reasons, the $\ns$ in general should be a function of $\eta$ even in a single event, which tends to weaken the centrality correlation between different rapidities. This also means that a simple combination of particles from two very different rapidity regions may not improve the centrality resolution if the longitudinal fluctuations are large. We plan to extend our model framework to explore this direction in the future. 


We believe that the study of the inter-event longitudinal fluctuations as a way to infer the bulk characteristic of the entire A+A event will be an important direction of heavy-ion research. Subevent correlation or subevent cumulant method is a valuable tool to disentangle physics happening at different time scales. Initial studies on flow and multiplicity fluctuations have been performed at RHIC~\cite{Back:2006id,Abelev:2009ag} and the LHC~\cite{Khachatryan:2015oea,Aaboud:2016jnr,Aaboud:2017blb,Aaboud:2017tql} but with rather limited $\eta$ range, comparing to their respective beam rapidities. At RHIC, STAR experiment has embarked on a very significant forward upgrade program, which extends the rapidity coverage for particle identification from $|\eta|<0.9$ to $|\eta|<1.5$~\cite{STAR1}, as well as instrumenting the forward region $2.5<\eta<4$ with tracking detector and calorimeter~\cite{STAR}. A forward upgrade has also been planned for the sPHENIX experiment~\cite{Adare:2015kwa}. Experiments at LHC also proposed forward upgrades~\cite{LHC}, mostly notably the upgrades from ATLAS and CMS to extend rapidity coverage of tracking from $|\eta|<2.5$ to $|\eta|<4$.  Another interesting possiblity is to directly measure $\npart$ in each event, therefore the $p(\npart)$, by detecting all spectator fragments using a dedicated ``centrality detector''~\cite{Tarafdar:2014oua}, which should provide strong model-independent constraints on the centrality fluctuations~\footnote{This is an extention to the commonly used Zero-Degree Calorimeter at RHIC and LHC, which detects only a small fraction of all spectators.}. These upgrades will allow us to work towards a complete picture the bulk characteristic of the entire event in the coming years.

We appreciate valuable comments from Xiaofeng Luo and Volker Koch. This research is supported by National Science Foundation under grant number PHY-1613294. 
\appendix
\section{From multiplicity fluctuation to centrality fluctuation}\label{sec:a1}
In this section, we discuss how the centrality selection in a subevent is related to the resulting centrality fluctuation in the same subevent. For large $\ns$,  $p(N;\ns)$ can be approximated by a narrow Gaussian distribution using central-limit theorem~\cite{Begun:2006uu}: 
\begin{eqnarray}
\label{eq:a1}
p(N;\ns) \approx \frac{1}{\sqrt{2\pi \sigma^2 \ns}} e^{-\frac{(N-\bar{n}\ns)^2}{2\sigma^2\ns}}, \bar{N}=\bar{n}\ns
\end{eqnarray}
Therefore the moments of multiplicity fluctuation can be written as:
\begin{eqnarray}
\label{eq:a2}
\lr{(N-\bar{N})^k}_{\ns} =  \int (N-\bar{N})^k p(N;\ns) dN \approx \frac{\sqrt{\bar{n}}}{\sqrt{2\pi\sigma^2}} \bar{N}^{k+1/2} \int t^k e^{-t^2\frac{\bar{n}\bar{N}}{2\sigma^2}} dt,\; t \equiv \frac{N-\bar{N}}{\bar{N}}\;.
\end{eqnarray}
Similarly the $\ns$ fluctuation for at fixed $N$, which we denote it also as $\bar{N}$, can be approximated by
\begin{eqnarray}
\nonumber
\lr{(\ns-\bar{\ns})^k}_{N=\bar{N}} &=& \int (\ns-\bar{\ns})^k p(\bar{N};\ns) p(\ns) d\ns \\\nonumber
&\approx& \frac{1}{\bar{n}^k}\frac{\sqrt{\bar{n}}}{\sqrt{2\pi\sigma^2}} \bar{N}^{k+1/2} \int t^k (1+t)^{-1/2} e^{-\frac{t^2}{1+t}\frac{\bar{n}\bar{N}}{2\sigma^2}} \frac{1}{\bar{n}}p(\ns) dt\\\label{eq:a3}
&\approx& \frac{1}{\bar{n}^k}\frac{\sqrt{\bar{n}}}{\sqrt{2\pi\sigma^2}} \bar{N}^{k+1/2}\int x^k(1+x/2)^k e^{-x^2\frac{\bar{n}\bar{N}}{2\sigma^2}} \frac{1}{\bar{n}}p(\frac{\bar{N}}{\bar{n}}(1+x)) dx, x \equiv t^2/(1+t), t \equiv \frac{\bar{n}\ns-\bar{N}}{\bar{N}}
\end{eqnarray}
In these equations, the range where the integrand is significant is limited to $|t|, |x|<<1$. 

In mid-central collisions, where $p(\ns)$ can be treated as constant, the following approximation holds for even $k$:
\begin{eqnarray}
\label{eq:a4}
\lr{(\ns-\bar{\ns})^k}_{N=\bar{N}} \approx \bar{n}^k\lr{(N-\bar{N})^k}_{\ns}, k=2,4,..
\end{eqnarray}
From this, we obtain a useful relation for the scaled variance
\begin{eqnarray}
\label{eq:a5}
k_{2}\approx \bar{n} k_{2}^{\mathrm{v}}\;,
\end{eqnarray}
which relates the multiplicity fluctuation for each source with the corresponding centrality fluctuation obtained by fixing $N$ in the same subevent. The approximation Eq.~\ref{eq:a1}, however, is not sufficient for calculating the third- and higher-order cumulants, since they are zero for Gaussian fluctuation. We need to consider non-Gaussian feature of $p(N;\ns)$.

If $p(n)$ follows NBD distribution, some useful relations can be derived between $\lr{(\delta\ns)^k}_{N=\bar{N}}$ and $\lr{(\delta N)^k}_{\ns}$. Using the relation $n\pnbd(n;m,p) = m\pnbd(m;n,1-p)$, Eqs.~\ref{eq:a2} and \ref{eq:a3} become:
\begin{eqnarray}
\label{eq:a2b}
\lr{(\delta N)^k}_{\ns} &=&  \int (\delta N)^k \pnbd(N;m\ns,p) dN = \int (\delta N)^k \pnbd(N;\frac{1-p}{p}\bar{N},p) dN \\\nonumber
\lr{(\delta\ns)^k}_{N=\bar{N}} &=& \int (\delta\ns)^k \pnbd(\bar{N};m\ns,p) p(\ns) d\ns \\\nonumber
&=& \int (\delta\ns)^k \frac{m\ns}{\bar{N}}\pnbd(\frac{1-p}{p}\bar{n}\ns;\bar{N},1-p) p(\ns) d\ns \\\nonumber
&=& \frac{1}{\bar{n}^k}\int (\delta X)^k \frac{1-p}{p} \pnbd(\frac{1-p}{p}X;\bar{X},1-p)\frac{p(\frac{X}{\bar{n}})X}{\bar{n}\bar{X}} dX\\\label{eq:a3b}
&=& \frac{1}{\bar{n}^k} \frac{p^k}{(1-p)^k}\int (\delta Y)^k  \pnbd(Y;\frac{p}{1-p}\bar{Y},1-p) \frac{p(\frac{p}{1-p}\frac{Y}{\bar{n}})Y}{\bar{n}\bar{Y}} dY
\end{eqnarray}
where the following substitutions have been used: $X=\bar{n}\ns$, $\bar{X}=N=\bar{n}\bar{\ns}$, and $Y=\frac{1-p}{p}X$. 

If $p(\ns)\ns$ is a slowly varying function and therefore assumed to be constant in the significant integration range, the integration in the last part of Eq.~\ref{eq:a3b} is the same as that appearing in the Eq.~\ref{eq:a2b} except that parameter $p$ is replaced with $1-p$. If $p=0.5$, $k_m$ are the same as $\bar{n}^{m-1}k_m^{\mathrm{v}}$. But in general, we arrive the following relation between multiplicity fluctuation calculate for $p$ and induced centrality fluctuation calculated for $1-p$:
\begin{eqnarray}
\label{eq:a6}
\bar{n}^{m-1} k_m^{\mathrm{v}} (1-p) = \frac{p^{m-1}}{(1-p)^{m-1}}k_m (p)
\end{eqnarray}
Substituting $k_m$ with Eq.~\ref{eq:8}, we obtain the following relations for the first three cumulants:
\begin{eqnarray}
\nonumber
&&r_2 = \frac{\bar{n}k_2^{\mathrm{v}}}{k_2} = \frac{\bar{n}^2\lr{(\delta\ns)^2}}{\lr{(\delta N)^2}} =1\\\nonumber
&&r_3 = \frac{\bar{n}^2k_3^{\mathrm{v}}}{k_3} = \frac{\bar{n}^3\lr{(\delta\ns)^3}}{\lr{(\delta N)^3}} =\frac{2-p}{1+p}\;,\;\; \frac{1}{2} \leq r_3\leq2\\\label{eq:a7}
&&r_4 = \frac{\bar{n}^3k_4^{\mathrm{v}}}{k_4} = \frac{\bar{n}^4\left(\lr{(\delta\ns)^4}-3\lr{(\delta\ns)^2}^2\right)}{\lr{(\delta N)^4}-3\lr{(\delta N)^2}^2} =\frac{p^2+6(1-p)}{(1-p)^2+6p}\;,\;\; \frac{1}{6} \leq r_4\leq6
\end{eqnarray}
The fact that $r_2=1$ simply confirms Eq.~\ref{eq:a5}, which is valid for any $p(n)$ distribution. However, depending on the parameter $p$ of the NBD distribution, the range of $r_3$ is between 1/2 and 2; while the range of $r_4$ is between 1/6 and 6. Since the NBD distribution approaches Poisson distribution for $p\rightarrow0$,  we expect $r_3=2$ and $r_4=6$ if the particle production for each source follows a Poisson distribution. On the other hand, when $p\rightarrow1$, the NBD distribution approaches a Gamma distribution $p(n;m,\theta)= \frac{1}{\Gamma(m)\theta^m} n^{m-1}e^{-n/\theta}$ with $\theta=\frac{1}{1-p}\rightarrow\infty$ and $m=k$. In this limit the $r_3=1/2$ and $r_4=1/6$.

In regions of centrality where $p(\ns)$ is a rapidly changing function, e.g. very peripheral or ultra-central regions, there is no simple relation between $k_m^{\mathrm{v}}$ and $k_m$. In the ultra-central region however, one could use the Gaussian approximation Eq.~\ref{eq:a1} to estimate the centrality fluctuation:
\begin{eqnarray}
\label{eq:a8}
\lr{(\delta\ns)^k}\approx \int (\delta\ns)^k\frac{1}{\sqrt{2\pi\hat{\sigma}^2\ns}}  e^{-\frac{(\ns-\bar{\ns})^2}{2\hat{\sigma}^2\ns}} p(\ns) d\ns\;.
\end{eqnarray}
This shows that the centrality fluctuation is insensitive to the details of $p(n)$ for each source, rather it is only sensitive to the relative width $\hat{\sigma}$ of $p(n)$. For example, two $\pnbd(n;m,p)$ distributions with different average multiplicity but same relative width should have the same higher-order moments and cumulants for centrality fluctuations.

\section{Expressions for mixed and subevent cumulants}\label{sec:a2}

In the independent source model framework, eccentricity depends only on the fluctuation of position of sources, and not the multiplicity within each source, therefore the mixed correlations between multiplicity and eccentricity can be written as:
\begin{eqnarray}
\label{eq:ab1}
\lr{N\epsilon_n^2}&=&\bar{n}\lr{\ns \epsilon_n^2}\\\label{eq:ab2}
\lr{(\delta N)^2\epsilon_n^2} &=& \lr{(\delta n)^2}\lr{\ns\epsilon_n^2} + \bar{n}^2 \lr{(\delta \ns)^2\epsilon_n^2}\\\label{eq:ab3}
\lr{(\delta N)^3\epsilon_n^2} &=& \lr{(\delta n)^3}\lr{\ns\epsilon_n^2} + 3\bar{n}\lr{(\delta n)^2}\lr{(\delta \ns)^2\epsilon_n^2} + \bar{n}^3 \lr{(\delta \ns)^3\epsilon_n^2}
\end{eqnarray}
The form are very similar to the multiplicity cumulants Eq.~\ref{eq:7} (see also Ref.~\cite{Begun:2016sop}). For example the two terms in the right hand side of Eq.~\ref{eq:ab2} represent the contribution of correlation within each source, and correlation between different sources, respectively. From this, one can derive the expression for mixed correlators containing higher-order multiplicity correlations:
\begin{eqnarray}
\nonumber
&&F(k_1,\epsilon_n)=F(k_1^{\mathrm{v}},\epsilon_n)\\\nonumber
&&F(k_2,\epsilon_n)=  \frac{k_2F(k_1^{\mathrm{v}},\epsilon_2)+\bar{n}k_2^{\mathrm{v}}F(k_2^{\mathrm{v}},\epsilon_n)}{k_2+\bar{n}k_2^{\mathrm{v}}}\\\nonumber
&&F(k_3,\epsilon_n) = \frac{k_3F(k_1^{\mathrm{v}},\epsilon_n)+3k_2\bar{n}k_2^{\mathrm{v}}F(k_2^{\mathrm{v}},\epsilon_n)+\bar{n}^2k_3^{\mathrm{v}}F(k_3^{\mathrm{v}},\epsilon_n)}{k_3+3k_2\bar{n}k_2^{\mathrm{v}}+\bar{n}^2k_3^{\mathrm{v}}}\\\label{eq:ab4}
&&F(k_4,\epsilon_n) = \frac{k_4F(k_1^{\mathrm{v}},\epsilon_n)+(4k_3+3k_2^2)\bar{n}k_2^{\mathrm{v}}F(k_2^{\mathrm{v}},\epsilon_n)+6k_2\bar{n}^2k_3^{\mathrm{v}}F(k_3^{\mathrm{v}},\epsilon_n)+\bar{n}^3k_4^{\mathrm{v}}F(k_4^{\mathrm{v}},\epsilon_n)}{k_4+(4k_3+3k_2^2)\bar{n}k_2^{\mathrm{v}}+6k_2\bar{n}^2k_3^{\mathrm{v}}+\bar{n}^3k_4^{\mathrm{v}}}\;.
\end{eqnarray}

Similarly, we can also derive the expressions for subevent correlation:
\begin{eqnarray}\nonumber
&&\lr{\delta N_a\delta N_b} =\lr{\ns}\lr{\delta n_a\delta n_b}+\lr{\delta \ns^2}\bar{n}_a\bar{n}_b\;\\\nonumber
&&\lr{\delta N_a\delta N_b\delta N_c} = \lr{\delta n_a\delta n_b\delta n_c}\lr{\ns} + (\bar{n}_a\lr{\delta n_b\delta n_c}+\bar{n}_b\lr{\delta n_a\delta n_c}+\bar{n}_c\lr{\delta n_a\delta n_b})\lr{(\delta \ns)^2} + \bar{n}_a\bar{n}_b\bar{n}_c \lr{(\delta \ns)^3}\;\\\label{eq:ab5}
&&\lr{(\delta N_a)^2\delta N_b} = \lr{(\delta n_a)^2\delta n_b}\lr{\ns} + (2\bar{n}_a\lr{\delta n_a\delta n_b}+\bar{n}_b\lr{(\delta n_a)^2})\lr{(\delta \ns)^2} + \bar{n}_a^2\bar{n}_b \lr{(\delta \ns)^3}\;.
\end{eqnarray}
They lead to the following expressions for some selected normalized subevent cumulants:
\begin{eqnarray}\nonumber
&&C_{2,ab} = \frac{\lr{\delta N_a\delta N_b}}{\bar{N}_a\bar{N}_b}= \frac{1}{\lr{\ns}}\frac{\lr{\delta n_a\delta n_b}}{\bar{n}_a\bar{n}_b}+\frac{\lr{\delta \ns^2}}{\lr{\delta \ns}^2}= \frac{c_{2,ab}+k_2^{\mathrm{v}}}{\lr{\ns}}\\\nonumber
&&C_{3,abc} = \frac{\lr{\delta N_a\delta N_b\delta N_c}}{\bar{N}_a\bar{N}_b\bar{N}_c}=\frac{c_{3,abc}+(c_{2,ab}+c_{2,ac}+c_{2,bc})k_{2}^{\mathrm{v}}+k_{3}^{\mathrm{v}}}{\lr{\ns}^2}\\\nonumber
&&C_{3,a^2b} = \frac{\lr{(\delta N_a)^2\delta N_b}}{\bar{N}_a^2\bar{N}_b} =\frac{c_{3,a^2b}+(2c_{2,ab}+c_{2,a^2})k_{2}^{\mathrm{v}}+k_{3}^{\mathrm{v}}}{\lr{\ns}^2}\\\nonumber
&&C_{4,abcd}=\frac{c_{4,abcd}+(c_{2,ab}c_{2,cd}+pe.+c_{3,abc}+pe.)k_2^{\mathrm{v}}+(c_{2,ab}+pe.)k_3^{\mathrm{v}}+k_4^{\mathrm{v}}}{\lr{\ns}^3}\\\label{eq:ab6}
&&C_{4,a^2b^2}=\frac{c_{4,a^2b^2}+(c_{2,a^2}c_{2,b^2}+2c_{2,ab}^2+2c_{3,a^2b}+2c_{3,ab^2})k_2^{\mathrm{v}}+(c_{2,a^2}+c_{2,b^2}+4c_{2,ab})k_3^{\mathrm{v}}+k_4^{\mathrm{v}}}{\lr{\ns}^3} 
\end{eqnarray}

The subevent method can also be generalized to multiplicity-eccentricity mixed cumulants, for example
\begin{eqnarray}
\label{eq:ab7}
&&F(K_{2,ab},\epsilon_n) =\frac{\lr{\delta N_a\delta N_b\epsilon_n^2}}{\lr{\delta N_a\delta N_b}{\lr{\epsilon_n^2}}}-1 =  \frac{c_{2,ab}F(k_1^{\mathrm{v}},\epsilon_2)+k_2^{\mathrm{v}}F(k_2^{\mathrm{v}},\epsilon_n)}{c_{2,ab}+k_2^{\mathrm{v}}}\\\label{eq:ab8}
&&F(K_{3,abc},\epsilon_n) =\frac{\lr{\delta N_a\delta N_b\delta N_c\epsilon_n^2}}{\lr{\delta N_a\delta N_b\delta N_c}{\lr{\epsilon_n^2}}}-1= \frac{c_{3,abc}F(k_1^{\mathrm{v}},\epsilon_n)+(c_{2,ab}+c_{2,ac}+c_{2,bc})k_2^{\mathrm{v}}F(k_2^{\mathrm{v}},\epsilon_n)+k_3^{\mathrm{v}}F(k_3^{\mathrm{v}},\epsilon_n)}{c_{3,abc}+(c_{2,ab}+c_{2,ac}+c_{2,bc})k_{2}^{\mathrm{v}}+k_{3}^{\mathrm{v}}}
\end{eqnarray}
If the higher-order centrality fluctuation dominates, one expects $F(K_{2,ab},\epsilon_n)\approx F(k_2^{\mathrm{v}},\epsilon_n)$ and $F(K_{3,abc},\epsilon_n)\approx F(k_3^{\mathrm{v}},\epsilon_n)$.

Finally, in the independent source model framework, we can also relate the cumulants to some of the strongly intensive quantities studied earlier~\cite{Gorenstein:2011vq,Andronov:2018bln}. For example, correlators describing correlations between the multiplicity in two subevents can be written as~\cite{Gazdzicki:2013ana}:
\begin{eqnarray}\nonumber
&&\Sigma[N_a,N_b] \equiv \frac{1}{\lr{N_a}+\lr{N_b}}[\lr{N_a}K_{2,b}+\lr{N_b}K_{2,a}-2(\lr{N_aN_b}-\lr{N_a}\lr{N_b})]=\frac{\bar{n}_ak_{2,b}+\bar{n}_bk_{2,a}-2\lr{\delta n_a\delta n_b}}{\bar{n}_a+\bar{n}_b}\\\label{eq:ab9}
&&\Delta[N_a,N_b] \equiv \frac{1}{\lr{N_a}+\lr{N_b}}[\lr{N_b}K_{2,a}-\lr{N_a}K_{2,b}]=\frac{\bar{n}_bk_{2,a}-\bar{n}_ak_{2,b}}{\bar{n}_a+\bar{n}_b}
\end{eqnarray}

\section{Other results}\label{sec:a3}

\begin{figure}[h!]
\begin{center}
\includegraphics[width=1\linewidth]{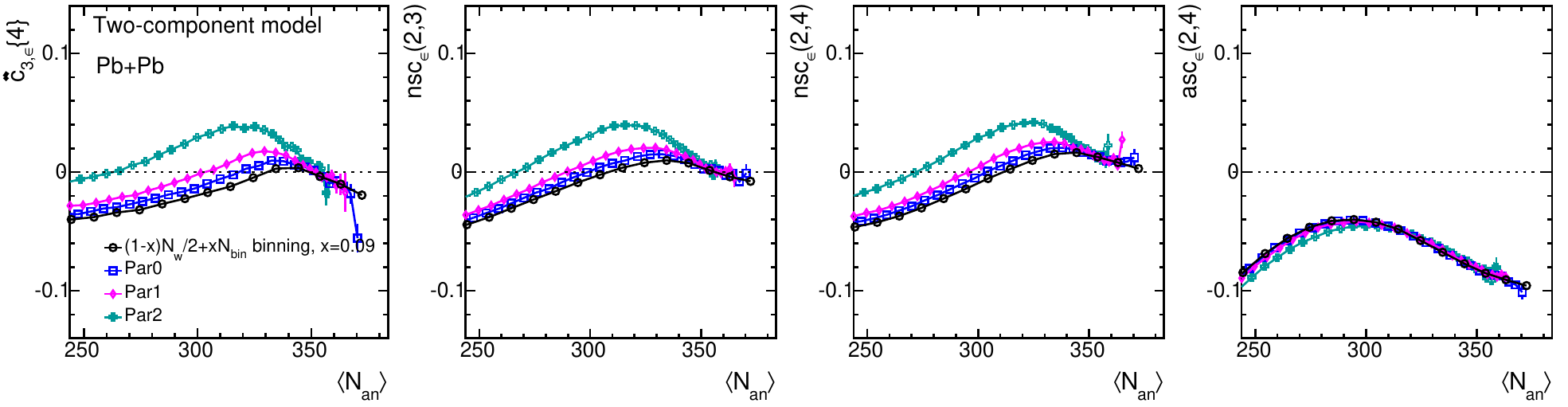}
\end{center}
\caption{\label{fig:a1}  The normalized cumulants $\hat{c}_{3,\epsilon}\{4\}$ (left), $\mathrm{nsc}(2,3)$ (second to the left), $\mathrm{nsc}(2,4)$ (second to the right) and  $\mathrm{asc}(2,4)$ (right) for the three parameter sets in Table.~\ref{tab:1} for the two-component model. They are calculated in narrow particle multiplicity bins then combined and mapped to average number of $\ntwoc$.}
\end{figure}

\begin{figure}[h!]
\begin{center}
\includegraphics[width=1\linewidth]{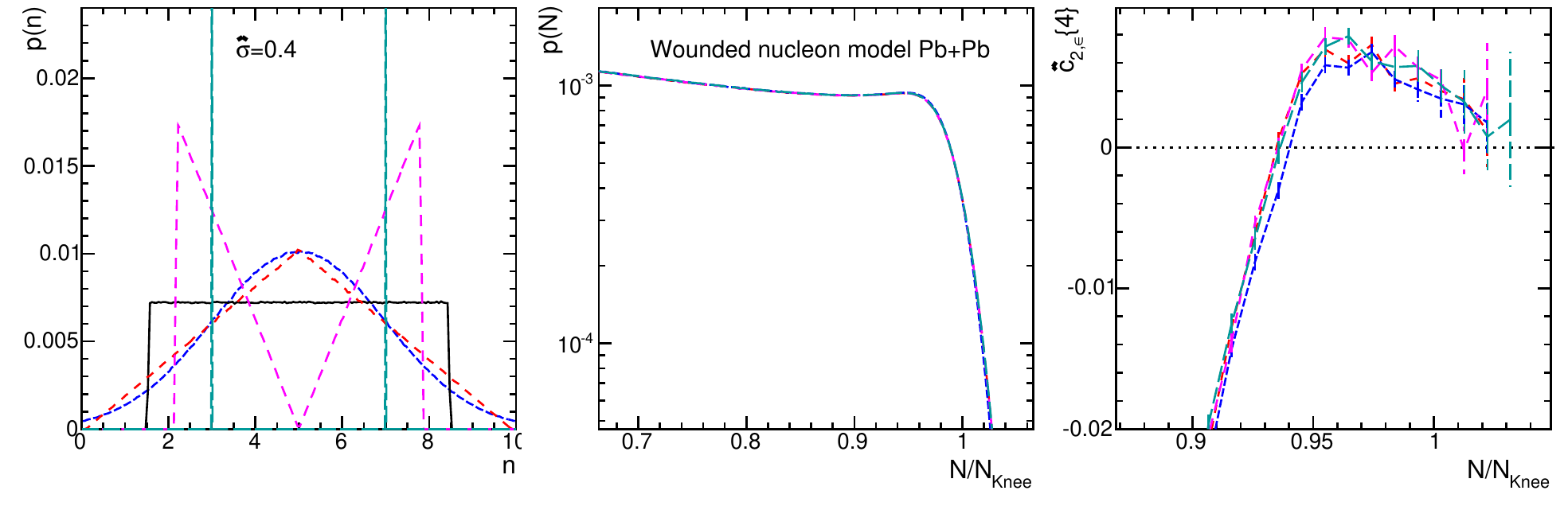}
\end{center}
\caption{\label{fig:a2} The multiplicity distribution for each source $p(n)$ (left panel), and the corresponding total multiplicity distribution $p(N)$ (middle) and four-particle normalized cumulant $\hat{c}_{2,\epsilon}\{4\}$ (right panel). They are calculated in the wounded nucleon model with $\hat{\sigma}=0.4$ for Pb+Pb collisions. The large statistical error bar is due to limited event statistics in the simulation.}
\end{figure}

\begin{figure}[t!]
\begin{center}
\includegraphics[width=1\linewidth]{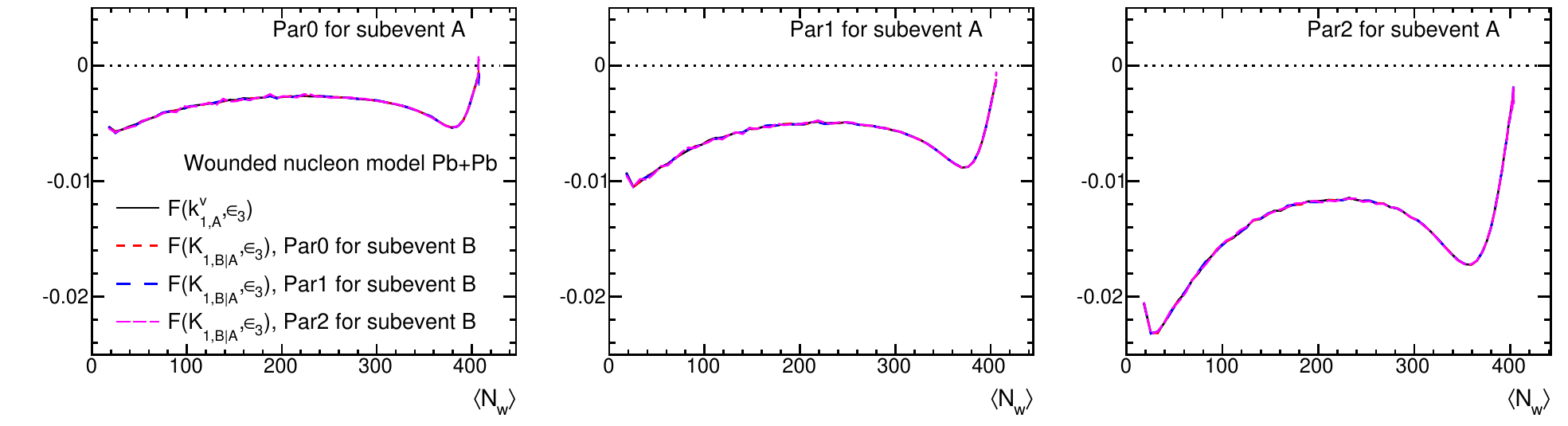}
\end{center}
\caption{\label{fig:a3} The multiplicity-eccentricity mixed cumulants $F(K_{1,B|A},\epsilon_3)$ calculated with three parameter sets for subevent B for centrality defined in subevent A, and compared with $F(k_{1,A}^{\mathrm{v}},\epsilon_3)$.  All four curves are found to be on top of each other. The three panels corresponds three different NBD parameter sets for subevent A. They are all calculated for the wounded nucleon model. }
\end{figure}

\begin{figure}[h!]
\begin{center}
\includegraphics[width=1\linewidth]{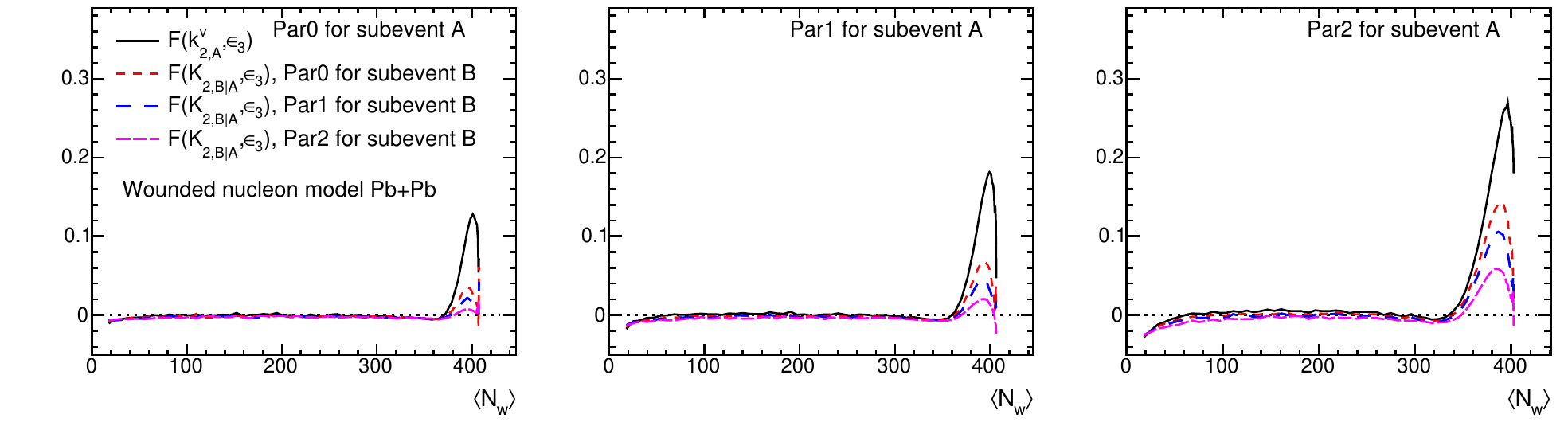}
\end{center}
\caption{\label{fig:a4} The multiplicity-eccentricity mixed cumulants $F(K_{2,B|A},\epsilon_3)$ calculated with three parameter sets for subevent B for centrality defined in subevent A, and compared with $F(k_{2,A}^{\mathrm{v}},\epsilon_3)$. The three panels corresponds three NBD parameter sets for subevent A. They are all calculated for the wounded nucleon model.}
\end{figure}

Figure~\ref{fig:a1} shows centrality dependence of several normalized cumulant observables for the two-component model. They are complementary to those shown in Figure~\ref{fig:c3}.
Figure~\ref{fig:c4} has shown that the eccentricity cumulants depend only on the relative width of $p(n)$ of the negative binomial distribution. Other forms of $p(n)$ are tried as well, including Gaussian, triangle, flat or even double delta distributions as shown in the left panel of Figure~\ref{fig:a2}. These functions have been chosen to have identical $\hat{\sigma}$ and $\bar{n}$. The resulting multiplicity distributions, as well as the normalized eccentricity cumulant $c_{2,\epsilon}\{4\}$ are shown in the middle and right panel of Figure~\ref{fig:a2}, respectively. The results are found to be insensitive to the functional form $p(n)$ the centrality range shown. Figures~\ref{fig:a3}--\ref{fig:a4} show the multiplicity-$\epsilon_3$ mixed cumulants, they are complementary to those shown in Figures~\ref{fig:d2}--\ref{fig:d3} for $\epsilon_2$.

\pagebreak
\bibliography{cumu4_v0}{}

\begin{thebibliography}{70}%
\makeatletter
\providecommand \@ifxundefined [1]{%
 \@ifx{#1\undefined}
}%
\providecommand \@ifnum [1]{%
 \ifnum #1\expandafter \@firstoftwo
 \else \expandafter \@secondoftwo
 \fi
}%
\providecommand \@ifx [1]{%
 \ifx #1\expandafter \@firstoftwo
 \else \expandafter \@secondoftwo
 \fi
}%
\providecommand \natexlab [1]{#1}%
\providecommand \enquote  [1]{``#1''}%
\providecommand \bibnamefont  [1]{#1}%
\providecommand \bibfnamefont [1]{#1}%
\providecommand \citenamefont [1]{#1}%
\providecommand \href@noop [0]{\@secondoftwo}%
\providecommand \href [0]{\begingroup \@sanitize@url \@href}%
\providecommand \@href[1]{\@@startlink{#1}\@@href}%
\providecommand \@@href[1]{\endgroup#1\@@endlink}%
\providecommand \@sanitize@url [0]{\catcode `\\12\catcode `\$12\catcode
  `\&12\catcode `\#12\catcode `\^12\catcode `\_12\catcode `\%12\relax}%
\providecommand \@@startlink[1]{}%
\providecommand \@@endlink[0]{}%
\providecommand \url  [0]{\begingroup\@sanitize@url \@url }%
\providecommand \@url [1]{\endgroup\@href {#1}{\urlprefix }}%
\providecommand \urlprefix  [0]{URL }%
\providecommand \Eprint [0]{\href }%
\providecommand \doibase [0]{http://dx.doi.org/}%
\providecommand \selectlanguage [0]{\@gobble}%
\providecommand \bibinfo  [0]{\@secondoftwo}%
\providecommand \bibfield  [0]{\@secondoftwo}%
\providecommand \translation [1]{[#1]}%
\providecommand \BibitemOpen [0]{}%
\providecommand \bibitemStop [0]{}%
\providecommand \bibitemNoStop [0]{.\EOS\space}%
\providecommand \EOS [0]{\spacefactor3000\relax}%
\providecommand \BibitemShut  [1]{\csname bibitem#1\endcsname}%
\let\auto@bib@innerbib\@empty
\bibitem [{\citenamefont {Miller}\ \emph {et~al.}(2007)\citenamefont {Miller},
  \citenamefont {Reygers}, \citenamefont {Sanders},\ and\ \citenamefont
  {Steinberg}}]{Miller:2007ri}%
  \BibitemOpen
  \bibfield  {author} {\bibinfo {author} {\bibfnamefont {M.~L.}\ \bibnamefont
  {Miller}}, \bibinfo {author} {\bibfnamefont {K.}~\bibnamefont {Reygers}},
  \bibinfo {author} {\bibfnamefont {S.~J.}\ \bibnamefont {Sanders}}, \ and\
  \bibinfo {author} {\bibfnamefont {P.}~\bibnamefont {Steinberg}},\ }\href
  {\doibase 10.1146/annurev.nucl.57.090506.123020} {\bibfield  {journal}
  {\bibinfo  {journal} {Ann. Rev. Nucl. Part. Sci.}\ }\textbf {\bibinfo
  {volume} {57}},\ \bibinfo {pages} {205} (\bibinfo {year} {2007})},\ \Eprint
  {http://arxiv.org/abs/nucl-ex/0701025} {arXiv:nucl-ex/0701025 [nucl-ex]}
  \BibitemShut {NoStop}%
\bibitem [{\citenamefont {Loizides}\ \emph {et~al.}(2015)\citenamefont
  {Loizides}, \citenamefont {Nagle},\ and\ \citenamefont
  {Steinberg}}]{Loizides:2014vua}%
  \BibitemOpen
  \bibfield  {author} {\bibinfo {author} {\bibfnamefont {C.}~\bibnamefont
  {Loizides}}, \bibinfo {author} {\bibfnamefont {J.}~\bibnamefont {Nagle}}, \
  and\ \bibinfo {author} {\bibfnamefont {P.}~\bibnamefont {Steinberg}},\ }\href
  {\doibase 10.1016/j.softx.2015.05.001} {\bibfield  {journal} {\bibinfo
  {journal} {SoftwareX}\ }\textbf {\bibinfo {volume} {1}},\ \bibinfo {pages}
  {13} (\bibinfo {year} {2015})},\ \Eprint {http://arxiv.org/abs/1408.2549}
  {arXiv:1408.2549 [nucl-ex]} \BibitemShut {NoStop}%
\bibitem [{\citenamefont {Adler}\ \emph {et~al.}(2014)\citenamefont {Adler}
  \emph {et~al.}}]{Adler:2013aqf}%
  \BibitemOpen
  \bibfield  {author} {\bibinfo {author} {\bibfnamefont {S.~S.}\ \bibnamefont
  {Adler}} \emph {et~al.} (\bibinfo {collaboration} {PHENIX}),\ }\href
  {\doibase 10.1103/PhysRevC.89.044905} {\bibfield  {journal} {\bibinfo
  {journal} {Phys. Rev. C}\ }\textbf {\bibinfo {volume} {89}},\ \bibinfo
  {pages} {044905} (\bibinfo {year} {2014})},\ \Eprint
  {http://arxiv.org/abs/1312.6676} {arXiv:1312.6676 [nucl-ex]} \BibitemShut
  {NoStop}%
\bibitem [{\citenamefont {Jeon}\ and\ \citenamefont {Koch}()}]{Jeon:2003gk}%
  \BibitemOpen
  \bibfield  {author} {\bibinfo {author} {\bibfnamefont {S.}~\bibnamefont
  {Jeon}}\ and\ \bibinfo {author} {\bibfnamefont {V.}~\bibnamefont {Koch}},\
  }\href@noop {} {\ }\Eprint {http://arxiv.org/abs/hep-ph/0304012}
  {arXiv:hep-ph/0304012 [hep-ph]} \BibitemShut {NoStop}%
\bibitem [{\citenamefont {Skokov}\ \emph {et~al.}(2013)\citenamefont {Skokov},
  \citenamefont {Friman},\ and\ \citenamefont {Redlich}}]{Skokov:2012ds}%
  \BibitemOpen
  \bibfield  {author} {\bibinfo {author} {\bibfnamefont {V.}~\bibnamefont
  {Skokov}}, \bibinfo {author} {\bibfnamefont {B.}~\bibnamefont {Friman}}, \
  and\ \bibinfo {author} {\bibfnamefont {K.}~\bibnamefont {Redlich}},\ }\href
  {\doibase 10.1103/PhysRevC.88.034911} {\bibfield  {journal} {\bibinfo
  {journal} {Phys. Rev. C}\ }\textbf {\bibinfo {volume} {88}},\ \bibinfo
  {pages} {034911} (\bibinfo {year} {2013})},\ \Eprint
  {http://arxiv.org/abs/1205.4756} {arXiv:1205.4756 [hep-ph]} \BibitemShut
  {NoStop}%
\bibitem [{\citenamefont {Luo}\ \emph {et~al.}(2013)\citenamefont {Luo},
  \citenamefont {Xu}, \citenamefont {Mohanty},\ and\ \citenamefont
  {Xu}}]{Luo:2013bmi}%
  \BibitemOpen
  \bibfield  {author} {\bibinfo {author} {\bibfnamefont {X.}~\bibnamefont
  {Luo}}, \bibinfo {author} {\bibfnamefont {J.}~\bibnamefont {Xu}}, \bibinfo
  {author} {\bibfnamefont {B.}~\bibnamefont {Mohanty}}, \ and\ \bibinfo
  {author} {\bibfnamefont {N.}~\bibnamefont {Xu}},\ }\href {\doibase
  10.1088/0954-3899/40/10/105104} {\bibfield  {journal} {\bibinfo  {journal}
  {J. Phys. G}\ }\textbf {\bibinfo {volume} {40}},\ \bibinfo {pages} {105104}
  (\bibinfo {year} {2013})},\ \Eprint {http://arxiv.org/abs/1302.2332}
  {arXiv:1302.2332 [nucl-ex]} \BibitemShut {NoStop}%
\bibitem [{\citenamefont {Xu}(2016{\natexlab{a}})}]{Xu:2016qzd}%
  \BibitemOpen
  \bibfield  {author} {\bibinfo {author} {\bibfnamefont {H.-j.}\ \bibnamefont
  {Xu}},\ }\href {\doibase 10.1103/PhysRevC.94.054903} {\bibfield  {journal}
  {\bibinfo  {journal} {Phys. Rev. C}\ }\textbf {\bibinfo {volume} {94}},\
  \bibinfo {pages} {054903} (\bibinfo {year} {2016}{\natexlab{a}})},\ \Eprint
  {http://arxiv.org/abs/1602.07089} {arXiv:1602.07089 [nucl-th]} \BibitemShut
  {NoStop}%
\bibitem [{\citenamefont {Bzdak}\ \emph
  {et~al.}(2017{\natexlab{a}})\citenamefont {Bzdak}, \citenamefont {Koch},\
  and\ \citenamefont {Skokov}}]{Bzdak:2016jxo}%
  \BibitemOpen
  \bibfield  {author} {\bibinfo {author} {\bibfnamefont {A.}~\bibnamefont
  {Bzdak}}, \bibinfo {author} {\bibfnamefont {V.}~\bibnamefont {Koch}}, \ and\
  \bibinfo {author} {\bibfnamefont {V.}~\bibnamefont {Skokov}},\ }\href
  {\doibase 10.1140/epjc/s10052-017-4847-0} {\bibfield  {journal} {\bibinfo
  {journal} {Eur. Phys. J. C}\ }\textbf {\bibinfo {volume} {77}},\ \bibinfo
  {pages} {288} (\bibinfo {year} {2017}{\natexlab{a}})},\ \Eprint
  {http://arxiv.org/abs/1612.05128} {arXiv:1612.05128 [nucl-th]} \BibitemShut
  {NoStop}%
\bibitem [{\citenamefont {Aggarwal}\ \emph {et~al.}(2010)\citenamefont
  {Aggarwal} \emph {et~al.}}]{Aggarwal:2010wy}%
  \BibitemOpen
  \bibfield  {author} {\bibinfo {author} {\bibfnamefont {M.~M.}\ \bibnamefont
  {Aggarwal}} \emph {et~al.} (\bibinfo {collaboration} {STAR}),\ }\href
  {\doibase 10.1103/PhysRevLett.105.022302} {\bibfield  {journal} {\bibinfo
  {journal} {Phys. Rev. Lett.}\ }\textbf {\bibinfo {volume} {105}},\ \bibinfo
  {pages} {022302} (\bibinfo {year} {2010})},\ \Eprint
  {http://arxiv.org/abs/1004.4959} {arXiv:1004.4959 [nucl-ex]} \BibitemShut
  {NoStop}%
\bibitem [{\citenamefont {Adamczyk}\ \emph
  {et~al.}(2014{\natexlab{a}})\citenamefont {Adamczyk} \emph
  {et~al.}}]{Adamczyk:2013dal}%
  \BibitemOpen
  \bibfield  {author} {\bibinfo {author} {\bibfnamefont {L.}~\bibnamefont
  {Adamczyk}} \emph {et~al.} (\bibinfo {collaboration} {STAR}),\ }\href
  {\doibase 10.1103/PhysRevLett.112.032302} {\bibfield  {journal} {\bibinfo
  {journal} {Phys. Rev. Lett.}\ }\textbf {\bibinfo {volume} {112}},\ \bibinfo
  {pages} {032302} (\bibinfo {year} {2014}{\natexlab{a}})},\ \Eprint
  {http://arxiv.org/abs/1309.5681} {arXiv:1309.5681 [nucl-ex]} \BibitemShut
  {NoStop}%
\bibitem [{\citenamefont {Adamczyk}\ \emph
  {et~al.}(2014{\natexlab{b}})\citenamefont {Adamczyk} \emph
  {et~al.}}]{Adamczyk:2014fia}%
  \BibitemOpen
  \bibfield  {author} {\bibinfo {author} {\bibfnamefont {L.}~\bibnamefont
  {Adamczyk}} \emph {et~al.} (\bibinfo {collaboration} {STAR}),\ }\href
  {\doibase 10.1103/PhysRevLett.113.092301} {\bibfield  {journal} {\bibinfo
  {journal} {Phys. Rev. Lett.}\ }\textbf {\bibinfo {volume} {113}},\ \bibinfo
  {pages} {092301} (\bibinfo {year} {2014}{\natexlab{b}})},\ \Eprint
  {http://arxiv.org/abs/1402.1558} {arXiv:1402.1558 [nucl-ex]} \BibitemShut
  {NoStop}%
\bibitem [{\citenamefont {Luo}\ and\ \citenamefont {Xu}(2017)}]{Luo:2017faz}%
  \BibitemOpen
  \bibfield  {author} {\bibinfo {author} {\bibfnamefont {X.}~\bibnamefont
  {Luo}}\ and\ \bibinfo {author} {\bibfnamefont {N.}~\bibnamefont {Xu}},\
  }\href {\doibase 10.1007/s41365-017-0257-0} {\bibfield  {journal} {\bibinfo
  {journal} {Nucl. Sci. Tech.}\ }\textbf {\bibinfo {volume} {28}},\ \bibinfo
  {pages} {112} (\bibinfo {year} {2017})},\ \Eprint
  {http://arxiv.org/abs/1701.02105} {arXiv:1701.02105 [nucl-ex]} \BibitemShut
  {NoStop}%
\bibitem [{\citenamefont {Li}\ \emph {et~al.}(2018)\citenamefont {Li},
  \citenamefont {Xu},\ and\ \citenamefont {Song}}]{Li:2017via}%
  \BibitemOpen
  \bibfield  {author} {\bibinfo {author} {\bibfnamefont {J.}~\bibnamefont
  {Li}}, \bibinfo {author} {\bibfnamefont {H.-j.}\ \bibnamefont {Xu}}, \ and\
  \bibinfo {author} {\bibfnamefont {H.}~\bibnamefont {Song}},\ }\href {\doibase
  10.1103/PhysRevC.97.014902} {\bibfield  {journal} {\bibinfo  {journal} {Phys.
  Rev. C}\ }\textbf {\bibinfo {volume} {97}},\ \bibinfo {pages} {014902}
  (\bibinfo {year} {2018})},\ \Eprint {http://arxiv.org/abs/1707.09742}
  {arXiv:1707.09742 [nucl-th]} \BibitemShut {NoStop}%
\bibitem [{\citenamefont {Qiu}\ and\ \citenamefont {Heinz}(2011)}]{Qiu:2011iv}%
  \BibitemOpen
  \bibfield  {author} {\bibinfo {author} {\bibfnamefont {Z.}~\bibnamefont
  {Qiu}}\ and\ \bibinfo {author} {\bibfnamefont {U.~W.}\ \bibnamefont
  {Heinz}},\ }\href {\doibase 10.1103/PhysRevC.84.024911} {\bibfield  {journal}
  {\bibinfo  {journal} {Phys. Rev. C}\ }\textbf {\bibinfo {volume} {84}},\
  \bibinfo {pages} {024911} (\bibinfo {year} {2011})},\ \Eprint
  {http://arxiv.org/abs/1104.0650} {arXiv:1104.0650 [nucl-th]} \BibitemShut
  {NoStop}%
\bibitem [{\citenamefont {Gardim}\ \emph {et~al.}(2012)\citenamefont {Gardim},
  \citenamefont {Grassi}, \citenamefont {Luzum},\ and\ \citenamefont
  {Ollitrault}}]{Gardim:2011xv}%
  \BibitemOpen
  \bibfield  {author} {\bibinfo {author} {\bibfnamefont {F.~G.}\ \bibnamefont
  {Gardim}}, \bibinfo {author} {\bibfnamefont {F.}~\bibnamefont {Grassi}},
  \bibinfo {author} {\bibfnamefont {M.}~\bibnamefont {Luzum}}, \ and\ \bibinfo
  {author} {\bibfnamefont {J.-Y.}\ \bibnamefont {Ollitrault}},\ }\href
  {\doibase 10.1103/PhysRevC.85.024908} {\bibfield  {journal} {\bibinfo
  {journal} {Phys. Rev. C}\ }\textbf {\bibinfo {volume} {85}},\ \bibinfo
  {pages} {024908} (\bibinfo {year} {2012})},\ \Eprint
  {http://arxiv.org/abs/1111.6538} {arXiv:1111.6538 [nucl-th]} \BibitemShut
  {NoStop}%
\bibitem [{\citenamefont {Niemi}\ \emph {et~al.}(2013)\citenamefont {Niemi},
  \citenamefont {Denicol}, \citenamefont {Holopainen},\ and\ \citenamefont
  {Huovinen}}]{Niemi:2012aj}%
  \BibitemOpen
  \bibfield  {author} {\bibinfo {author} {\bibfnamefont {H.}~\bibnamefont
  {Niemi}}, \bibinfo {author} {\bibfnamefont {G.~S.}\ \bibnamefont {Denicol}},
  \bibinfo {author} {\bibfnamefont {H.}~\bibnamefont {Holopainen}}, \ and\
  \bibinfo {author} {\bibfnamefont {P.}~\bibnamefont {Huovinen}},\ }\href
  {\doibase 10.1103/PhysRevC.87.054901} {\bibfield  {journal} {\bibinfo
  {journal} {Phys. Rev. C}\ }\textbf {\bibinfo {volume} {87}},\ \bibinfo
  {pages} {054901} (\bibinfo {year} {2013})},\ \Eprint
  {http://arxiv.org/abs/1212.1008} {arXiv:1212.1008 [nucl-th]} \BibitemShut
  {NoStop}%
\bibitem [{\citenamefont {Asakawa}\ and\ \citenamefont
  {Kitazawa}(2016)}]{Asakawa:2015ybt}%
  \BibitemOpen
  \bibfield  {author} {\bibinfo {author} {\bibfnamefont {M.}~\bibnamefont
  {Asakawa}}\ and\ \bibinfo {author} {\bibfnamefont {M.}~\bibnamefont
  {Kitazawa}},\ }\href {\doibase 10.1016/j.ppnp.2016.04.002} {\bibfield
  {journal} {\bibinfo  {journal} {Prog. Part. Nucl. Phys.}\ }\textbf {\bibinfo
  {volume} {90}},\ \bibinfo {pages} {299} (\bibinfo {year} {2016})},\ \Eprint
  {http://arxiv.org/abs/1512.05038} {arXiv:1512.05038 [nucl-th]} \BibitemShut
  {NoStop}%
\bibitem [{\citenamefont {Braun-Munzinger}\ \emph {et~al.}(2017)\citenamefont
  {Braun-Munzinger}, \citenamefont {Rustamov},\ and\ \citenamefont
  {Stachel}}]{Braun-Munzinger:2016yjz}%
  \BibitemOpen
  \bibfield  {author} {\bibinfo {author} {\bibfnamefont {P.}~\bibnamefont
  {Braun-Munzinger}}, \bibinfo {author} {\bibfnamefont {A.}~\bibnamefont
  {Rustamov}}, \ and\ \bibinfo {author} {\bibfnamefont {J.}~\bibnamefont
  {Stachel}},\ }\href {\doibase 10.1016/j.nuclphysa.2017.01.011} {\bibfield
  {journal} {\bibinfo  {journal} {Nucl. Phys. A}\ }\textbf {\bibinfo {volume}
  {960}},\ \bibinfo {pages} {114} (\bibinfo {year} {2017})},\ \Eprint
  {http://arxiv.org/abs/1612.00702} {arXiv:1612.00702 [nucl-th]} \BibitemShut
  {NoStop}%
\bibitem [{\citenamefont {Kitazawa}\ and\ \citenamefont
  {Luo}(2017)}]{Kitazawa:2017ljq}%
  \BibitemOpen
  \bibfield  {author} {\bibinfo {author} {\bibfnamefont {M.}~\bibnamefont
  {Kitazawa}}\ and\ \bibinfo {author} {\bibfnamefont {X.}~\bibnamefont {Luo}},\
  }\href {\doibase 10.1103/PhysRevC.96.024910} {\bibfield  {journal} {\bibinfo
  {journal} {Phys. Rev. C}\ }\textbf {\bibinfo {volume} {96}},\ \bibinfo
  {pages} {024910} (\bibinfo {year} {2017})},\ \Eprint
  {http://arxiv.org/abs/1704.04909} {arXiv:1704.04909 [nucl-th]} \BibitemShut
  {NoStop}%
\bibitem [{\citenamefont {Xu}(2016{\natexlab{b}})}]{Xu:2016jaz}%
  \BibitemOpen
  \bibfield  {author} {\bibinfo {author} {\bibfnamefont {H.-j.}\ \bibnamefont
  {Xu}},\ }\href@noop {} {\  (\bibinfo {year} {2016}{\natexlab{b}})},\ \Eprint
  {http://arxiv.org/abs/1602.06378} {arXiv:1602.06378 [nucl-th]} \BibitemShut
  {NoStop}%
\bibitem [{\citenamefont {{ATLAS Collaboration}}(2017)}]{ATLAS:2017zcm}%
  \BibitemOpen
  \bibfield  {author} {\bibinfo {author} {\bibnamefont {{ATLAS
  Collaboration}}},\ }\href {http://cds.cern.ch/record/2285570} {\bibfield
  {journal} {\bibinfo  {journal} {ATLAS-CONF-2017-066}\ } (\bibinfo {year}
  {2017})}\BibitemShut {NoStop}%
\bibitem [{\citenamefont {Jia}\ \emph {et~al.}(2017)\citenamefont {Jia},
  \citenamefont {Zhou},\ and\ \citenamefont {Trzupek}}]{Jia:2017hbm}%
  \BibitemOpen
  \bibfield  {author} {\bibinfo {author} {\bibfnamefont {J.}~\bibnamefont
  {Jia}}, \bibinfo {author} {\bibfnamefont {M.}~\bibnamefont {Zhou}}, \ and\
  \bibinfo {author} {\bibfnamefont {A.}~\bibnamefont {Trzupek}},\ }\href
  {\doibase 10.1103/PhysRevC.96.034906} {\bibfield  {journal} {\bibinfo
  {journal} {Phys. Rev. C}\ }\textbf {\bibinfo {volume} {96}},\ \bibinfo
  {pages} {034906} (\bibinfo {year} {2017})},\ \Eprint
  {http://arxiv.org/abs/1701.03830} {arXiv:1701.03830 [nucl-th]} \BibitemShut
  {NoStop}%
\bibitem [{\citenamefont {Aaboud}\ \emph
  {et~al.}(2018{\natexlab{a}})\citenamefont {Aaboud} \emph
  {et~al.}}]{Aaboud:2017blb}%
  \BibitemOpen
  \bibfield  {author} {\bibinfo {author} {\bibfnamefont {M.}~\bibnamefont
  {Aaboud}} \emph {et~al.} (\bibinfo {collaboration} {ATLAS}),\ }\href
  {\doibase 10.1103/PhysRevC.97.024904} {\bibfield  {journal} {\bibinfo
  {journal} {Phys. Rev. C}\ }\textbf {\bibinfo {volume} {97}},\ \bibinfo
  {pages} {024904} (\bibinfo {year} {2018}{\natexlab{a}})},\ \Eprint
  {http://arxiv.org/abs/1708.03559} {arXiv:1708.03559 [hep-ex]} \BibitemShut
  {NoStop}%
\bibitem [{\citenamefont {Huo}\ \emph {et~al.}(2018)\citenamefont {Huo},
  \citenamefont {Gajdošová}, \citenamefont {Jia},\ and\ \citenamefont
  {Zhou}}]{Huo:2017nms}%
  \BibitemOpen
  \bibfield  {author} {\bibinfo {author} {\bibfnamefont {P.}~\bibnamefont
  {Huo}}, \bibinfo {author} {\bibfnamefont {K.}~\bibnamefont {Gajdošová}},
  \bibinfo {author} {\bibfnamefont {J.}~\bibnamefont {Jia}}, \ and\ \bibinfo
  {author} {\bibfnamefont {Y.}~\bibnamefont {Zhou}},\ }\href {\doibase
  10.1016/j.physletb.2017.12.035} {\bibfield  {journal} {\bibinfo  {journal}
  {Phys. Lett. B}\ }\textbf {\bibinfo {volume} {777}},\ \bibinfo {pages} {201}
  (\bibinfo {year} {2018})},\ \Eprint {http://arxiv.org/abs/1710.07567}
  {arXiv:1710.07567 [nucl-ex]} \BibitemShut {NoStop}%
\bibitem [{\citenamefont {Nie}\ \emph {et~al.}(2018)\citenamefont {Nie},
  \citenamefont {Huo}, \citenamefont {Jia},\ and\ \citenamefont
  {Ma}}]{Nie:2018xog}%
  \BibitemOpen
  \bibfield  {author} {\bibinfo {author} {\bibfnamefont {M.-W.}\ \bibnamefont
  {Nie}}, \bibinfo {author} {\bibfnamefont {P.}~\bibnamefont {Huo}}, \bibinfo
  {author} {\bibfnamefont {J.}~\bibnamefont {Jia}}, \ and\ \bibinfo {author}
  {\bibfnamefont {G.-L.}\ \bibnamefont {Ma}},\ }\href {\doibase
  10.1103/PhysRevC.98.034903} {\bibfield  {journal} {\bibinfo  {journal} {Phys.
  Rev. C}\ }\textbf {\bibinfo {volume} {98}},\ \bibinfo {pages} {034903}
  (\bibinfo {year} {2018})},\ \Eprint {http://arxiv.org/abs/1802.00374}
  {arXiv:1802.00374 [hep-ph]} \BibitemShut {NoStop}%
\bibitem [{\citenamefont {Giovannini}\ and\ \citenamefont
  {Van~Hove}(1986)}]{Giovannini:1985mz}%
  \BibitemOpen
  \bibfield  {author} {\bibinfo {author} {\bibfnamefont {A.}~\bibnamefont
  {Giovannini}}\ and\ \bibinfo {author} {\bibfnamefont {L.}~\bibnamefont
  {Van~Hove}},\ }\href {\doibase 10.1007/BF01557602} {\bibfield  {journal}
  {\bibinfo  {journal} {Z. Phys. C}\ }\textbf {\bibinfo {volume} {30}},\
  \bibinfo {pages} {391} (\bibinfo {year} {1986})}\BibitemShut {NoStop}%
\bibitem [{\citenamefont {Ghosh}(2012)}]{Ghosh:2012xh}%
  \BibitemOpen
  \bibfield  {author} {\bibinfo {author} {\bibfnamefont {P.}~\bibnamefont
  {Ghosh}},\ }\href {\doibase 10.1103/PhysRevD.85.054017} {\bibfield  {journal}
  {\bibinfo  {journal} {Phys. Rev. D}\ }\textbf {\bibinfo {volume} {85}},\
  \bibinfo {pages} {054017} (\bibinfo {year} {2012})},\ \Eprint
  {http://arxiv.org/abs/1202.4221} {arXiv:1202.4221 [hep-ph]} \BibitemShut
  {NoStop}%
\bibitem [{\citenamefont {Abelev}\ \emph {et~al.}(2013)\citenamefont {Abelev}
  \emph {et~al.}}]{Abelev:2013qoq}%
  \BibitemOpen
  \bibfield  {author} {\bibinfo {author} {\bibfnamefont {B.}~\bibnamefont
  {Abelev}} \emph {et~al.} (\bibinfo {collaboration} {ALICE}),\ }\href
  {\doibase 10.1103/PhysRevC.88.044909} {\bibfield  {journal} {\bibinfo
  {journal} {Phys. Rev. C}\ }\textbf {\bibinfo {volume} {88}},\ \bibinfo
  {pages} {044909} (\bibinfo {year} {2013})},\ \Eprint
  {http://arxiv.org/abs/1301.4361} {arXiv:1301.4361 [nucl-ex]} \BibitemShut
  {NoStop}%
\bibitem [{\citenamefont {Aad}\ \emph {et~al.}(2016)\citenamefont {Aad} \emph
  {et~al.}}]{Aad:2016xww}%
  \BibitemOpen
  \bibfield  {author} {\bibinfo {author} {\bibfnamefont {G.}~\bibnamefont
  {Aad}} \emph {et~al.} (\bibinfo {collaboration} {ATLAS}),\ }\href {\doibase
  10.1140/epjc/s10052-016-4203-9} {\bibfield  {journal} {\bibinfo  {journal}
  {Eur. Phys. J. C}\ }\textbf {\bibinfo {volume} {76}},\ \bibinfo {pages} {403}
  (\bibinfo {year} {2016})},\ \Eprint {http://arxiv.org/abs/1603.02439}
  {arXiv:1603.02439 [hep-ex]} \BibitemShut {NoStop}%
\bibitem [{\citenamefont {Heiselberg}(2001)}]{Heiselberg:2000fk}%
  \BibitemOpen
  \bibfield  {author} {\bibinfo {author} {\bibfnamefont {H.}~\bibnamefont
  {Heiselberg}},\ }\href {\doibase 10.1016/S0370-1573(00)00140-X} {\bibfield
  {journal} {\bibinfo  {journal} {Phys. Rept.}\ }\textbf {\bibinfo {volume}
  {351}},\ \bibinfo {pages} {161} (\bibinfo {year} {2001})},\ \Eprint
  {http://arxiv.org/abs/nucl-th/0003046} {arXiv:nucl-th/0003046 [nucl-th]}
  \BibitemShut {NoStop}%
\bibitem [{\citenamefont {Das}\ \emph {et~al.}(2018)\citenamefont {Das},
  \citenamefont {Giacalone}, \citenamefont {Monard},\ and\ \citenamefont
  {Ollitrault}}]{Das:2017ned}%
  \BibitemOpen
  \bibfield  {author} {\bibinfo {author} {\bibfnamefont {S.~J.}\ \bibnamefont
  {Das}}, \bibinfo {author} {\bibfnamefont {G.}~\bibnamefont {Giacalone}},
  \bibinfo {author} {\bibfnamefont {P.-A.}\ \bibnamefont {Monard}}, \ and\
  \bibinfo {author} {\bibfnamefont {J.-Y.}\ \bibnamefont {Ollitrault}},\ }\href
  {\doibase 10.1103/PhysRevC.97.014905} {\bibfield  {journal} {\bibinfo
  {journal} {Phys. Rev. C}\ }\textbf {\bibinfo {volume} {97}},\ \bibinfo
  {pages} {014905} (\bibinfo {year} {2018})},\ \Eprint
  {http://arxiv.org/abs/1708.00081} {arXiv:1708.00081 [nucl-th]} \BibitemShut
  {NoStop}%
\bibitem [{\citenamefont {Bozek}\ and\ \citenamefont
  {Broniowski}(2016)}]{Bozek:2015bna}%
  \BibitemOpen
  \bibfield  {author} {\bibinfo {author} {\bibfnamefont {P.}~\bibnamefont
  {Bozek}}\ and\ \bibinfo {author} {\bibfnamefont {W.}~\bibnamefont
  {Broniowski}},\ }\href {\doibase 10.1016/j.physletb.2015.11.054} {\bibfield
  {journal} {\bibinfo  {journal} {Phys. Lett. B}\ }\textbf {\bibinfo {volume}
  {752}},\ \bibinfo {pages} {206} (\bibinfo {year} {2016})},\ \Eprint
  {http://arxiv.org/abs/1506.02817} {arXiv:1506.02817 [nucl-th]} \BibitemShut
  {NoStop}%
\bibitem [{\citenamefont {Pang}\ \emph {et~al.}(2016)\citenamefont {Pang},
  \citenamefont {Petersen}, \citenamefont {Qin}, \citenamefont {Roy},\ and\
  \citenamefont {Wang}}]{Pang:2015zrq}%
  \BibitemOpen
  \bibfield  {author} {\bibinfo {author} {\bibfnamefont {L.-G.}\ \bibnamefont
  {Pang}}, \bibinfo {author} {\bibfnamefont {H.}~\bibnamefont {Petersen}},
  \bibinfo {author} {\bibfnamefont {G.-Y.}\ \bibnamefont {Qin}}, \bibinfo
  {author} {\bibfnamefont {V.}~\bibnamefont {Roy}}, \ and\ \bibinfo {author}
  {\bibfnamefont {X.-N.}\ \bibnamefont {Wang}},\ }\href {\doibase
  10.1140/epja/i2016-16097-x} {\bibfield  {journal} {\bibinfo  {journal} {Eur.
  Phys. J. A}\ }\textbf {\bibinfo {volume} {52}},\ \bibinfo {pages} {97}
  (\bibinfo {year} {2016})},\ \Eprint {http://arxiv.org/abs/1511.04131}
  {arXiv:1511.04131 [nucl-th]} \BibitemShut {NoStop}%
\bibitem [{\citenamefont {Schenke}\ and\ \citenamefont
  {Schlichting}(2016)}]{Schenke:2016ksl}%
  \BibitemOpen
  \bibfield  {author} {\bibinfo {author} {\bibfnamefont {B.}~\bibnamefont
  {Schenke}}\ and\ \bibinfo {author} {\bibfnamefont {S.}~\bibnamefont
  {Schlichting}},\ }\href {\doibase 10.1103/PhysRevC.94.044907} {\bibfield
  {journal} {\bibinfo  {journal} {Phys. Rev. C}\ }\textbf {\bibinfo {volume}
  {94}},\ \bibinfo {pages} {044907} (\bibinfo {year} {2016})},\ \Eprint
  {http://arxiv.org/abs/1605.07158} {arXiv:1605.07158 [hep-ph]} \BibitemShut
  {NoStop}%
\bibitem [{\citenamefont {Ke}\ \emph {et~al.}(2017)\citenamefont {Ke},
  \citenamefont {Moreland}, \citenamefont {Bernhard},\ and\ \citenamefont
  {Bass}}]{Ke:2016jrd}%
  \BibitemOpen
  \bibfield  {author} {\bibinfo {author} {\bibfnamefont {W.}~\bibnamefont
  {Ke}}, \bibinfo {author} {\bibfnamefont {J.~S.}\ \bibnamefont {Moreland}},
  \bibinfo {author} {\bibfnamefont {J.~E.}\ \bibnamefont {Bernhard}}, \ and\
  \bibinfo {author} {\bibfnamefont {S.~A.}\ \bibnamefont {Bass}},\ }\href
  {\doibase 10.1103/PhysRevC.96.044912} {\bibfield  {journal} {\bibinfo
  {journal} {Phys. Rev.}\ }\textbf {\bibinfo {volume} {C96}},\ \bibinfo {pages}
  {044912} (\bibinfo {year} {2017})},\ \Eprint
  {http://arxiv.org/abs/1610.08490} {arXiv:1610.08490 [nucl-th]} \BibitemShut
  {NoStop}%
\bibitem [{\citenamefont {Shen}\ and\ \citenamefont
  {Schenke}(2018)}]{Shen:2017bsr}%
  \BibitemOpen
  \bibfield  {author} {\bibinfo {author} {\bibfnamefont {C.}~\bibnamefont
  {Shen}}\ and\ \bibinfo {author} {\bibfnamefont {B.}~\bibnamefont {Schenke}},\
  }\href {\doibase 10.1103/PhysRevC.97.024907} {\bibfield  {journal} {\bibinfo
  {journal} {Phys. Rev.}\ }\textbf {\bibinfo {volume} {C97}},\ \bibinfo {pages}
  {024907} (\bibinfo {year} {2018})},\ \Eprint
  {http://arxiv.org/abs/1710.00881} {arXiv:1710.00881 [nucl-th]} \BibitemShut
  {NoStop}%
\bibitem [{\citenamefont {Jia}\ \emph {et~al.}(2016)\citenamefont {Jia},
  \citenamefont {Radhakrishnan},\ and\ \citenamefont {Zhou}}]{Jia:2015jga}%
  \BibitemOpen
  \bibfield  {author} {\bibinfo {author} {\bibfnamefont {J.}~\bibnamefont
  {Jia}}, \bibinfo {author} {\bibfnamefont {S.}~\bibnamefont {Radhakrishnan}},
  \ and\ \bibinfo {author} {\bibfnamefont {M.}~\bibnamefont {Zhou}},\ }\href
  {\doibase 10.1103/PhysRevC.93.044905} {\bibfield  {journal} {\bibinfo
  {journal} {Phys. Rev. C}\ }\textbf {\bibinfo {volume} {93}},\ \bibinfo
  {pages} {044905} (\bibinfo {year} {2016})},\ \Eprint
  {http://arxiv.org/abs/1506.03496} {arXiv:1506.03496 [nucl-th]} \BibitemShut
  {NoStop}%
\bibitem [{\citenamefont {Aggarwal}\ \emph {et~al.}(2002)\citenamefont
  {Aggarwal} \emph {et~al.}}]{Aggarwal:2001aa}%
  \BibitemOpen
  \bibfield  {author} {\bibinfo {author} {\bibfnamefont {M.~M.}\ \bibnamefont
  {Aggarwal}} \emph {et~al.} (\bibinfo {collaboration} {WA98}),\ }\href
  {\doibase 10.1103/PhysRevC.65.054912} {\bibfield  {journal} {\bibinfo
  {journal} {Phys. Rev. C}\ }\textbf {\bibinfo {volume} {65}},\ \bibinfo
  {pages} {054912} (\bibinfo {year} {2002})},\ \Eprint
  {http://arxiv.org/abs/nucl-ex/0108029} {arXiv:nucl-ex/0108029 [nucl-ex]}
  \BibitemShut {NoStop}%
\bibitem [{\citenamefont {Adare}\ \emph {et~al.}(2008)\citenamefont {Adare}
  \emph {et~al.}}]{Adare:2008ns}%
  \BibitemOpen
  \bibfield  {author} {\bibinfo {author} {\bibfnamefont {A.}~\bibnamefont
  {Adare}} \emph {et~al.} (\bibinfo {collaboration} {PHENIX}),\ }\href
  {\doibase 10.1103/PhysRevC.78.044902} {\bibfield  {journal} {\bibinfo
  {journal} {Phys. Rev. C}\ }\textbf {\bibinfo {volume} {78}},\ \bibinfo
  {pages} {044902} (\bibinfo {year} {2008})},\ \Eprint
  {http://arxiv.org/abs/0805.1521} {arXiv:0805.1521 [nucl-ex]} \BibitemShut
  {NoStop}%
\bibitem [{\citenamefont {Konchakovski}\ \emph {et~al.}(2006)\citenamefont
  {Konchakovski}, \citenamefont {Haussler}, \citenamefont {Gorenstein},
  \citenamefont {Bratkovskaya}, \citenamefont {Bleicher},\ and\ \citenamefont
  {Stoecker}}]{Konchakovski:2005hq}%
  \BibitemOpen
  \bibfield  {author} {\bibinfo {author} {\bibfnamefont {V.~P.}\ \bibnamefont
  {Konchakovski}}, \bibinfo {author} {\bibfnamefont {S.}~\bibnamefont
  {Haussler}}, \bibinfo {author} {\bibfnamefont {M.~I.}\ \bibnamefont
  {Gorenstein}}, \bibinfo {author} {\bibfnamefont {E.~L.}\ \bibnamefont
  {Bratkovskaya}}, \bibinfo {author} {\bibfnamefont {M.}~\bibnamefont
  {Bleicher}}, \ and\ \bibinfo {author} {\bibfnamefont {H.}~\bibnamefont
  {Stoecker}},\ }\href {\doibase 10.1103/PhysRevC.73.034902} {\bibfield
  {journal} {\bibinfo  {journal} {Phys. Rev. C}\ }\textbf {\bibinfo {volume}
  {73}},\ \bibinfo {pages} {034902} (\bibinfo {year} {2006})},\ \Eprint
  {http://arxiv.org/abs/nucl-th/0511083} {arXiv:nucl-th/0511083 [nucl-th]}
  \BibitemShut {NoStop}%
\bibitem [{\citenamefont {Begun}\ \emph {et~al.}(2007)\citenamefont {Begun},
  \citenamefont {Gazdzicki}, \citenamefont {Gorenstein}, \citenamefont {Hauer},
  \citenamefont {Konchakovski},\ and\ \citenamefont {Lungwitz}}]{Begun:2006uu}%
  \BibitemOpen
  \bibfield  {author} {\bibinfo {author} {\bibfnamefont {V.~V.}\ \bibnamefont
  {Begun}}, \bibinfo {author} {\bibfnamefont {M.}~\bibnamefont {Gazdzicki}},
  \bibinfo {author} {\bibfnamefont {M.~I.}\ \bibnamefont {Gorenstein}},
  \bibinfo {author} {\bibfnamefont {M.}~\bibnamefont {Hauer}}, \bibinfo
  {author} {\bibfnamefont {V.~P.}\ \bibnamefont {Konchakovski}}, \ and\
  \bibinfo {author} {\bibfnamefont {B.}~\bibnamefont {Lungwitz}},\ }\href
  {\doibase 10.1103/PhysRevC.76.024902} {\bibfield  {journal} {\bibinfo
  {journal} {Phys. Rev. C}\ }\textbf {\bibinfo {volume} {76}},\ \bibinfo
  {pages} {024902} (\bibinfo {year} {2007})},\ \Eprint
  {http://arxiv.org/abs/nucl-th/0611075} {arXiv:nucl-th/0611075 [nucl-th]}
  \BibitemShut {NoStop}%
\bibitem [{\citenamefont {Aad}\ \emph {et~al.}(2014)\citenamefont {Aad} \emph
  {et~al.}}]{Aad:2014fla}%
  \BibitemOpen
  \bibfield  {author} {\bibinfo {author} {\bibfnamefont {G.}~\bibnamefont
  {Aad}} \emph {et~al.} (\bibinfo {collaboration} {ATLAS}),\ }\href {\doibase
  10.1103/PhysRevC.90.024905} {\bibfield  {journal} {\bibinfo  {journal} {Phys.
  Rev. C}\ }\textbf {\bibinfo {volume} {90}},\ \bibinfo {pages} {024905}
  (\bibinfo {year} {2014})},\ \Eprint {http://arxiv.org/abs/1403.0489}
  {arXiv:1403.0489 [hep-ex]} \BibitemShut {NoStop}%
\bibitem [{\citenamefont {Aad}\ \emph {et~al.}(2015)\citenamefont {Aad} \emph
  {et~al.}}]{Aad:2015lwa}%
  \BibitemOpen
  \bibfield  {author} {\bibinfo {author} {\bibfnamefont {G.}~\bibnamefont
  {Aad}} \emph {et~al.} (\bibinfo {collaboration} {ATLAS}),\ }\href {\doibase
  10.1103/PhysRevC.92.034903} {\bibfield  {journal} {\bibinfo  {journal} {Phys.
  Rev. C}\ }\textbf {\bibinfo {volume} {92}},\ \bibinfo {pages} {034903}
  (\bibinfo {year} {2015})},\ \Eprint {http://arxiv.org/abs/1504.01289}
  {arXiv:1504.01289 [hep-ex]} \BibitemShut {NoStop}%
\bibitem [{\citenamefont {Giacalone}\ \emph {et~al.}(2017)\citenamefont
  {Giacalone}, \citenamefont {Noronha-Hostler},\ and\ \citenamefont
  {Ollitrault}}]{Giacalone:2017uqx}%
  \BibitemOpen
  \bibfield  {author} {\bibinfo {author} {\bibfnamefont {G.}~\bibnamefont
  {Giacalone}}, \bibinfo {author} {\bibfnamefont {J.}~\bibnamefont
  {Noronha-Hostler}}, \ and\ \bibinfo {author} {\bibfnamefont {J.-Y.}\
  \bibnamefont {Ollitrault}},\ }\href {\doibase 10.1103/PhysRevC.95.054910}
  {\bibfield  {journal} {\bibinfo  {journal} {Phys. Rev. C}\ }\textbf {\bibinfo
  {volume} {95}},\ \bibinfo {pages} {054910} (\bibinfo {year} {2017})},\
  \Eprint {http://arxiv.org/abs/1702.01730} {arXiv:1702.01730 [nucl-th]}
  \BibitemShut {NoStop}%
\bibitem [{\citenamefont {Bilandzic}\ \emph {et~al.}(2014)\citenamefont
  {Bilandzic}, \citenamefont {Christensen}, \citenamefont {Gulbrandsen},
  \citenamefont {Hansen},\ and\ \citenamefont {Zhou}}]{Bilandzic:2013kga}%
  \BibitemOpen
  \bibfield  {author} {\bibinfo {author} {\bibfnamefont {A.}~\bibnamefont
  {Bilandzic}}, \bibinfo {author} {\bibfnamefont {C.~H.}\ \bibnamefont
  {Christensen}}, \bibinfo {author} {\bibfnamefont {K.}~\bibnamefont
  {Gulbrandsen}}, \bibinfo {author} {\bibfnamefont {A.}~\bibnamefont {Hansen}},
  \ and\ \bibinfo {author} {\bibfnamefont {Y.}~\bibnamefont {Zhou}},\ }\href
  {\doibase 10.1103/PhysRevC.89.064904} {\bibfield  {journal} {\bibinfo
  {journal} {Phys. Rev. C}\ }\textbf {\bibinfo {volume} {89}},\ \bibinfo
  {pages} {064904} (\bibinfo {year} {2014})},\ \Eprint
  {http://arxiv.org/abs/1312.3572} {arXiv:1312.3572 [nucl-ex]} \BibitemShut
  {NoStop}%
\bibitem [{\citenamefont {Jia}\ and\ \citenamefont
  {Mohapatra}(2013)}]{Jia:2012ma}%
  \BibitemOpen
  \bibfield  {author} {\bibinfo {author} {\bibfnamefont {J.}~\bibnamefont
  {Jia}}\ and\ \bibinfo {author} {\bibfnamefont {S.}~\bibnamefont
  {Mohapatra}},\ }\href {\doibase 10.1140/epjc/s10052-013-2510-y} {\bibfield
  {journal} {\bibinfo  {journal} {Eur. Phys. J. C}\ }\textbf {\bibinfo {volume}
  {73}},\ \bibinfo {pages} {2510} (\bibinfo {year} {2013})},\ \Eprint
  {http://arxiv.org/abs/1203.5095} {arXiv:1203.5095 [nucl-th]} \BibitemShut
  {NoStop}%
\bibitem [{\citenamefont {Huo}\ \emph {et~al.}(2014)\citenamefont {Huo},
  \citenamefont {Jia},\ and\ \citenamefont {Mohapatra}}]{Huo:2013qma}%
  \BibitemOpen
  \bibfield  {author} {\bibinfo {author} {\bibfnamefont {P.}~\bibnamefont
  {Huo}}, \bibinfo {author} {\bibfnamefont {J.}~\bibnamefont {Jia}}, \ and\
  \bibinfo {author} {\bibfnamefont {S.}~\bibnamefont {Mohapatra}},\ }\href
  {\doibase 10.1103/PhysRevC.90.024910} {\bibfield  {journal} {\bibinfo
  {journal} {Phys. Rev. C}\ }\textbf {\bibinfo {volume} {90}},\ \bibinfo
  {pages} {024910} (\bibinfo {year} {2014})},\ \Eprint
  {http://arxiv.org/abs/1311.7091} {arXiv:1311.7091 [nucl-ex]} \BibitemShut
  {NoStop}%
\bibitem [{\citenamefont {Aaboud}\ \emph
  {et~al.}(2018{\natexlab{b}})\citenamefont {Aaboud} \emph
  {et~al.}}]{Aaboud:2017tql}%
  \BibitemOpen
  \bibfield  {author} {\bibinfo {author} {\bibfnamefont {M.}~\bibnamefont
  {Aaboud}} \emph {et~al.} (\bibinfo {collaboration} {ATLAS}),\ }\href
  {\doibase 10.1140/epjc/s10052-018-5605-7} {\bibfield  {journal} {\bibinfo
  {journal} {Eur. Phys. J. C}\ }\textbf {\bibinfo {volume} {78}},\ \bibinfo
  {pages} {142} (\bibinfo {year} {2018}{\natexlab{b}})},\ \Eprint
  {http://arxiv.org/abs/1709.02301} {arXiv:1709.02301 [nucl-ex]} \BibitemShut
  {NoStop}%
\bibitem [{\citenamefont {Bzdak}\ \emph
  {et~al.}(2017{\natexlab{b}})\citenamefont {Bzdak}, \citenamefont {Koch},\
  and\ \citenamefont {Strodthoff}}]{Bzdak:2016sxg}%
  \BibitemOpen
  \bibfield  {author} {\bibinfo {author} {\bibfnamefont {A.}~\bibnamefont
  {Bzdak}}, \bibinfo {author} {\bibfnamefont {V.}~\bibnamefont {Koch}}, \ and\
  \bibinfo {author} {\bibfnamefont {N.}~\bibnamefont {Strodthoff}},\ }\href
  {\doibase 10.1103/PhysRevC.95.054906} {\bibfield  {journal} {\bibinfo
  {journal} {Phys. Rev. C}\ }\textbf {\bibinfo {volume} {95}},\ \bibinfo
  {pages} {054906} (\bibinfo {year} {2017}{\natexlab{b}})},\ \Eprint
  {http://arxiv.org/abs/1607.07375} {arXiv:1607.07375 [nucl-th]} \BibitemShut
  {NoStop}%
\bibitem [{\citenamefont {Chatterjee}\ and\ \citenamefont
  {Bozek}(2017)}]{Chatterjee:2017mhc}%
  \BibitemOpen
  \bibfield  {author} {\bibinfo {author} {\bibfnamefont {S.}~\bibnamefont
  {Chatterjee}}\ and\ \bibinfo {author} {\bibfnamefont {P.}~\bibnamefont
  {Bozek}},\ }\href {\doibase 10.1103/PhysRevC.96.014906} {\bibfield  {journal}
  {\bibinfo  {journal} {Phys. Rev. C}\ }\textbf {\bibinfo {volume} {96}},\
  \bibinfo {pages} {014906} (\bibinfo {year} {2017})},\ \Eprint
  {http://arxiv.org/abs/1704.02777} {arXiv:1704.02777 [nucl-th]} \BibitemShut
  {NoStop}%
\bibitem [{\citenamefont {Aaboud}\ \emph {et~al.}(2017)\citenamefont {Aaboud}
  \emph {et~al.}}]{Aaboud:2016jnr}%
  \BibitemOpen
  \bibfield  {author} {\bibinfo {author} {\bibfnamefont {M.}~\bibnamefont
  {Aaboud}} \emph {et~al.} (\bibinfo {collaboration} {ATLAS}),\ }\href
  {\doibase 10.1103/PhysRevC.95.064914} {\bibfield  {journal} {\bibinfo
  {journal} {Phys. Rev. C}\ }\textbf {\bibinfo {volume} {95}},\ \bibinfo
  {pages} {064914} (\bibinfo {year} {2017})},\ \Eprint
  {http://arxiv.org/abs/1606.08170} {arXiv:1606.08170 [hep-ex]} \BibitemShut
  {NoStop}%
\bibitem [{\citenamefont {Bzdak}\ \emph {et~al.}(2016)\citenamefont {Bzdak},
  \citenamefont {Holzmann},\ and\ \citenamefont {Koch}}]{Bzdak:2016qdc}%
  \BibitemOpen
  \bibfield  {author} {\bibinfo {author} {\bibfnamefont {A.}~\bibnamefont
  {Bzdak}}, \bibinfo {author} {\bibfnamefont {R.}~\bibnamefont {Holzmann}}, \
  and\ \bibinfo {author} {\bibfnamefont {V.}~\bibnamefont {Koch}},\ }\href
  {\doibase 10.1103/PhysRevC.94.064907} {\bibfield  {journal} {\bibinfo
  {journal} {Phys. Rev. C}\ }\textbf {\bibinfo {volume} {94}},\ \bibinfo
  {pages} {064907} (\bibinfo {year} {2016})},\ \Eprint
  {http://arxiv.org/abs/1603.09057} {arXiv:1603.09057 [nucl-th]} \BibitemShut
  {NoStop}%
\bibitem [{\citenamefont {He}\ and\ \citenamefont {Luo}(2018)}]{He:2018mri}%
  \BibitemOpen
  \bibfield  {author} {\bibinfo {author} {\bibfnamefont {S.}~\bibnamefont
  {He}}\ and\ \bibinfo {author} {\bibfnamefont {X.}~\bibnamefont {Luo}},\
  }\href {\doibase 10.1088/1674-1137/42/10/104001} {\bibfield  {journal}
  {\bibinfo  {journal} {Chin. Phys. C}\ }\textbf {\bibinfo {volume} {42}},\
  \bibinfo {pages} {104001} (\bibinfo {year} {2018})},\ \Eprint
  {http://arxiv.org/abs/1802.02911} {arXiv:1802.02911 [physics.data-an]}
  \BibitemShut {NoStop}%
\bibitem [{\citenamefont {Gyulassy}\ and\ \citenamefont
  {Wang}(1994)}]{Gyulassy:1994ew}%
  \BibitemOpen
  \bibfield  {author} {\bibinfo {author} {\bibfnamefont {M.}~\bibnamefont
  {Gyulassy}}\ and\ \bibinfo {author} {\bibfnamefont {X.-N.}\ \bibnamefont
  {Wang}},\ }\href {\doibase 10.1016/0010-4655(94)90057-4} {\bibfield
  {journal} {\bibinfo  {journal} {Comput.~Phys.~Commun.}\ }\textbf {\bibinfo
  {volume} {83}},\ \bibinfo {pages} {307} (\bibinfo {year} {1994})},\ \Eprint
  {http://arxiv.org/abs/nucl-th/9502021} {arXiv:nucl-th/9502021} \BibitemShut
  {NoStop}%
\bibitem [{\citenamefont {Lin}\ \emph {et~al.}(2005)\citenamefont {Lin},
  \citenamefont {Ko}, \citenamefont {Li}, \citenamefont {Zhang},\ and\
  \citenamefont {Pal}}]{Lin:2004en}%
  \BibitemOpen
  \bibfield  {author} {\bibinfo {author} {\bibfnamefont {Z.-W.}\ \bibnamefont
  {Lin}}, \bibinfo {author} {\bibfnamefont {C.~M.}\ \bibnamefont {Ko}},
  \bibinfo {author} {\bibfnamefont {B.-A.}\ \bibnamefont {Li}}, \bibinfo
  {author} {\bibfnamefont {B.}~\bibnamefont {Zhang}}, \ and\ \bibinfo {author}
  {\bibfnamefont {S.}~\bibnamefont {Pal}},\ }\href {\doibase
  10.1103/PhysRevC.72.064901} {\bibfield  {journal} {\bibinfo  {journal}
  {Phys.~Rev. C}\ }\textbf {\bibinfo {volume} {72}},\ \bibinfo {pages} {064901}
  (\bibinfo {year} {2005})},\ \Eprint {http://arxiv.org/abs/nucl-th/0411110}
  {arXiv:nucl-th/0411110 [nucl-th]} \BibitemShut {NoStop}%
\bibitem [{\citenamefont {Broniowski}\ \emph {et~al.}(2009)\citenamefont
  {Broniowski}, \citenamefont {Chojnacki},\ and\ \citenamefont
  {Obara}}]{Broniowski:2009fm}%
  \BibitemOpen
  \bibfield  {author} {\bibinfo {author} {\bibfnamefont {W.}~\bibnamefont
  {Broniowski}}, \bibinfo {author} {\bibfnamefont {M.}~\bibnamefont
  {Chojnacki}}, \ and\ \bibinfo {author} {\bibfnamefont {L.}~\bibnamefont
  {Obara}},\ }\href {\doibase 10.1103/PhysRevC.80.051902} {\bibfield  {journal}
  {\bibinfo  {journal} {Phys. Rev. C}\ }\textbf {\bibinfo {volume} {80}},\
  \bibinfo {pages} {051902} (\bibinfo {year} {2009})},\ \Eprint
  {http://arxiv.org/abs/0907.3216} {arXiv:0907.3216 [nucl-th]} \BibitemShut
  {NoStop}%
\bibitem [{\citenamefont {Bozek}(2016)}]{Bozek:2016yoj}%
  \BibitemOpen
  \bibfield  {author} {\bibinfo {author} {\bibfnamefont {P.}~\bibnamefont
  {Bozek}},\ }\href {\doibase 10.1103/PhysRevC.93.044908} {\bibfield  {journal}
  {\bibinfo  {journal} {Phys. Rev. C}\ }\textbf {\bibinfo {volume} {93}},\
  \bibinfo {pages} {044908} (\bibinfo {year} {2016})},\ \Eprint
  {http://arxiv.org/abs/1601.04513} {arXiv:1601.04513 [nucl-th]} \BibitemShut
  {NoStop}%
\bibitem [{\citenamefont {Jia}\ and\ \citenamefont {Huo}(2014)}]{Jia:2014ysa}%
  \BibitemOpen
  \bibfield  {author} {\bibinfo {author} {\bibfnamefont {J.}~\bibnamefont
  {Jia}}\ and\ \bibinfo {author} {\bibfnamefont {P.}~\bibnamefont {Huo}},\
  }\href {\doibase 10.1103/PhysRevC.90.034915} {\bibfield  {journal} {\bibinfo
  {journal} {Phys. Rev. C}\ }\textbf {\bibinfo {volume} {90}},\ \bibinfo
  {pages} {034915} (\bibinfo {year} {2014})},\ \Eprint
  {http://arxiv.org/abs/1403.6077} {arXiv:1403.6077 [nucl-th]} \BibitemShut
  {NoStop}%
\bibitem [{\citenamefont {{PHOBOS Collaboration}}\ \emph
  {et~al.}(2006)\citenamefont {{PHOBOS Collaboration}}, \citenamefont {Back}
  \emph {et~al.}}]{Back:2006id}%
  \BibitemOpen
  \bibfield  {author} {\bibinfo {author} {\bibnamefont {{PHOBOS
  Collaboration}}}, \bibinfo {author} {\bibfnamefont {B.}~\bibnamefont {Back}},
   \emph {et~al.},\ }\href {\doibase 10.1103/PhysRevC.74.011901} {\bibfield
  {journal} {\bibinfo  {journal} {Phys.~Rev.~C}\ }\textbf {\bibinfo {volume}
  {74}},\ \bibinfo {pages} {011901} (\bibinfo {year} {2006})},\ \Eprint
  {http://arxiv.org/abs/nucl-ex/0603026} {nucl-ex/0603026} \BibitemShut
  {NoStop}%
\bibitem [{\citenamefont {{STAR Collaboration}}\ \emph
  {et~al.}(2009)\citenamefont {{STAR Collaboration}}, \citenamefont {Abelev}
  \emph {et~al.}}]{Abelev:2009ag}%
  \BibitemOpen
  \bibfield  {author} {\bibinfo {author} {\bibnamefont {{STAR Collaboration}}},
  \bibinfo {author} {\bibfnamefont {B.}~\bibnamefont {Abelev}},  \emph
  {et~al.},\ }\href {\doibase 10.1103/PhysRevLett.103.172301} {\bibfield
  {journal} {\bibinfo  {journal} {Phys.~Rev.~Lett.}\ }\textbf {\bibinfo
  {volume} {103}},\ \bibinfo {pages} {172301} (\bibinfo {year} {2009})},\
  \Eprint {http://arxiv.org/abs/0905.0237} {0905.0237} \BibitemShut {NoStop}%
\bibitem [{\citenamefont {{CMS Collaboration}}(2015)}]{Khachatryan:2015oea}%
  \BibitemOpen
  \bibfield  {author} {\bibinfo {author} {\bibnamefont {{CMS Collaboration}}},\
  }\href {\doibase 10.1103/PhysRevC.92.034911} {\bibfield  {journal} {\bibinfo
  {journal} {Phys. Rev. C}\ }\textbf {\bibinfo {volume} {92}},\ \bibinfo
  {pages} {034911} (\bibinfo {year} {2015})},\ \Eprint
  {http://arxiv.org/abs/1503.01692} {arXiv:1503.01692 [nucl-ex]} \BibitemShut
  {NoStop}%
\bibitem [{\citenamefont {{STAR Collaboration, Technical Design Report for the
  iTPC Upgrade}}()}]{STAR1}%
  \BibitemOpen
  \bibfield  {author} {\bibinfo {author} {\bibnamefont {{STAR Collaboration,
  Technical Design Report for the iTPC Upgrade}}},\ }\href@noop {} {\enquote
  {\bibinfo {title}
  {\url{https://drupal.star.bnl.gov/STAR/starnotes/public/sn0644}},}\
  }\BibitemShut {NoStop}%
\bibitem [{\citenamefont {{STAR Collaboration, The STAR Forward Calorimeter
  System and Forward Tracking System}}()}]{STAR}%
  \BibitemOpen
  \bibfield  {author} {\bibinfo {author} {\bibnamefont {{STAR Collaboration,
  The STAR Forward Calorimeter System and Forward Tracking System}}},\
  }\href@noop {} {\enquote {\bibinfo {title}
  {\url{https://drupal.star.bnl.gov/STAR/starnotes/public/sn0648}},}\
  }\BibitemShut {NoStop}%
\bibitem [{\citenamefont {Adare}\ \emph {et~al.}(2015)\citenamefont {Adare}
  \emph {et~al.}}]{Adare:2015kwa}%
  \BibitemOpen
  \bibfield  {author} {\bibinfo {author} {\bibfnamefont {A.}~\bibnamefont
  {Adare}} \emph {et~al.},\ }\href@noop {} {\  (\bibinfo {year} {2015})},\
  \Eprint {http://arxiv.org/abs/1501.06197} {arXiv:1501.06197 [nucl-ex]}
  \BibitemShut {NoStop}%
\bibitem [{\citenamefont {{Workshop on the physics of HL-LHC, and perspectives
  at HE-LHC}}()}]{LHC}%
  \BibitemOpen
  \bibfield  {author} {\bibinfo {author} {\bibnamefont {{Workshop on the
  physics of HL-LHC, and perspectives at HE-LHC}}},\ }\href@noop {} {\enquote
  {\bibinfo {title} {\url{https://indico.cern.ch/event/647676/}},}\
  }\BibitemShut {NoStop}%
\bibitem [{\citenamefont {Tarafdar}\ \emph {et~al.}(2014)\citenamefont
  {Tarafdar}, \citenamefont {Citron},\ and\ \citenamefont
  {Milov}}]{Tarafdar:2014oua}%
  \BibitemOpen
  \bibfield  {author} {\bibinfo {author} {\bibfnamefont {S.}~\bibnamefont
  {Tarafdar}}, \bibinfo {author} {\bibfnamefont {Z.}~\bibnamefont {Citron}}, \
  and\ \bibinfo {author} {\bibfnamefont {A.}~\bibnamefont {Milov}},\ }\href
  {\doibase 10.1016/j.nima.2014.09.060} {\bibfield  {journal} {\bibinfo
  {journal} {Nucl. Instrum. Meth. A}\ }\textbf {\bibinfo {volume} {768}},\
  \bibinfo {pages} {170} (\bibinfo {year} {2014})},\ \Eprint
  {http://arxiv.org/abs/1405.4555} {arXiv:1405.4555 [nucl-ex]} \BibitemShut
  {NoStop}%
\bibitem [{\citenamefont {Begun}(2016)}]{Begun:2016sop}%
  \BibitemOpen
  \bibfield  {author} {\bibinfo {author} {\bibfnamefont {V.}~\bibnamefont
  {Begun}},\ }\href@noop {} {\  (\bibinfo {year} {2016})},\ \Eprint
  {http://arxiv.org/abs/1606.05358} {arXiv:1606.05358 [nucl-th]} \BibitemShut
  {NoStop}%
\bibitem [{\citenamefont {Gorenstein}\ and\ \citenamefont
  {Gazdzicki}(2011)}]{Gorenstein:2011vq}%
  \BibitemOpen
  \bibfield  {author} {\bibinfo {author} {\bibfnamefont {M.~I.}\ \bibnamefont
  {Gorenstein}}\ and\ \bibinfo {author} {\bibfnamefont {M.}~\bibnamefont
  {Gazdzicki}},\ }\href {\doibase 10.1103/PhysRevC.84.014904} {\bibfield
  {journal} {\bibinfo  {journal} {Phys. Rev. C}\ }\textbf {\bibinfo {volume}
  {84}},\ \bibinfo {pages} {014904} (\bibinfo {year} {2011})},\ \Eprint
  {http://arxiv.org/abs/1101.4865} {arXiv:1101.4865 [nucl-th]} \BibitemShut
  {NoStop}%
\bibitem [{\citenamefont {Andronov}(2018)}]{Andronov:2018bln}%
  \BibitemOpen
  \bibfield  {author} {\bibinfo {author} {\bibfnamefont {E.}~\bibnamefont
  {Andronov}} (\bibinfo {collaboration} {NA61/SHINE}),\ }\bibfield  {booktitle}
  {\emph {\bibinfo {booktitle} {{Proceedings,ICPPA 2017: Moscow, Russia,
  October 2, 2017}}},\ }\href@noop {} {\  (\bibinfo {year} {2018})},\ \Eprint
  {http://arxiv.org/abs/1801.03711} {arXiv:1801.03711 [nucl-ex]} \BibitemShut
  {NoStop}%
\bibitem [{\citenamefont {Gazdzicki}\ \emph {et~al.}(2013)\citenamefont
  {Gazdzicki}, \citenamefont {Gorenstein},\ and\ \citenamefont
  {Mackowiak-Pawlowska}}]{Gazdzicki:2013ana}%
  \BibitemOpen
  \bibfield  {author} {\bibinfo {author} {\bibfnamefont {M.}~\bibnamefont
  {Gazdzicki}}, \bibinfo {author} {\bibfnamefont {M.~I.}\ \bibnamefont
  {Gorenstein}}, \ and\ \bibinfo {author} {\bibfnamefont {M.}~\bibnamefont
  {Mackowiak-Pawlowska}},\ }\href {\doibase 10.1103/PhysRevC.88.024907}
  {\bibfield  {journal} {\bibinfo  {journal} {Phys. Rev. C}\ }\textbf {\bibinfo
  {volume} {88}},\ \bibinfo {pages} {024907} (\bibinfo {year} {2013})},\
  \Eprint {http://arxiv.org/abs/1303.0871} {arXiv:1303.0871 [nucl-th]}
  \BibitemShut {NoStop}%
\end{thebibliography}%
\bibliographystyle{apsrev4-1}

\end{document}